%% file: b-anti-b_crosssections.tex
\documentclass[%
reprint,
superscriptaddress,
titlepage,
amsmath,amssymb,
aps, prc,
longbibliography,
floatfix,
noeprint,
nopreprintnumbers
]{revtex4-2}
\usepackage{graphicx}
\usepackage{rotating}
\usepackage{wrapfig}
\usepackage{dcolumn}
\usepackage{bm}
\usepackage{physics}
\usepackage[colorlinks=true, linkcolor=blue, citecolor=blue, urlcolor=blue]{hyperref}

\usepackage{url}
\usepackage[mathlines]{lineno}
\usepackage{xcolor}
\usepackage{booktabs}
\usepackage{caption}
\usepackage{subcaption} 
\usepackage{multirow}
\usepackage{cleveref} 
\usepackage{orcidlink}

\usepackage{algorithm}
\usepackage{algpseudocode}
\usepackage{letltxmacro}
\LetLtxMacro{\oldalgorithmic}{\algorithmic}
\LetLtxMacro{\endoldalgorithmic}{\endalgorithmic}
\renewenvironment{algorithmic}[1][0]{%
  \hrulefill\par
  \oldalgorithmic[#1]}
  {\endoldalgorithmic\par
   \vspace*{-.5\baselineskip}
   \hrulefill\par
  }
\newcounter{algo}
\renewcommand{\thealgo}{\arabic{algo}}
\usepackage{placeins}

\usepackage{cancel}
\newcommand{\gx}{\textsc{GlueX} }

\newcommand{\antip}{\mbox{$\overline{p}$}}
\newcommand{\antips}{\mbox{$\overline{p}$} }
\newcommand{\antiLambda}{\mbox{$\overline{\Lambda}$}}
\newcommand{\antiLambdas}{\mbox{$\overline{\Lambda}$} }
\newcommand{\antiL}{\mbox{$\overline{\Lambda}$}}
\newcommand{\antiB}{\mbox{$\overline{B}$}}
\newcommand{\pbar}{\mbox{$\overline{p}$} }

\newcommand{\Lbar}{\mbox{$\overline{\Lambda}$} }

\newcommand{\antiCascade}{\mbox{$\overline{\Xi}$}}
\newcommand{\antiSigma}{\mbox{$\overline{\Sigma}$}}
\newcommand{\antineutron}{\mbox{$\overline{n}$}}
\newcommand{\antiN}{\mbox{$\overline{N}$}}
\newcommand{\antis}{\mbox{$\overline{s}$}}
\newcommand{\antid}{\mbox{$\overline{d}$}}
\newcommand{\antiu}{\mbox{$\overline{u}$}}
\newcommand{\tprime}{\mbox{$t^\prime$}}

\newcommand{\pbarp}{\mbox{$p\antip$}}

\newcommand{\pbarL}{\mbox{$p\antiLambda$}}
\newcommand{\LambarLam}{\mbox{$\Lambda\antiLambda$}}
\newcommand{\pantip}{\mbox{$p\antip$} }
\newcommand{\santis}{\mbox{$s\antis$} }
\newcommand{\dantid}{\mbox{$d\antid$} }
\newcommand{\pantiL}{\mbox{$p\antiLambda$} }
\newcommand{\LamantiLam}{\mbox{$\Lambda\antiLambda$} }
\newcommand{\NantiN}{\mbox{$N\antiN$} }
\newcolumntype{L}{>{\hspace{1em}}l}

\def\ra {\rightarrow}
\def\BBbar {B\bar{B}}
\def\ppbar {p\bar{p}}

\def\lamlambar {\Lambda\bar{\Lambda}}
\def\plambar {p\bar{\Lambda}}

\begin{document}


\title{Baryon antibaryon photoproduction cross sections off the proton} 


%
\input{authors}

\date{March 30, 2026}

\begin{abstract}
The \gx experiment at Jefferson Lab has observed \pantip  and, for the first time, \LamantiLam and \pantiL photoproduction from a proton target at photon energies up to 11.6~GeV. 
The angular distributions are forward peaked for all produced pairs, consistent with Regge-like $t$-channel exchange. 
Asymmetric wide-angle antibaryon distributions show the presence of additional processes.  
In a phenomenological model, we find consistency with a double $t$-channel exchange process where antibaryons are created only at the middle vertex. 
The model matches all observed distributions with a small number of free parameters. 
In the hyperon channels, we observe a clear distinction between photoproduction of the $\LamantiLam$ and $\pantiL$ systems but general similarity to the $\pantip$ system. 
We report both total cross sections and cross sections differential with respect to momentum transfer and the invariant masses of the created particle pairs.  
No narrow resonant structures were found in these reaction channels.  The suppression of $s\antis$ quark pairs relative to $d \antid$ quark pairs is similar to what has been seen in other reactions.

\end{abstract}
\maketitle

\section{Introduction}\label{sec:introduction}

We present a survey of measurements for three photoproduction reactions on a proton target that result in the creation of baryon antibaryon pairs.   They are
\begin{eqnarray}
    \gamma + p &\rightarrow \{p + \antip \} + p,
    \label{reaction1} \\     
    \gamma + p &\rightarrow \{ \Lambda + \antiLambda \} + p,
    \label{reaction2} \\    
    \gamma + p &\rightarrow \{p + \antiLambda \} + \Lambda,
    \label{reaction3}      
\end{eqnarray}
where the particles in brackets are created recoiling against the third particle. We explore the data for evidence of the reaction mechanisms that drive these processes, with the expectation that they are similar, insofar as all involve the creation of three light quark-antiquark pairs.  The latter two reactions involve creation of \santis quark pairs leading to the lightest hyperons, \LamantiLam pairs, which can then be contrasted with the nonstrange \dantid case leading to \pantip pairs.  

\begin{figure*}[tph]
    \centering
       \includegraphics[width=\textwidth]{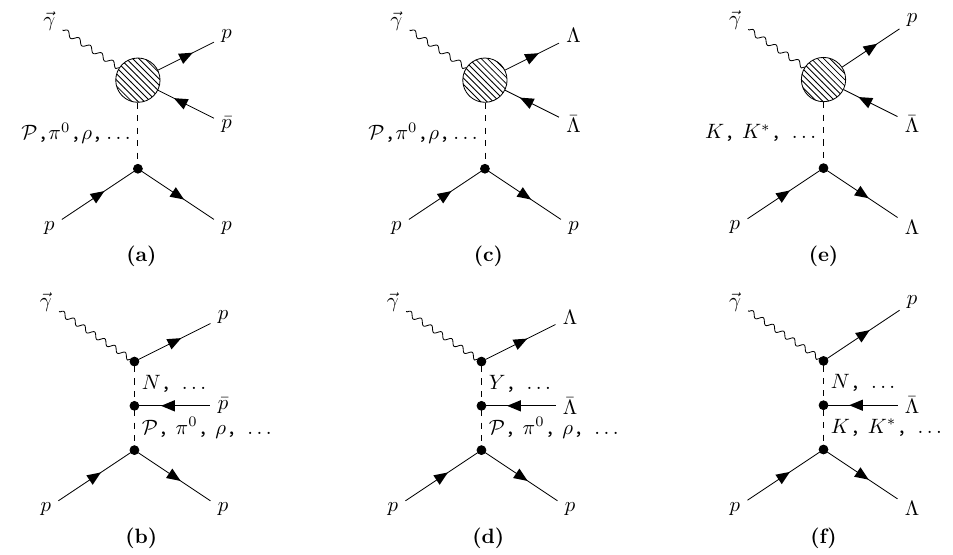}
\caption{
Possible production mechanisms for $\gamma p \rightarrow \{ \overline{p}p \} p$ in panels (a), (b);  $\gamma p \rightarrow \{ \overline{\Lambda}\Lambda \} p$ in panels (c), (d); $\gamma p \rightarrow \{p \overline{\Lambda} \} \Lambda$ in panels (e), (f). These involve either single Regge-like exchange (a), (c), (e) or double Regge-like exchange (b), (d), (f). Note the different exchange particles available in the  nonstrange (a), (c) and strange (e) single-exchange cases. 
\label{fig:feynman}}
\end{figure*}

This is the first rather complete exploration of these reactions over a wide range of kinematic space, and hence is the first to allow a phenomenological approach to decompose the data according to apparent subprocesses, as shown in Fig.~\ref{fig:feynman}.  
All three reactions will be shown to have the expected single Regge-like exchange component but also an indispensable double exchange component, which we will model in a simple phenomenological way.  

Early measurements of the exclusive \pantip photoproduction reaction on the proton were made first at SLAC in 1973~\cite{Bingham:1973fu}, followed by measurements at Daresbury, CERN and DESY~\cite{Barber:1979ah,  Aston:1980gri,  Bodenkamp:1981ui, Bodenkamp:1984dg} until 1984. There is some unpublished work from CLAS at Jefferson Lab
~\cite{bstokesPhD2006, wphelpsPhD2017} and~\cite{Phelps:2016huv}.
Previous work was mostly for beam energies below 6 GeV.  This work reports results collected with beam photon energies from 3.8 to 11.5 GeV. 

Various early measurements ~\cite{Barber:1979ah, Bodenkamp:1984dg, BENKHEIRI1977483, PhysRevLett.42.1593, BRUCKNER1987463} hunted for ``baryonium" in the near-threshold region above $2 m_\text{proton}c^2 = 1.8745$~GeV in different production and decay channels.  No such states were conclusively confirmed~\cite{MONTANET1980201}, and the present work will revisit this question briefly with much better photoproduction statistics.  


There are no full-scale theoretical calculations of these reactions available.  Therefore, the approach taken here was to create an entirely phenomenological model intended to pick out the prominent aspects of the reaction mechanism without a full quantum-mechanical treatment.

Baryon-antibaryon interaction theory has been extensively studied in work related to the \NantiN system at low energies, with a focus on the nature of the nucleon-anti-nucleon potential, final-state interactions (FSI), and possible bound states.  Historic reviews include Refs.~\cite{Amsler1991, DOVER199287, RICHARD2000361, Klempt:2002ap, PDG2024}.  
For example, single-pion exchange between nucleons is expected to lead to an attractive interaction in the $S=I=0$ channel that is stronger than the analogous interaction in the deuteron~\cite{Zou:2003zn}.  Annihilation processes are a dominant feature of low-energy \NantiN scattering, leading to a complex interaction picture approached with models of final-state interactions~\cite{Sibirtsev:2005, Kang:2015yka}.   
Recent work has developed the \NantiN potential at $N^3LO$ in effective field theory~\cite{Haidenbauer:2022baz, Dai2017ont}, leading to very good agreement with low-energy \pantip scattering phase shifts, scattering data, and spin observables.  The \LamantiLam interaction potential is closely related to the \pantip case, as developed in connection with studies of $\pantip \to \LambarLam$ at PS185 at LEAR~\cite{Barnes:2000be}, and Ref.~\cite{Haidenbauer:1993ws} (and references therein).  Recently, the latter model was exploited to investigate possible threshold and final-state interaction effects in $e^+e^- \rightarrow (meson) + \LambarLam$~\cite{Haidenbauer:2023zcu}.  It was further exploited to investigate the \pantiL potential in relation to \pantip and \LamantiLam cases~\cite{Haidenbauer:2024smo}. Correlation observables in heavy-ion data, e.g. at ~ALICE, can also probe the  dynamics between baryon-antibaryon pairs~\cite{acharya2022investigating, Sarti2025}.  We anticipate that the present results may benefit these investigations.   

Little theoretical work directly applicable to the present photoproduction study has been published.  One relevant set of exploratory calculations was presented in Refs.~\cite{Gutsche:2016wix, Gutsche:2017xtm}, in which the possible contribution of scalar mesons in the single Regge exchanges of Fig.~\ref{fig:feynman} was investigated.   The ansatz was that formation of sub-threshold glueball states might have significant contributions to baryon-antibaryon decay channels.  These calculations differ strongly from the results shown in this paper, both in magnitude and in $t$-dependence, and are not discussed further.  

We present here the experimental methods and results for cross section measurements of the reactions listed in Eqs.~\ref{reaction1}, \ref{reaction2}, \ref{reaction3}.
Preliminary results of the current work were presented at the HYP 2018 conference~\cite{Schumacher:2018xnh}, the MENU 2019 conference~\cite{Li:2019rts}, and in a Ph.D. thesis~\cite{HaoLiPhD}.

This paper is organized as follows.  
The experimental setup for these measurements at \gx and the data selection procedures are presented in Sec.~\ref{sec:setup_and_dataselection}.  The reaction model that we developed to describe the observed data is described in Sec.~\ref{sec:reactionmodel}. 
The observed kinematic distributions and correlations among particles are then shown and related to our model in Sec.~\ref{sec:data_analysis}.
The method used to fit the data to our reaction model is described in Sec.~\ref{sec:reactionmodelfit}.
Cross sections and related results are presented in Sec.~\ref{sec:resultsanddiscussion}.
Discussion of systematic uncertainties is found in Sec.~\ref{sec:systematics}.
Some further discussion and interpretation of the results are given in Sec.~\ref{sec:furtherdiscussion}.
Conclusions are recapitulated in Sec.~\ref{sec:conclusions}.
Appendixes are included to explain certain details and to tabulate selected numerical results.

\section{Experimental Setup and Data Selection}
\label{sec:setup_and_dataselection}

Measurements were performed in Hall D at the Thomas Jefferson National Accelerator Facility using the \gx spectrometer system~\cite{GlueX:2020idb} in its standard configuration. Tagged polarized photons with beam energy up to 11.6 GeV were used. The reaction thresholds were $E_\gamma = 4 m_{\text{proton}}c^2 = 3.75$ GeV for the \pantip reaction and $E_\gamma = 4.88$ GeV for the \LamantiLam  reactions.  
Coherent bremsstrahlung yielded up to $\sim 35\%$ linear polarization between beam energies of 8.2 and 8.8 GeV; observables related to this beam polarization are not included in this article.  A 30~cm long liquid hydrogen target was near the center of a solenoidal $\sim 2$~Tesla magnet.  Within the bore of the magnet were a scintillator array around the target for particle timing, a cylindrical drift chamber for charged particle tracking, and a lead/scintillator electromagnetic calorimeter (BCAL) for both photon and charged particle detection.   
Charged particles at angles more forward than 10 degrees were detected using planar drift chambers, a time-of-flight scintillator wall, and a lead-glass electromagnetic calorimeter (FCAL).   Photon detection was not used in this measurement.
The laboratory polar angle acceptance for charged particle tracking ranged from $2^{\circ}$ to $128^{\circ}$.  In the overall center-of-momentum inertial frame (CM) of the interactions the angular and azimuthal angle coverage was close to complete.  The momentum resolution for charged particles was $\delta p/p\approx 1\% - 5\%$ for reconstructed pions greater than 100 MeV/c and for protons greater than 350 MeV/c.

The results presented here came from three data collection periods, from spring 2017, spring 2018 and fall 2018, comprising the \gx Phase~I dataset.   
Running conditions across these data-taking periods left a gap in beam energy coverage from 5.5 to 6.5 GeV.   The physics trigger for the measurements consisted of a deposited-energy threshold for all charged and neutral particles in the two calorimeters.  It required either more than 1.2 GeV in the BCAL alone or a sum of at least 1.0 GeV in the BCAL plus twice the energy deposited in the FCAL.
An $e^+e^-$ pair spectrometer upstream of the \gx spectrometer was used for luminosity monitoring, and the absolute luminosity at each beam energy was verified using a lead-glass total absorption counter.  The maximum overall trigger rate was about 40 kHz, whereas the instantaneous coherent bremsstrahlung beam photon flux in the tagging hodoscope was about $2 \times 10^7 \gamma$/sec.    

Detector calibration, track reconstruction, preliminary particle identification, and acceptance calculations were carried out consistently for the entire \gx Phase~I dataset.  Events for the $p$\pantip creation channel were ``skimmed'' using a kinematic fit, enforcing energy and momentum conservation and fitting the primary reaction vertex (7C).  Every combination of hadronic tracks within one hadronic event in \gx was tested against every in-time beam photon for consistency with the reaction hypothesis. A confidence level (CL) was associated with each trial combination.

\begin{figure}[htbp]
\begin{minipage}{.492\columnwidth}
    \includegraphics[width=\textwidth]{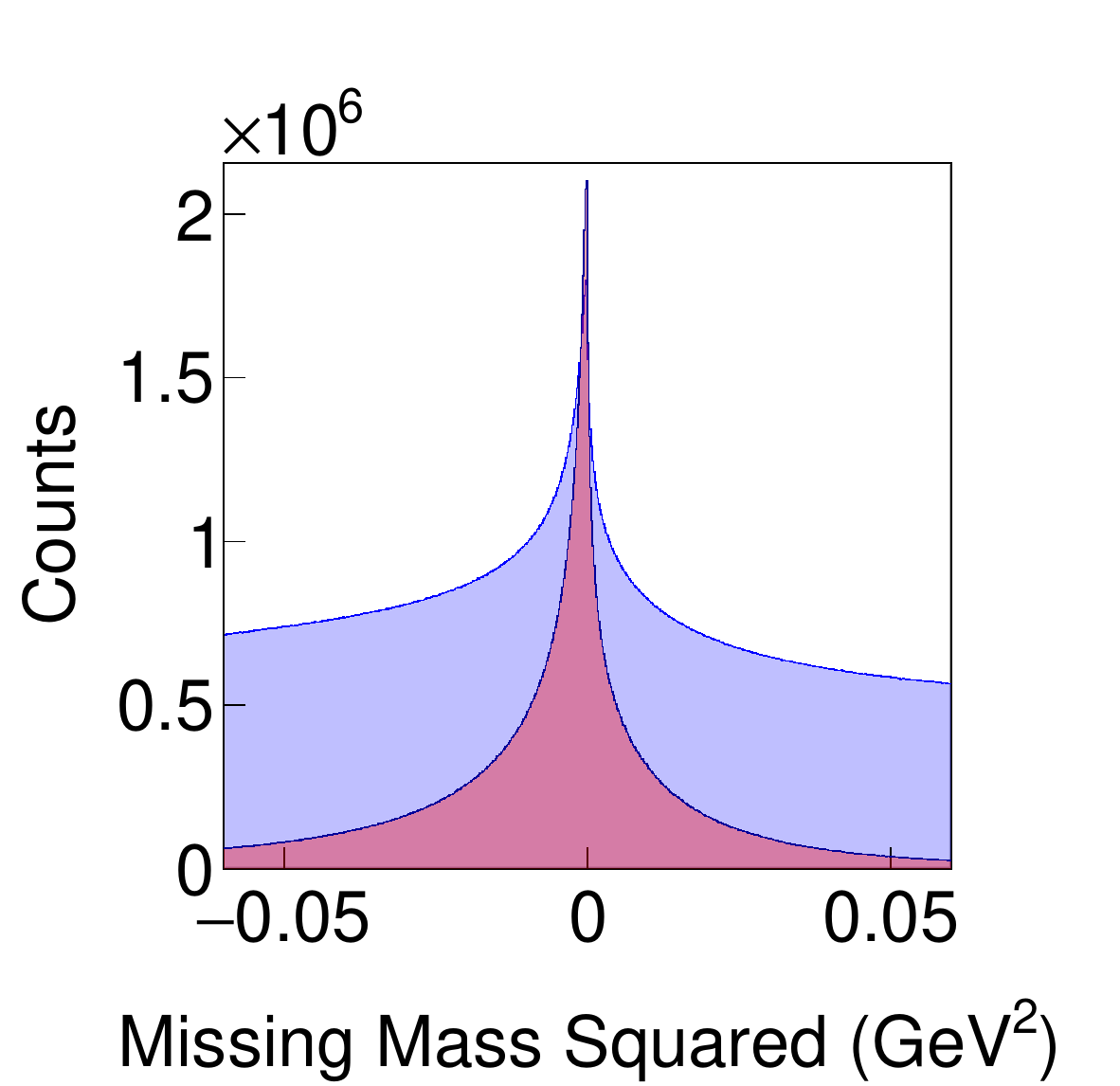}\\ (a)
\end{minipage}
\hfill
\begin{minipage}{.492\columnwidth}
    \includegraphics[width=\textwidth]{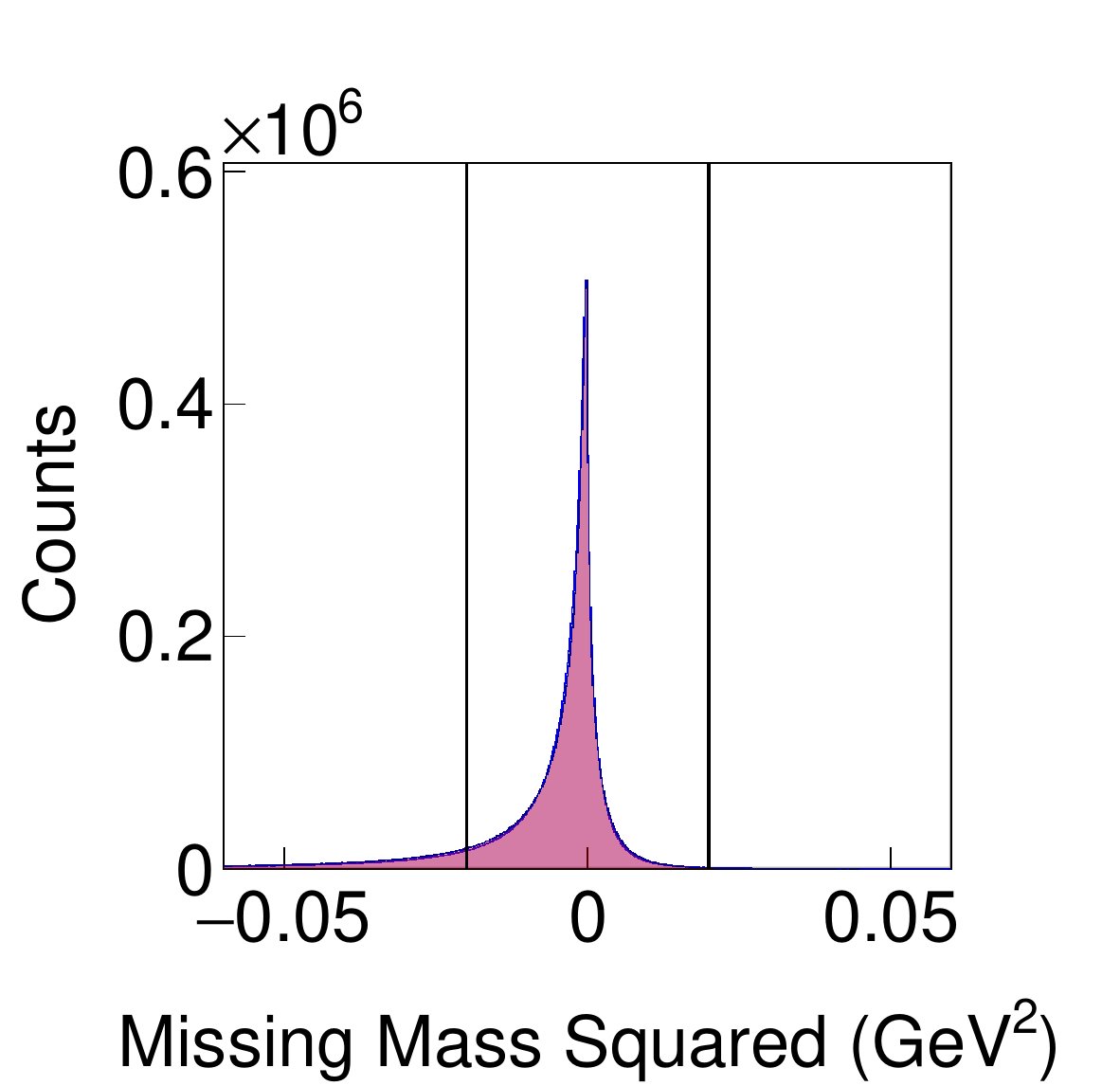}\\ (b)
\end{minipage}
\caption{\label{fig:MMS}
Missing-mass squared for kinematic fits to the  $\gamma p \rightarrow \ppbar p$ reaction. The initial skimmed data (blue) are compared with signal Monte Carlo (red) with (a) minimal cuts applied, and (b) final cuts applied, except for the selection  on the missing-mass squared itself (marked by black solid lines).} 
\end{figure}


Fig.~\ref{fig:MMS}(a) shows the missing-mass-squared (MMS) distribution of all events when no selection was applied to the confidence level of the kinematic fits, indicating background from nonexclusive events or incorrect reaction hypotheses or particle identification errors. The Monte Carlo distribution indicates what was expected for events that simulate the intended reaction, showing a sharp spike at MMS=0.  In Fig.~\ref{fig:MMS}(b) a CL selection was applied ($log_{10} CL > 10^{-5}$ in this measurement) as well as additional selections on particle time of flight and energy losses ($dE/dx$) in various detector elements.  An additional selection at $\pm.02$ GeV$^2$ was applied to trim events far from the zero of MMS.  

The momenta of charged particles were determined using drift chamber tracking in the solenoidal magnetic field.  Protons and antiprotons were unambiguously separated on the basis of the sense of their trajectory curvatures. 

The time of flight of particles between the hydrogen target and the scintillator wall located 5.4~m downstream was compared to the expected time of flight of particles of a given species.  With a timing resolution of close to 100~ps, the system could easily separate pions from protons up to about 3~GeV/c.  Small tails of pions that contaminated the proton bands up to 5~GeV/c were identified and eliminated using mild momentum-dependent timing cuts.   At even higher momenta, no remaining pion contamination was detected after all other cuts were applied.   

Protons at laboratory angles larger than $10^\circ$ entered the BCAL 0.64~m from the beam axis.  Again, a time-of-flight method was used to separate pions from protons, and no remaining pion contamination at higher momenta was found after all other cuts were applied.  Pions with momenta less than 1.0 GeV/c were separated from protons using the energy deposited in the thin timing scintillators that surrounded the liquid hydrogen target cell. 

When selecting events with hyperons, the data were skimmed for exclusive events containing the hyperon decay products for $\Lambda \rightarrow \pi^-p$ and $\antiL \rightarrow \pi^+ \antip$ in addition to a final state proton. 
The kinematic fit included additional  constraints on the detached vertices of the weak decay products (9C). The reconstructed $\Lambda(\bar{\Lambda})$ yield was corrected based on the branching fraction $\mathcal{B}(\Lambda\rightarrow \pi^{-}p) = (64.1\pm0.5)\%$ from the PDG \cite{ParticleDataGroup:2024cfk}.    

\begin{figure}[htbp]
\centering
\begin{minipage}{.485\columnwidth}
    \includegraphics[width=\textwidth]{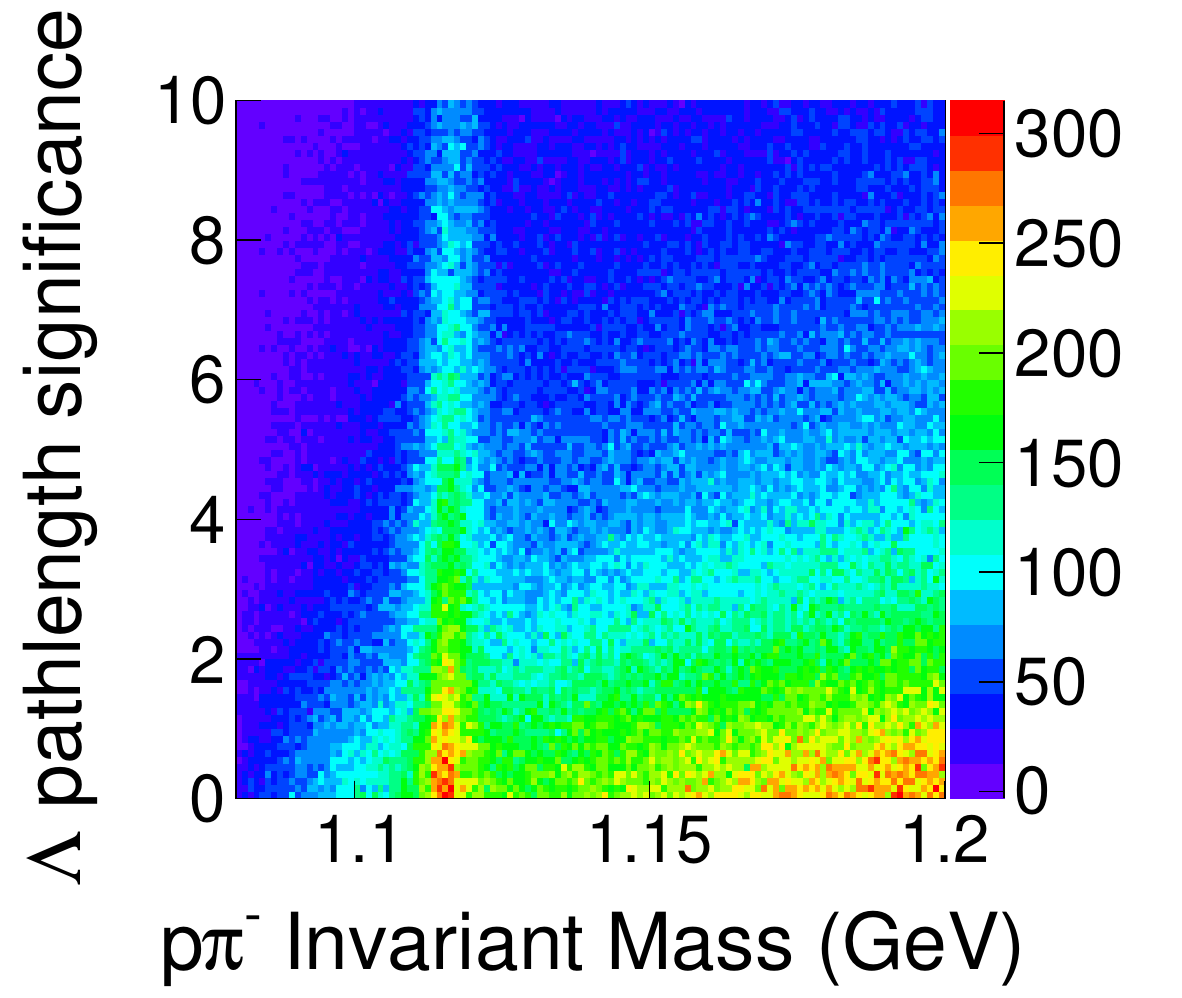}\\ (a)
\end{minipage}
\hfill
\begin{minipage}{.485\columnwidth}
    \includegraphics[width=\textwidth]{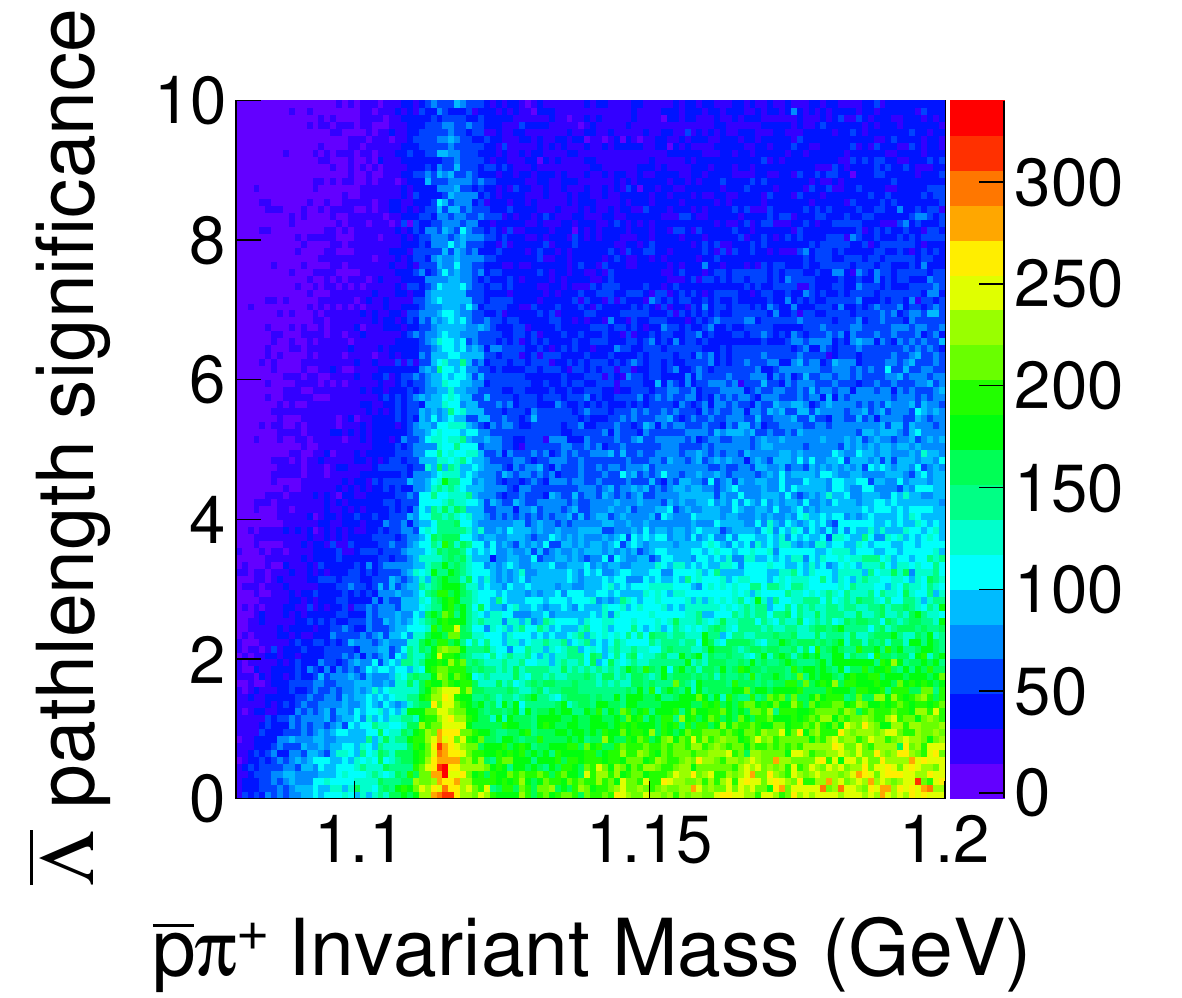}\\ (b)
\end{minipage}
\caption{
Correlation between the hyperon decay path-length significance (PLS) and the invariant mass distributions for (a) $\Lambda$  and (b) $\bar{\Lambda}$ particles.
The masses were unconstrained in the  $\gamma p \rightarrow \LamantiLam p$  kinematic fits.
\label{fig:lam_lambar_massVSpls}
} 
\end{figure}

The masses of the hyperons were not kinematically constrained to enable a good separation of nonstrange backgrounds from strangeness-containing events. 
The main background was strong decay events with $\{p  \pbar \pi^+ \pi^- \} p $ final states.  The spatial separation of the weak decays from the photoproduction vertex was exploited to reject the more copious nonstrange backgrounds.   A typical hyperon with laboratory momentum of 1~GeV/c has a mean decay distance of close to 7~cm, whereas the overall vertex resolution of the event reconstruction was about 2~cm.   The most effective selection criterion was found to be the ``path-length significance" (PLS) of the reconstructed hyperons.  This was defined as the laboratory path length of a hyperon between the primary vertex and its decay point divided by the uncertainty on the distance computed within the kinematic fit. This is illustrated in Fig.~\ref{fig:lam_lambar_massVSpls}, before any other selection cuts were applied.  Most of the background events had a small decay PLS  of about 1 or 2 units. To eliminate the nonstrange event background, we applied a selection at PLS~$> 5$.  

\begin{figure}[htb]
\centering
\begin{minipage}{.485\columnwidth}
    \centering
    \includegraphics[width=\linewidth]{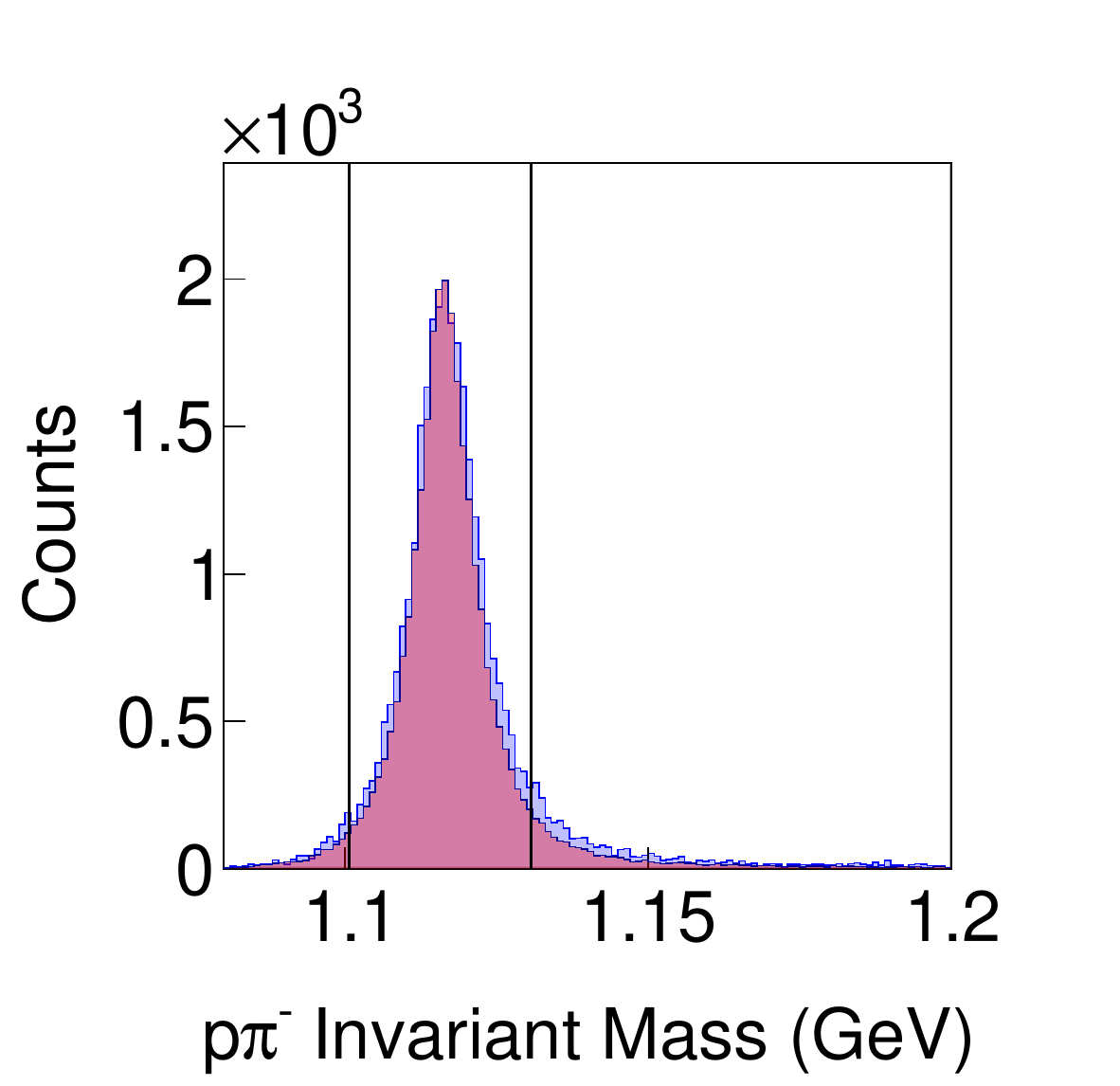}
    \\ \text{(a)}
    \label{fig:lam_mass}
\end{minipage}
\hfill
\begin{minipage}{.485\columnwidth}
    \centering
    \includegraphics[width=\linewidth]{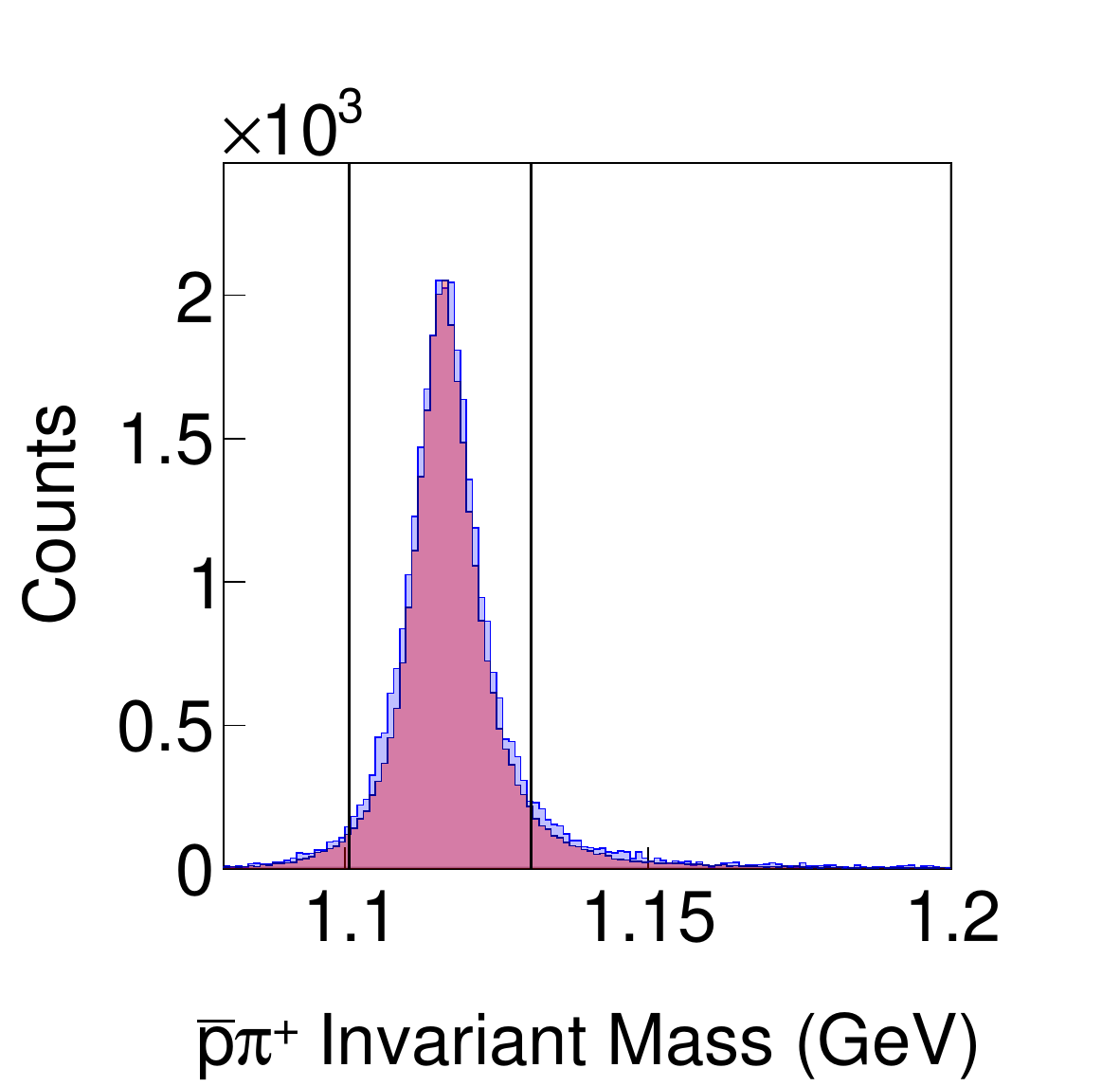}
    \\ \text{(b)}  
    \label{fig:lambar_mass}
\end{minipage}
\medskip 
\begin{minipage}{.485\columnwidth}
    \centering
    \includegraphics[width=\linewidth]{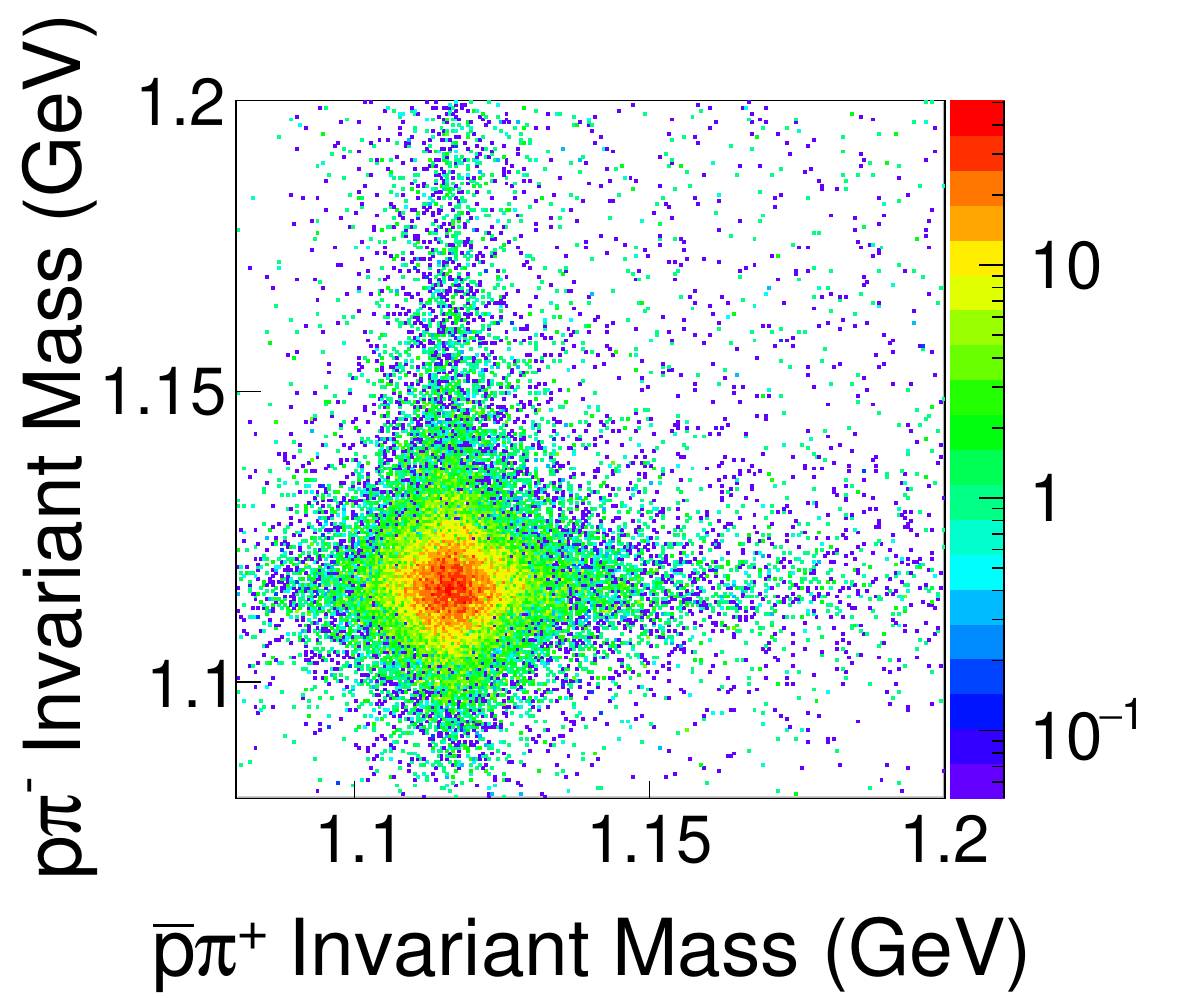}
    \\ \text{(c)}  
    \label{fig:data_mass}
\end{minipage}
\hfill
\begin{minipage}{.485\columnwidth}
    \centering
    \includegraphics[width=\linewidth]{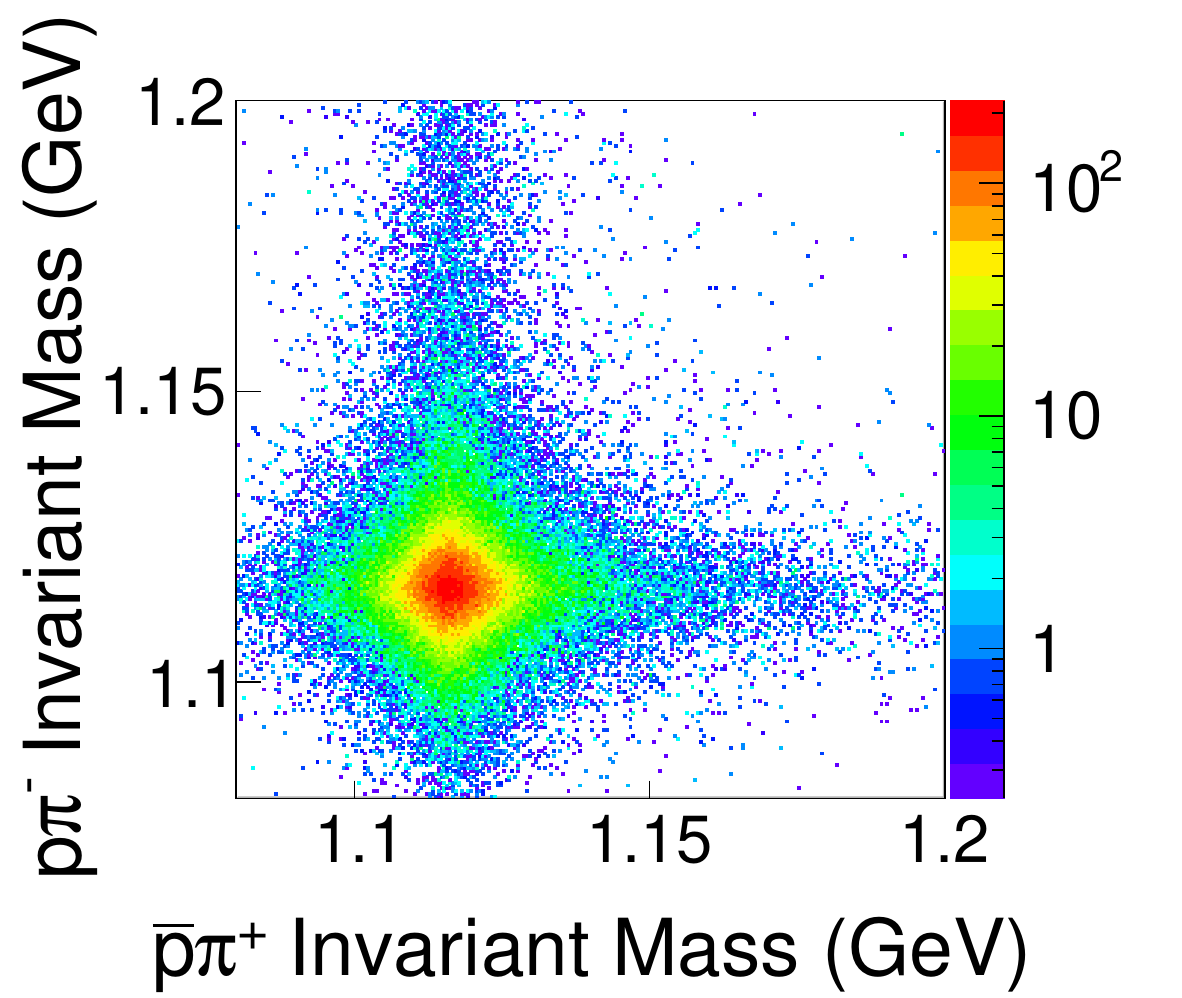}
    \\ \text{(d)}  
    \label{fig:mc_mass}
\end{minipage}
\caption{\label{fig:lam_lambar_mass_fullcut}
Top row: data (blue) is compared to signal Monte Carlo simulation (red). (a) Invariant mass of the reconstructed $\Lambda$, (b) Invariant mass of the reconstructed $\bar{\Lambda}$, (c) the correlation of two hyperon masses from data, (d) the correlation of two hyperon masses from MC simulation. The distributions are shown after all selections except for the selection on the invariant mass itself (denoted by the black solid lines).}
\end{figure}

Other selections for the hyperon channels were similar to the \pantip channel, as discussed above.   After these selections were made, the hyperon mass distributions appeared as shown in Fig.~\ref{fig:lam_lambar_mass_fullcut}, with scant background events surviving.  The correlation between the two mass distributions also confirms that no extra nonexclusive background remained. The comparison between data and signal Monte Carlo also shows good agreement. 

The horizontal and vertical ``tails'' in Fig.~\ref{fig:lam_lambar_mass_fullcut}(c) were thought to arise from events in which one of the four tracks ($\pi^-$, $\pi^+$, $p$, $\bar{p}$) suffered a slight deflection, causing its momentum to be slightly mismeasured. As seen in Fig.~\ref{fig:lam_lambar_mass_fullcut}(d) the Monte Carlo simulation also showed this effect.
To eliminate events far away from the nominal mass values, we applied a $\pm .015$~(GeV/c$^2$) mass selection window around the hyperon peak distributions.   No additional background subtraction using invariant mass distributions was necessary.

Fiducial cuts were applied to all tracks to maintain the match between data and simulations of data.   All selected tracks had laboratory angles greater than $1^\circ$. All selected proton and antiproton tracks had momenta greater than 350~MeV/c, and all pion tracks had momenta greater than 100~MeV/c.  In this analysis, after all selections, the dataset included about 10 million $p \antip p$ events and 0.4 million $\Lambda \antiL p$ events.

\section{Reaction Model}
\label{sec:reactionmodel}

We introduce here the dynamical picture that describes the data observed in this experiment.  The intent is to reproduce and explain the observed kinematic distributions with physically motivated intensity functions that allow an accurate estimation of the many-body acceptance of the apparatus. Driven by experimental data, we focus on building an intensity-based model to kinematically identify regions dominated by specific reaction dynamics.  

The reaction components are introduced using the diagrams in Fig.~\ref{fig:feynman}.   
Figures~\ref{fig:feynman}(a), 1(c), and 1(e) illustrate the Regge-like\cite{Collins_1977} single particle $t$-channel exchanges that may be expected to be dominant at \gx beam energies and low momentum transfer $-t$.  
We construe the hatched upper vertices to couple an exchanged Reggeon (a meson or Pomeron)  to the beam photon and the baryon-antibaryon pair directly.  The internal two-body correlation between the created baryon-antibaryon pair is decoupled from the external kinematics of the momentum transfer in the exchange. Furthermore, we assume isotropic angular distributions in the rest frame of the baryon-antibaryon pairs, without specifying any particular spin structure. Note that the case of strange meson ($K$, $K^*$) exchange in Fig.~\ref{fig:feynman}(e), with a recoiling $\Lambda$ hyperon, will show similarities in production mechanism when contrasted with the nonstrange exchanges in Figs.~\ref{fig:feynman}(a) and 1(c).

Figs.~\ref{fig:feynman}(b), 1(d), and 1(f) illustrate the possibility of double-exchange processes \cite{Collins_1977, byckling1973particle, PhysRevD.91.034007} that involve baryon exchanges in the upper leg and meson exchanges on the lower leg. 
The postulated double-exchange diagrams all have the {\it anti}-baryon produced at the middle vertex, which couples two exchanged states to one final-state antibaryon. The necessity for this will be shown in the discussion of what is observed in the data.  A satisfactory explanation for the lack of corresponding diagrams with the {\it baryon} produced at the middle vertex will remain unexplained in this study of experimental findings: there is a difference in the production mechanism between baryons and antibaryons.  The symmetry of producing antibaryons at the top or the middle vertices is not seen.

Baryon exchange ($u$ channel) was initially believed to also play a role in the reaction mechanism~\cite{Li:2019rts}, but subsequent modeling found it to be inconsistent with the data.   That is, the distributions of pairs produced via a $u$-channel  interaction peak symmetrically at backward CM angles (minimum $u$) at \gx energies, contrary to what the experiment shows (cf. Sec.~\ref{sec:data_analysis}).

\begin{figure}[hbt]
    \centering 
    \includegraphics[width=0.95\columnwidth]{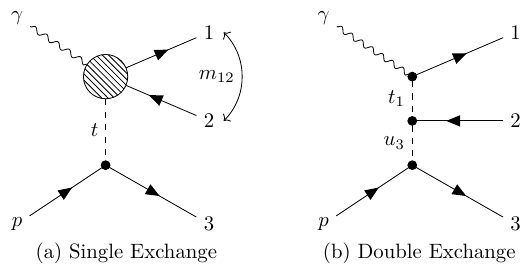}
    \caption{\label{fig:ppbar_combined_model}Diagrams of the (a) single $t$-channel exchange and (b) double $t$-channel exchange model formulated for the reaction $\gamma + p \ra 1+2 +3 $. Particle 2 is the antibaryon in each case.} 
    \end{figure}

The reaction channels were modeled on two categories of single and double exchanges shown in Fig.~\ref{fig:ppbar_combined_model}. For photoproduction off a proton target,   
$  \gamma + p \rightarrow 1+2+3,  $
we construe final-state particles 1 and 2 to be the produced baryon-antibaryon pair recoiling against particle 3.  We take the overall invariant energy of the reaction squared to be $s = W^2 =(p_\gamma + p_p)^2$.


For the dominant single $t$-channel exchange component in Fig.~\ref{fig:ppbar_combined_model}(a), the Mandelstam invariants are:
\begin{align*}
    m_{12}^2 = s_{12} &= (p_1 + p_2)^2, \\
    t = (p_\gamma - p_1-p_2)^2 &= (p_p - p_3)^2,
\end{align*}
where $m_{12}$ is the invariant mass of the produced pair and $t$ is the squared 4-momentum transfer to the target proton. 

For the double-$t$-channel exchange diagram in Fig.~\ref{fig:ppbar_combined_model}(b), the invariants are 4-momentum transfers at the top and bottom vertices:
\begin{equation*}
        t_1 = (p_{\gamma}-p_1)^2, \hspace{10pt}
        u_3 = (p_{p}-p_3)^2.
\end{equation*}

\subsection{Cross section formulation}\label{sec:crosssection}

    The fully differential cross section in terms of the invariant amplitude $\mathcal{M}_{fi}$ for this reaction can be written as \cite{chung2008spin}:
    \begin{equation}
        d\sigma = \frac{1}{4\mathcal{F}} \|\mathcal{M}_{fi}\|^2 d\Phi, \label{eq:dsigma}
    \end{equation}
    where $\mathcal{F}=p_i\sqrt{s}$ is the flux factor,
    and $d\Phi$ is the three-body phase space element, whose form depends on the choice of factorization.

    For the single $t$-channel exchange process depicted in Fig.~\ref{fig:ppbar_combined_model} (a), we factorize the three-body phase space $\Phi$ into a sequence of two-body processes.  The first process is described in the overall CM frame defined by the incident $\gamma$ and the target proton:
        $\gamma (p_i) +  p (-p_i) \rightarrow M_{12} (p_f) + 3 (-p_f)$,  
    where $p_i$ and $p_f$ are the initial and final state momenta in the CM frame.  The secondary two-body decay is described in the rest frame of the $M_{12}$ system, 
        $M_{12}\rightarrow 1 (q^*) + 2 (-q^*)$,
    where $q^*$ is the two-body decay momentum in the $M_{12}$ rest frame, using the asterisk notation to distinguish it from the overall CM frame. 
    The three-body phase space element can then be obtained:
    \begin{equation}
        d\Phi = \frac{4}{(4\pi)^5}\frac{p_f}{\sqrt{s}}d\Omega_3 dm_{12} \left\{ q^*  d\Omega^* \right\}. \label{eq:dphi3}
    \end{equation}
    Here, $d\Omega_3$ is the CM-frame solid angle differential of the $M_{12}$ system recoiling against particle ``3", and $m_{12}$ is the invariant mass of the system $M_{12}$. The quantities in brackets are evaluated in the $M_{12}$ rest system, where $d\Omega^*$ is the solid angle element of the two-body decay. The expressions for all the above-mentioned momenta are given explicitly in  Appendix~\ref{appendix:kinematics}.

    It is useful to note the overall three-body phase space factor,
    \begin{equation}
        \text{Phsp}(m_{12}) = \frac{p_f q^*}{p_i s} \label{eq:phsp}
    \end{equation} 
    that arises from $m_{12}$-dependent factors in Eqs.~\ref{eq:dsigma} and~\ref{eq:dphi3}, as shown in Eq.~\ref{eq:phsp_dsigma_dm12} (Appendix~\ref{appendix:kinematics}). It characterizes the ``model-independent" kinematics in the cross section distributions, arising solely from momentum and energy conservation. This is an important reference point, since departures observed in the data from this ``benchmark" may signal evidence of underlying dynamics. 
    
    The four-momentum transfer to the target proton in the $t$-channel is best characterized at \gx energies by the reduced four-momentum transfer squared,  
     \begin{align}
        t^{\prime} = t - t_{\text{min}} = 2p_i p_f(\cos\theta -1). \label{eq:reduced_t}
     \end{align}
     In photoproduction, the momentum transfer at scattering angle $\theta \rightarrow 0$, is
     $t_{\text{min}} = m_{12}^2- 2p_i\sqrt{m_{12}^2+p_f^2} + 2p_ip_f$, 
     which is subtracted from $t$. 
     The variable $t^\prime$ facilitates modeling of the three-body kinematics in these reactions, as discussed in Appendix~\ref{appendix:kinematics}.
     
    Using Eq.~\ref{eq:reduced_t}, we express $d\Omega_3$ as $d(\cos\theta)d\phi$ or $dt^{\prime}d\phi/(2p_i p_f)$, using $\phi$ as the azimuthal angle of the CM frame.
    Taking the reaction as uniform in $\phi$, integrating Eq.~\ref{eq:dsigma} over it gives the differential cross section for a single exchange in $m_{12}$ and $t^{\prime}$ as:
     \begin{align}
        \frac{d^2\sigma_{\text{single}}}{dt^{\prime}dm_{12}}
        =\frac{1}{(4\pi)^4}\frac{q^*}{4 s p_i^2} \int \|\mathcal{M}_{fi}(t^{\prime}, m_{12})\|^2 d\Omega^*. \label{eq:dsigma_dtprimedm}
    \end{align}
    In this analysis, the general invariant amplitude $\mathcal{M}_{fi}$ for the $2\rightarrow3$ process is formulated phenomenologically. We provide an economical description of its $m_{12}$ and $t$-dependent intensity using two key distributions introduced below.

   The double $t$-channel exchange diagram in Fig.~\ref{fig:ppbar_combined_model}(b) is treated as a double diffraction process. Its kinematic landscape is assumed to be shaped predominantly by the 4-momentum transfer $t_1$ at the top vertex.  The second 4-momentum transfer $u_3$ at the bottom vertex is kinematically constrained by the choice made for the top vertex. This leads to a factorization of the three-body phase space as $\gamma p \rightarrow 1 + M_{23}$ followed by $M_{23}\rightarrow 2 + 3$. The resulting phase space element whose closed form $\kappa_{double}(t_1, u_3)$, integrated over the full range of $s_{23}=(p_2+p_3)^2$, is derived in \cite{byckling1973particle}. In contrast to the isotropic decay $M_{12}\rightarrow 1 + 2$ in the single exchange process, here the two-body system $M_{23}$ is modeled as the final state of a diffraction in which the top exchange particle scatters off the target proton. We use the reduced 4-momentum transfer squared $t_1^{\prime} = t_1-t_{1min}$ and $ u_3^{\prime}=u_3-u_{3min}$  to characterize the differential cross section, including the invariant amplitude squared formulated with its exponential dependency on both 4-momentum transfers.    

    \subsection{t-dependence}\label{sec:t_dependence}
     We assume a dominant linear Regge-like trajectory that uses $t^\prime$ as the relevant dynamical variable,
        $ \alpha(t^{\prime}) = \alpha^0 + \alpha^{\prime}t^{\prime}$.
     In the high-energy limit ($s \gg t$) the $t$ distribution is expected to show exponential behavior \cite{Collins_1977}.
     For single $t$-channel exchange in Fig.~\ref{fig:ppbar_combined_model},
     we consider a parametrization with a piecewise function in $t^{\prime}$ of the form
          \begin{align}
          \mathcal{R}(t^{\prime})\equiv
            (1 - H_{\Delta}) \frac{t^{\prime}}{t_{\text{cutoff}}} +  H_{\Delta}  \exp\left(-2\alpha^{\prime}\ln\left(\frac{s}{s_0}\right) \Delta \right). \label{eq:reduced_t_modeling}
         \end{align}
     $H_{\Delta} = \mathbf{1}_{\Delta\geq0}$ is the Heaviside step function with $\Delta = t_{\text{cutoff}}-t^{\prime}$. For the exponential part of the formulation, $\alpha^{\prime}$ is the slope parameter, and the scale parameter $s_0$ is set to 1.0~GeV \cite{donnachie2000exclusive}\cite{PhysRevD.98.034020}. 
     The constant Regge intercept parameter, $\alpha_0$, is unused to simplify the parametrization, as it is only associated with the $s$-dependency of the cross section at energies much higher than those of the \gx experiment. 
     The linear increase in $t^\prime$ with the ``cutoff" parameter, $t_{\text{cutoff}}$, is introduced to provide the flexibility to account for the observed rise from small values in the acceptance-corrected \pantip cross sections.  
     The free parameters in this part of model are $\alpha^{\prime}$ and $t_{\text{cutoff}}$.

    \subsection{Mass clustering}

\begin{figure}[hbtp!]
      \centerline{\includegraphics[width=0.95\columnwidth]{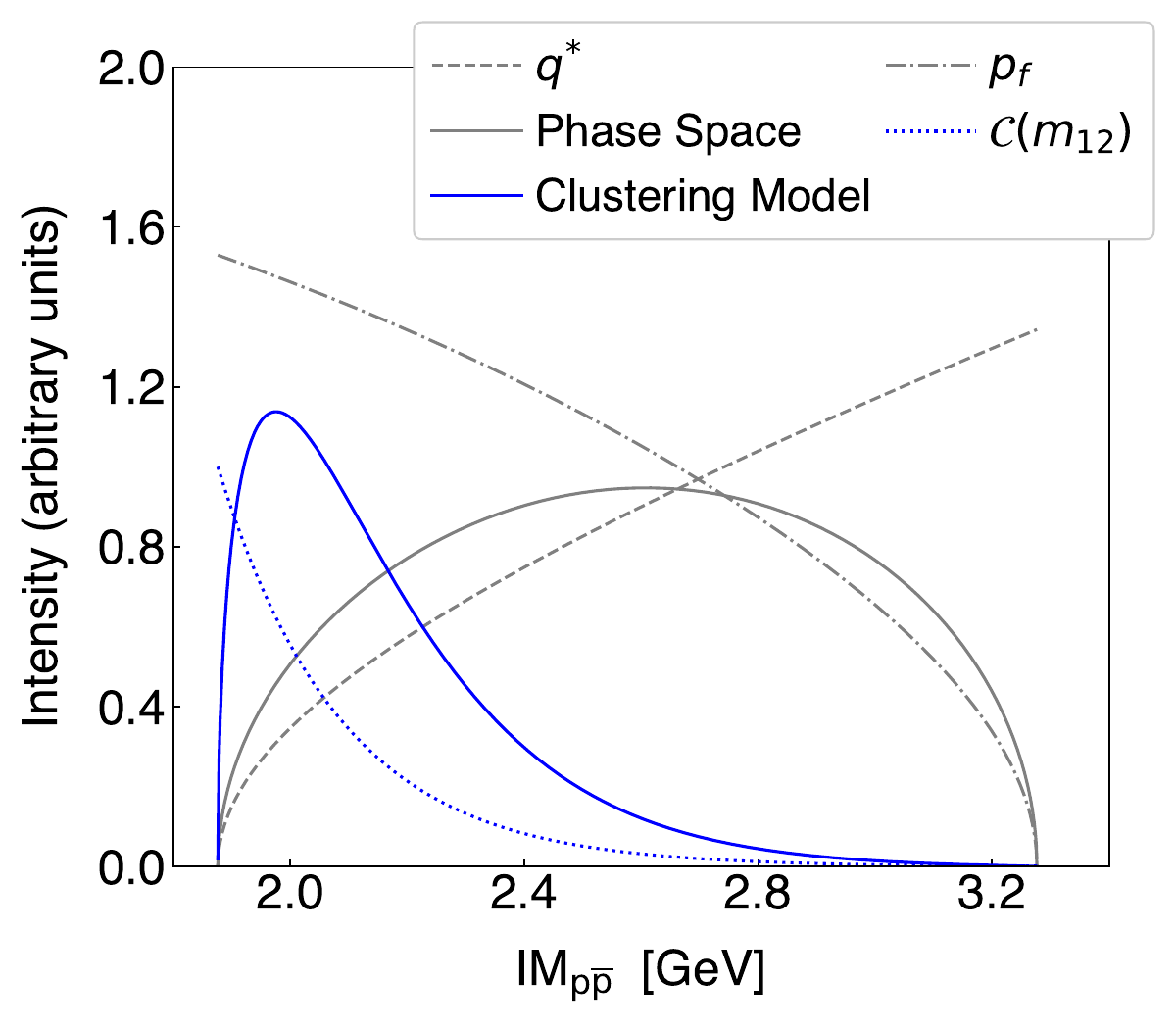}}
      \caption{\label{fig:q_pcm_vsIM} Intensities (in arbitrary units) of components of the 3-body kinematics for the reaction $\gamma p \rightarrow \{p\bar{p}\} p$ as a function of $m_{12}$  at $E_{\gamma} = 9.0$~GeV: the two-body decay momentum $q^*$ (gray dashed line) in the $\ppbar$ pair rest frame;  the two-body breakup momentum $p_{f}$ (grey dot-dashed line) between the recoil proton and $\ppbar$ system in the overall CM frame;
      and the full three-body phase space $\text{Phsp}(m_{12})$ (gray solid line)  (Eq.~\ref{eq:phsp}). The phase space is modified by an exponential mass profile introduced as $\mathcal{C}(m_{12})$ (blue dotted curve)(Eq.~\ref{eq:clustering_modeling}), leading to the differential cross section line shape of the ``clustering" model (blue solid curve)  given in Eq.~\ref{eq:dsigma_dm12}.}
    \end{figure}

    It will be seen in the data that the created pairs tend to cluster toward low invariant masses.   
    The attractive dynamic of the created baryon-antibaryon pair in the single-exchange picture~Fig.~\ref{fig:ppbar_combined_model}(a) is assumed to be independent of the rest of the reaction kinematics. This leads to the hypothesis that the invariant mass distribution of the system, after integrating over other kinematic variables, 
    will exhibit enhancement at low $m_{12}$.
    We introduce a ``clustering'' model with the mass profile
        \begin{equation}
            \mathcal{C}(m_{12}; c_m) \equiv \exp\left[-(m_{12} -m_{12}^{\min})/c_m\right],\label{eq:clustering_modeling}
        \end{equation}
    to characterize the produced pairs that cluster toward their mass threshold $m_{12}^{\min}$ (Eq.~\ref{eq:m_lowerbound}) more than determined by phase space only.  At fixed $s$, the three-body final state phase space is characterized by both $p_f$ (near threshold) and $q^*$ (near the upper bound of $m_{12}$). Illustrated in Fig.~\ref{fig:q_pcm_vsIM} for the $\ppbar$ system,
    the exponentially distributed mass profile of the baryon-antibaryon system invariant mass, 
     $\mathcal{C}(m_{12};c_m)$, combines with the phase space factor $\text{Phsp}(m_{12})$ to sculpt the overall invariant mass distributions, particularly in the low $m_{12}$ region. The free parameter in this part of the model is $c_m$.

    
\subsection{Full reaction model}
\label{sec:fullreactionmodels}
\subsubsection{Single $t$-channel exchange}
     The baryon-antibaryon creation step illustrated as the shaded blob in Fig.~\ref{fig:ppbar_combined_model}(a) is taken as independent of $t$ in our model of the single-exchange process. To describe the differential cross section in a factorizable manner, we cast the matrix element in Eq.~\ref{eq:dsigma_dtprimedm} in a form that includes the mass clustering factor $\mathcal{C}(m_{12})$ and the parametrization $\mathcal{R}(t^{\prime})$:
    \begin{equation}
        \|\mathcal{M}_{fi}(t^{\prime}, m_{12})\|^2 = \mathcal{N}_{\BBbar}
                \times \kappa(m_{12})
                \times \mathcal{C}(m_{12}) 
                \times \mathcal{R}(t^{\prime}),
        \label{eq:amp_squared}
    \end{equation}
    where $\mathcal{N}_{\BBbar}$ is the global scaling associated with the creation of the baryon($B$)-antibaryon($\bar{B}$) pair.
    The factor
    \begin{equation}
       \kappa(m_{12})=\left( \dfrac{ \int_{t^{\prime}_-(m_{12})}^{t^{\prime}_+} dt^{\prime} }{ \int_{t^{\prime}_-(m_{12})}^{t^{\prime}_+} \mathcal{R}(t^{\prime})\, dt^{\prime} } \right),\label{eq:kappa} 
    \end{equation}
    is introduced to normalize the $t$-channel $\mathcal{R}(t^{\prime})$ dependence, to preserve the three-body phase space (Eq.~\ref{eq:phsp}) when different effective masses of $m_{12}$ are combined into a $2 \rightarrow 2$ single exchange process.  The integration limits are given in Appendix~\ref{appendix:kinematics}. 

    We thus write the differential cross section for the single-exchange intensity as an explicit function of $m_{12}$ and $t^{\prime}$: 
    \begin{equation}    
    \begin{split}
        \frac{d^2\sigma_{\text{single}}}{dm_{12} dt^{\prime}}
        &= \frac{\mathcal{N}_{\BBbar}}{(4\pi)^3} 
        \times \left( \frac{\kappa \: q^*}{4 s \: p_i^2}\right) \times 
        \exp\left[-\frac{m_{12} -m_{12}^-}{c_m}\right] \\
        &\times \left[ (1 - H_{\Delta}) \frac{t^{\prime}}{t_{\text{cutoff}}} \right.\\ 
        &\left. + H_{\Delta}  \exp\left(-2\alpha^{\prime}\ln\left(\frac{s}{s_0}\right) \Delta \right)\right].
        \label{eq:sR_modeling}
    \end{split}  
    \raisetag{18pt}
    \end{equation}     
    
    \noindent
    When integrating $d^2\sigma_{\text{single}} / (dm_{12} dt^{\prime})$ over $t^{\prime}$, the introduction of $\mathcal{R}(t^{\prime})$ modifies the phase space dependence on $m_{12}$. The expression for $\kappa(m_{12})$  restores the correct three-body phase space (Eq.~\ref{eq:phsp}) behavior:
    \begin{equation}
        \frac{d\sigma_{\text{single}}}{dm_{12}} 
        =
        \frac{\mathcal{N}_{\BBbar}}{(4\pi)^3}  \times \text{Phsp}(m_{12}) \times\mathcal{C}(m_{12}) \label{eq:dsigma_dm12},
    \end{equation}
as illustrated in Fig.~\ref{fig:q_pcm_vsIM}.
      
\subsubsection{Double $t$-channel exchange}
The fully differential cross section of the double exchange  diagram in Fig.~\ref{fig:ppbar_combined_model}(b) may be written as a function of reduced four-momentum transfers $t_1^{\prime}$ and $u_3^{\prime}$, each dominated by a linear trajectory, $ \alpha_1(t_1^{\prime}) = \alpha_1^0 + \alpha_1^{\prime}t_1^{\prime}$ and $ \alpha_3(u_3^{\prime}) = \alpha_3^0 + \alpha_3^{\prime}u_3^{\prime}$, respectively:

    \begin{equation}    
    \begin{split}
     \frac{d^2\sigma_{\text{double}}}{dt_1^{\prime} du_3^{\prime}}
    &= 
    \frac{\kappa_{\text{double}}(t_1^{\prime}, u_3^{\prime})}{(4\pi)^3}\times 
    \left[ (1 - H_{\Delta_1})\frac{t_1^{\prime}}{t_{1~ \text{cutoff}}} \right.\\
    &\left. +  H_{\Delta_1} \exp(-2\alpha_1^{\prime}\ln\left(\frac{s}{s_0}\right) \Delta_1) \right] \\
    &\times 
    \exp[-2\alpha_3^{\prime}\ln\left(\frac{s}{s_0}\right)u_3^{\prime}],
    \label{eq:dR_modeling}
    \end{split}
    \raisetag{-18pt}
    \end{equation}

\noindent where the Heaviside step function $H_{\Delta_1} = \mathbf{1}_{\Delta_1\geq0}$ is similar to the single exchange case, with $\Delta_1 = t_{1~\text{cutoff}}-t_1^{\prime}$.  The three-body phase space factor $\kappa_{\text{double}}(t_1^{\prime}, u_3^{\prime})$, introduced in Sec.~\ref{sec:crosssection},  was implicitly taken into account in the Monte Carlo event generator (Appendix~\ref{appendix:optimization}).
The invariant masses of both baryon-antibaryon combinations in the final state, namely $s_{12}$ and $s_{23}$, are entirely determined by the invariants $(t_1^{\prime}, u_3^{\prime})$, and hence no further modeling is needed to describe them.

    \subsection{Simplified representation of the total cross section}
    \label{sec:totalcrosssectiontheory}   
    \vspace{2.0em}
    A simplified representation of the total cross section of the baryon-antibaryon photoproduced off the proton can be constructed from  Eq.~\ref{eq:sR_modeling} in the single exchange model.
    It can be obtained by further integrating Eq.~\ref{eq:dsigma_dm12} over the full $m_{12}$ range, yielding a compact two-parameter function of $s$:
    \begin{equation}
    \begin{split} 
        &\sigma_{\text{total}}(s; \mathcal{N}_{\BBbar}, c_{\text{total}})=\\
        &\frac{\mathcal{N}_{\BBbar}}{(4\pi)^3}
        \int^{m_{12}^{\max}(s)}_{m_{12}^{\min}}   
        \text{Phsp}(s, m_{12}) 
        ~ \mathcal{C}(m_{12}; c_{\text{total}}) ~ dm_{12}, \label{eq:totalcrosssectionmodel}  
    \end{split}
    \end{equation}
    where the functions $\text{Phsp}(s, m_{12})$ is the three-body phase space factor (Eq.~\ref{eq:phsp}) and $\mathcal{C}(m_{12}; c_{\text{total}})$ is the exponential mass profile (Eq.~\ref{eq:clustering_modeling}) from the clustering model. The lower and upper bounds of the integral are given in Eqs.~\ref{eq:m_lowerbound} and \ref{eq:m_bound}. 
    This total cross section model exhibits no sensitivity to the values of the parameters $\alpha^{\prime}$ and $t_{\text{cutoff}}$ due to the normalization introduced in Eq.~\ref{eq:kappa}. 
    This two-parameter model will be used to compare with the measured total cross section in Sec.~\ref{sec:totalcrosssection}. 

    We do not include a formulation of the double exchange $\sigma_{\text{double}}$ in this simplified model for the total cross section.  Doing so would involve integration over kinematics with cumbersome and complex dependencies on the exchange parameters. In the simplified representation, used only for the total cross section, the double exchange effects are ignored and subsumed into these two fitted parameters $\mathcal{N}_{\BBbar}$ and $c_{\text{total}}$. For this reason, the $c_{\text{total}}$ for the total cross section is not comparable to the $c_{m}$ values for the full model.
    
\section{Kinematic Correlations and Reaction Mechanisms}
\label{sec:data_analysis}

\subsection{Single-particle angular distributions}
\label{sec:single_particle_correlations}
The particle angular distributions in the overall CM frame reveal much about the reaction mechanisms.  The \gx data together with our model-based Monte Carlo (discussed in Sec.~\ref{sec:reactionmodelfit}) for the \pantip reaction are shown in Fig.~\ref{fig:ppbar_angular}, where $\cos \theta^{\text{CM}}$ refers to the polar CM angle of the named particle.   These distributions are summed over all beam energies and are not acceptance corrected.   In a pure phase space situation, acceptance-corrected distributions would each be perfectly flat.  
We have chosen to identify the proton that moves more backward, labeled $p_{bkwd}$ in the figure, as the probable recoiling target proton, while the proton that moves more forward, labeled $p_{fwd}$, is likely part of the created \pantip pair.

The strong forward peaking of one proton and of the antiproton is expected from a diffractive reaction mechanism such as illustrated by the single exchange diagram in Fig.~\ref{fig:feynman}(a), and similarly in (c) and (e), together with the backward peaking of the recoil proton plotted in Fig.~\ref{fig:ppbar_angular}(c).   
However, as seen in the middle panel Fig.~\ref{fig:ppbar_angular}(b), the $\bar{p}$ has a wider angular distribution in addition to the dominant forward peak.  We will describe this feature using a double-exchange mechanism. Our Monte Carlo simulation of this effect was optimized to match the data quite well, as explained in Sec.~\ref{sec:reactionmodelfit}. 
It should be noted that the double exchange mechanism accounts for the broad anti-particle distributions by placing the created {\it anti}-particle in the ``middle'' of the double-exchange diagram, rather than the created {\it particle}.   

In the $\gamma p \rightarrow \Lambda  \antiLambda p$  reaction, we found that the two distinct cases \LamantiLam and $p\antiLambda$ can be cleanly separated using information about the multi-particle correlations in the reaction, as discussed in Sec.~\ref{sec:vanhove}.  The two cases are shown in Figs.~\ref{fig:lamlambar_angular} and \ref{fig:plambar_angular}. These reactions have never been seen or studied in previous experiments.  The three particles are unique, so no ``forward" and ``backward" distinctions are needed, unlike in the \pantip channel.

Figure~\ref{fig:lamlambar_angular} shows the angular distributions for the events of the $\gamma p \rightarrow  \{ {\overline{\Lambda} \Lambda}\}  p$ reaction, summed over the full range of the photon beam energies and not corrected for acceptance.  Fig.~\ref{fig:lamlambar_angular}(a) shows the polar angular distribution of $\Lambda$ hyperons in the overall CM frame.  The forward hyperon peak is analogous to the proton reaction forward peak in Fig.~\ref{fig:ppbar_angular}(a), which can be explained using the single exchange mechanism. Fig.~\ref{fig:lamlambar_angular}(c) shows the proton angular distribution, exhibiting a backward recoiling proton due to single-exchange hyperon pair production, analogous to the \pantip reaction.  Fig.~\ref{fig:lamlambar_angular}(b), shows the distribution of \antiLambda~particles. They are mainly forward peaked, understood as due to the single exchange mechanism, but they also exhibit a broad tail to large angles, analogous to the \pantip case.
The unexpected presence of significant antibaryon production across all CM angles is again seen in this hyperon reaction, and we explain this on the basis of the same double exchange ansatz in our reaction model.   

\newcommand{\figwidth}{0.31\textwidth}
\begin{figure*}[htpb]
		\centering
		  \begin{minipage}[b]{\figwidth}
		    \includegraphics[width=\textwidth]{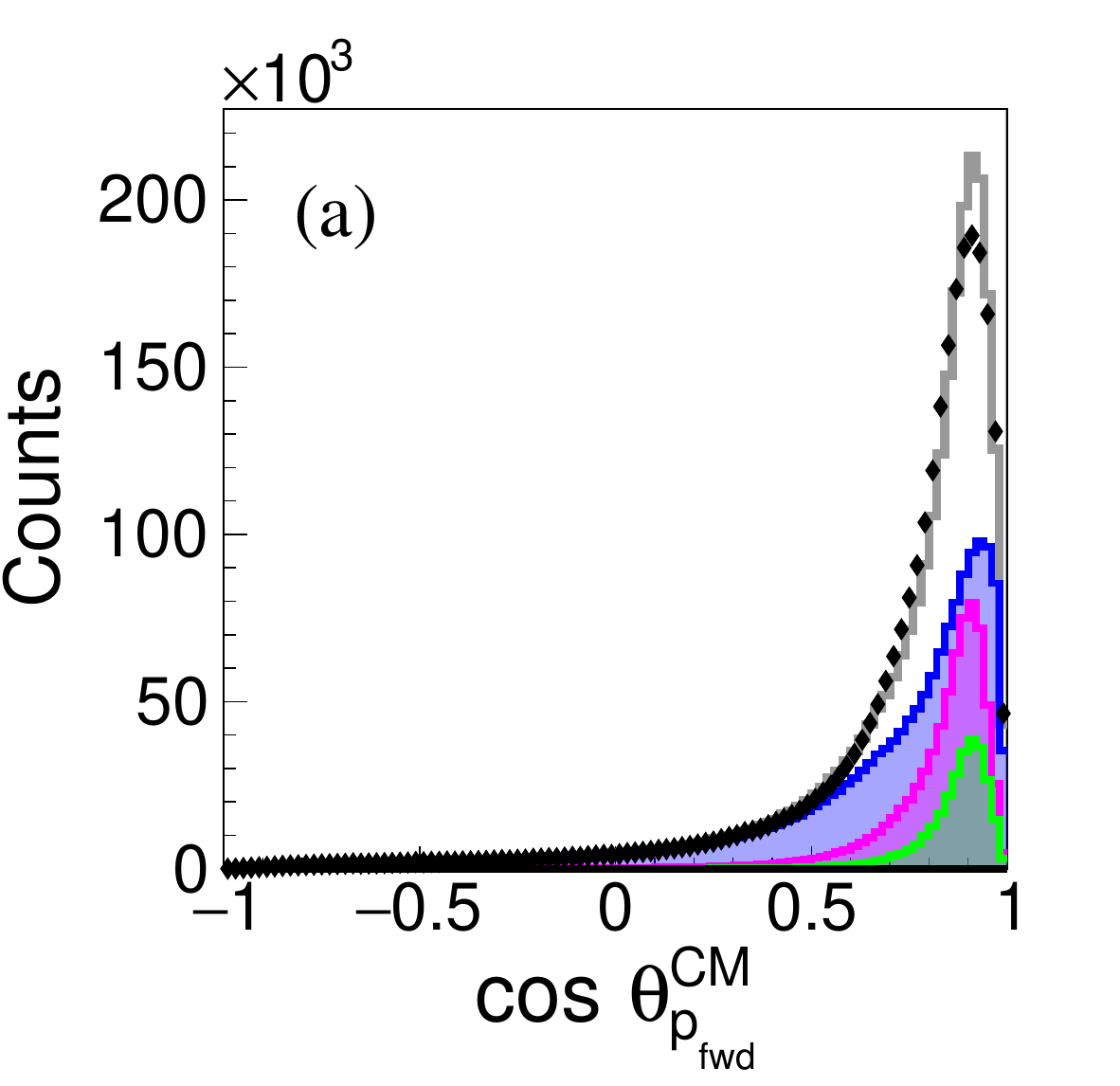} 
		  \end{minipage}
		  \begin{minipage}[b]{\figwidth}
		    \includegraphics[width=\textwidth]{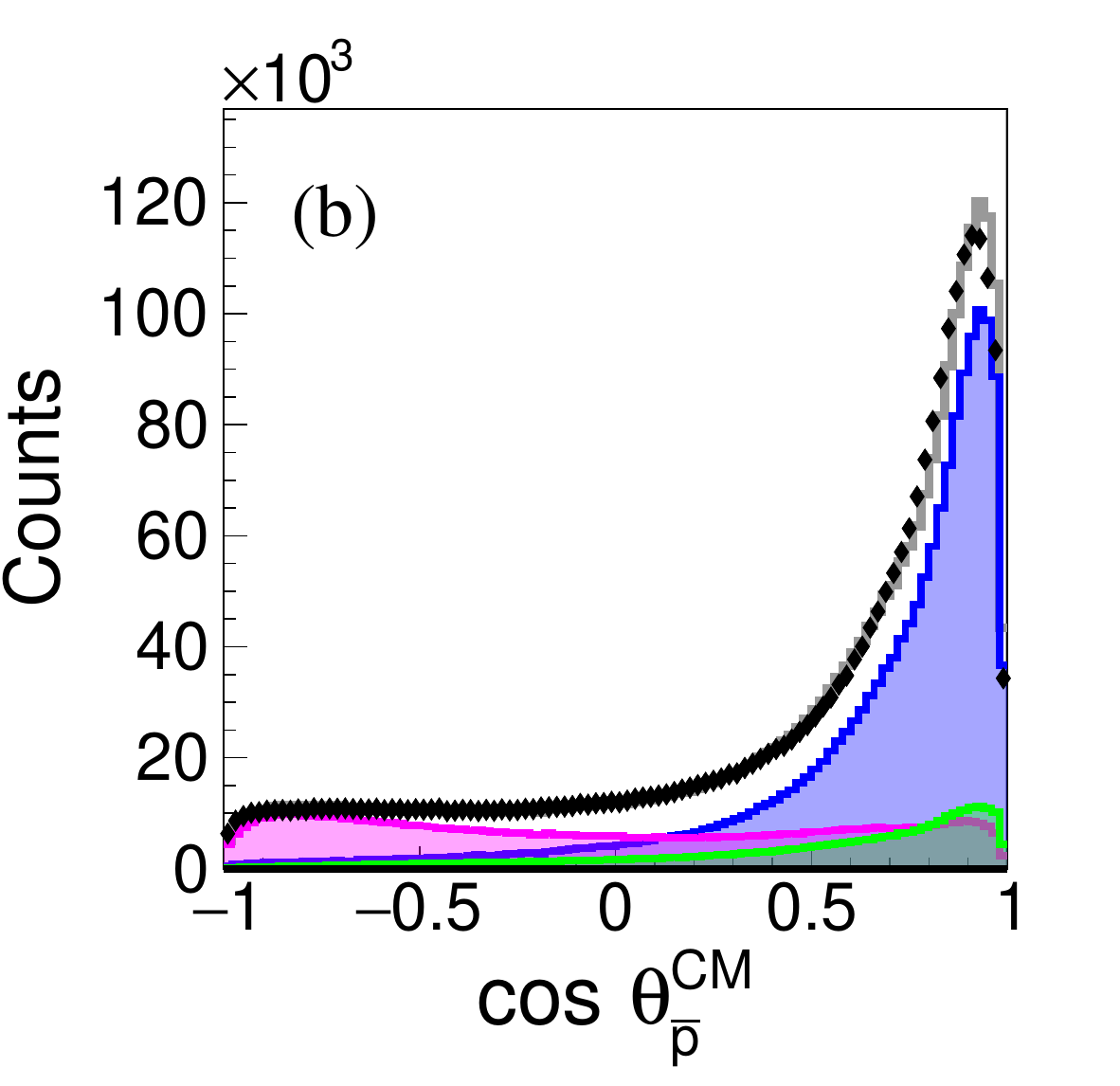} 
		  \end{minipage}
		  \begin{minipage}[b]{\figwidth}
		    \includegraphics[width=\textwidth]{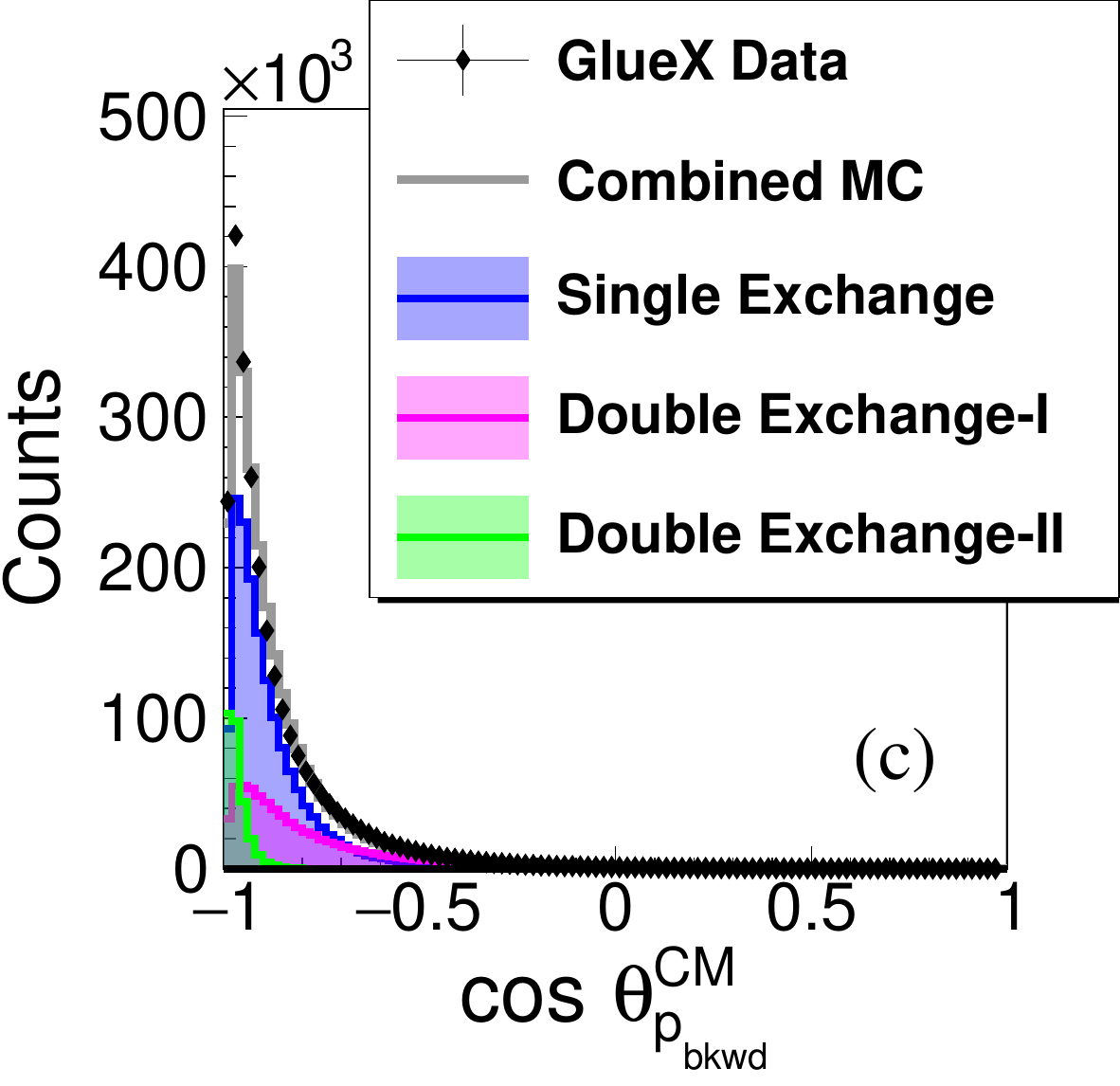} 
		  \end{minipage}
\caption{\label{fig:ppbar_angular} 
Angular distributions of the two protons (a), (c), and the antiproton (b) in the $\gamma p \rightarrow p \antip p $ reaction in the overall CM frame. Not acceptance corrected. The data are shown as black points and the summed simulation fit as a gray histogram.  Components of the Monte Carlo simulations are shown as: Single Exchange (blue filled), Double Exchange I  (magenta filled), Double Exchange II (green filled).    }
\end{figure*}

\begin{figure*}[htpb]
	\centering
	   \begin{minipage}[b]{\figwidth}
            \includegraphics[width=\textwidth]{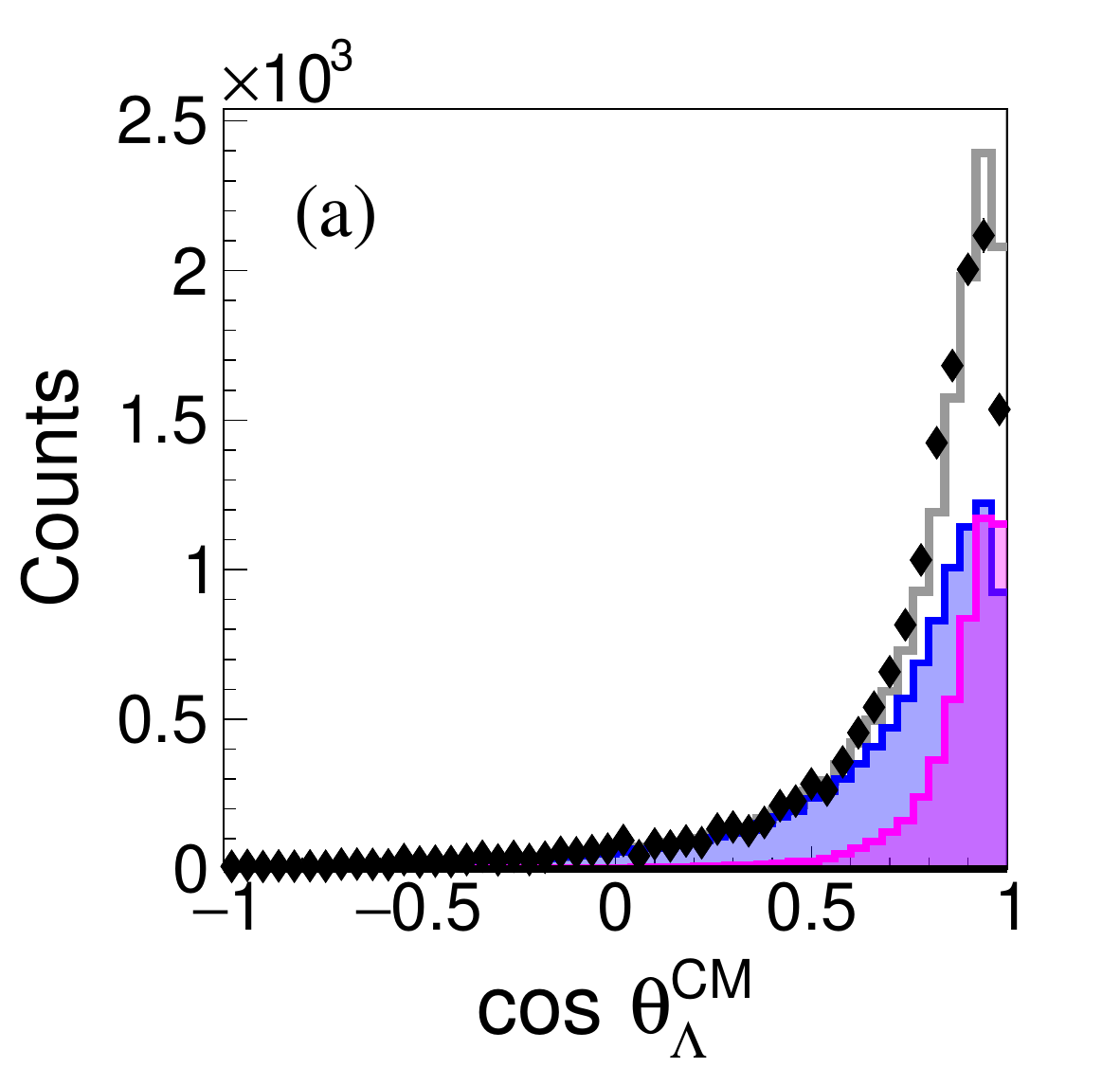}
	\end{minipage}
	\begin{minipage}[b]{\figwidth}
    		 \includegraphics[width=\textwidth]{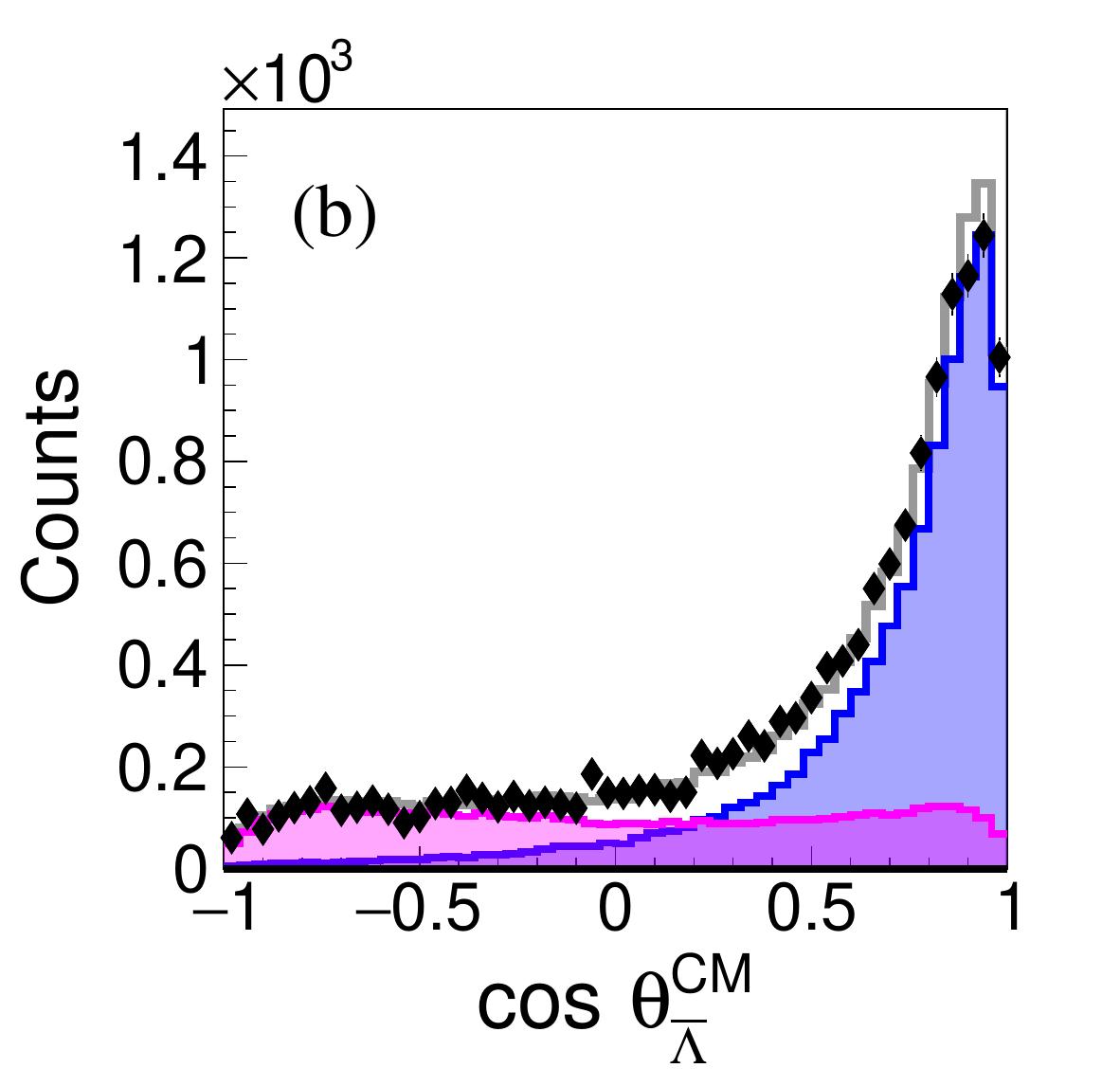}
	\end{minipage}
	\begin{minipage}[b]{\figwidth}
    		 \includegraphics[width=\textwidth]{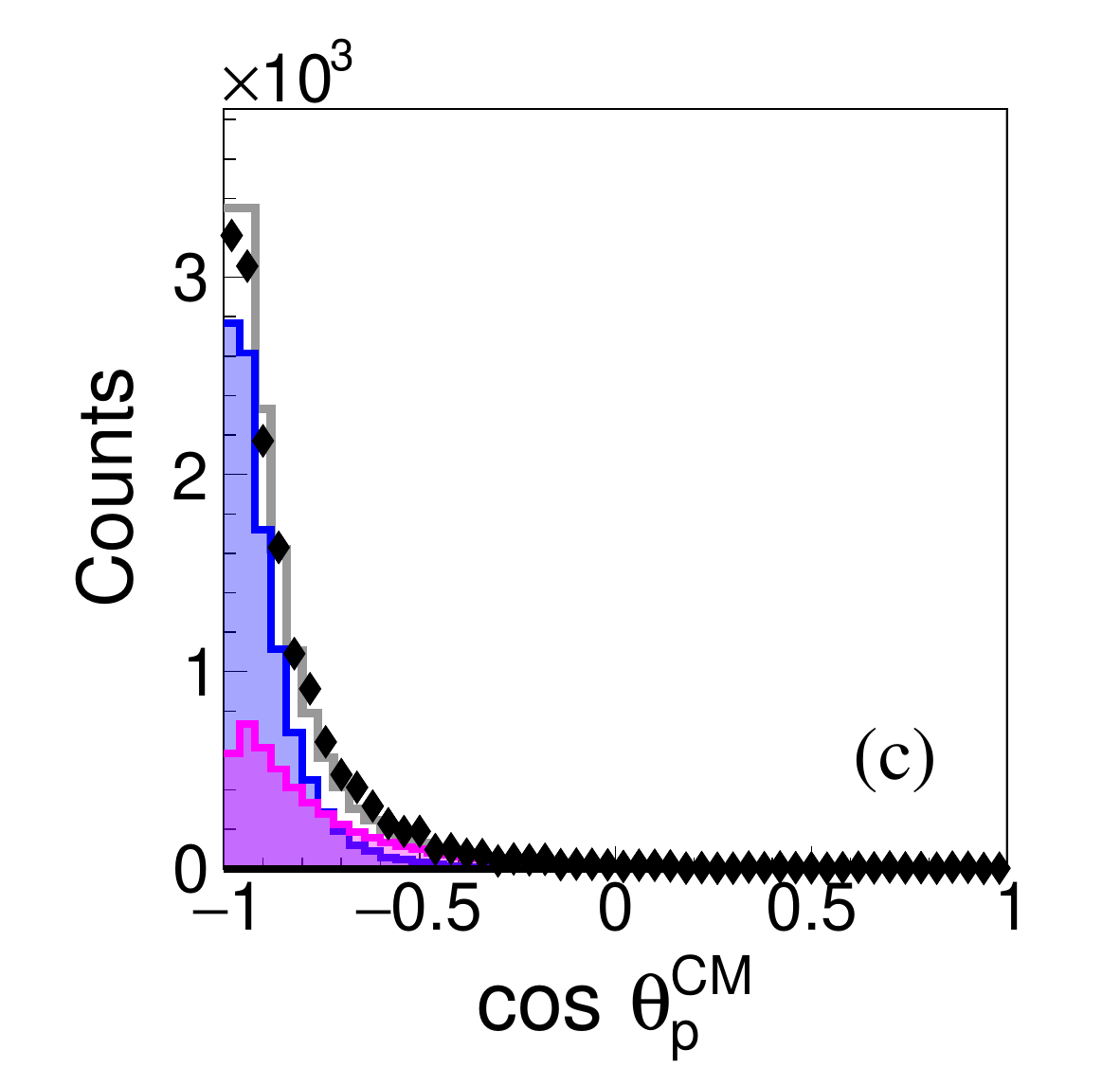}
	\end{minipage}
\caption{\label{fig:lamlambar_angular} Angular distributions of the three particles in the $\gamma p \rightarrow \{ \Lambda \bar{\Lambda} \} p $ reaction channel in the overall CM system. The points and curves are organized as in Fig.~\ref{fig:ppbar_angular}.  Unlike the proton reaction shown in Fig.~\ref{fig:ppbar_angular}, only one Double Exchange component is needed in the hyperon cases (magenta filled).}
\end{figure*}

\begin{figure*}[htpb]
	\centering
	   \begin{minipage}[b]{\figwidth}
             \includegraphics[width=\textwidth]{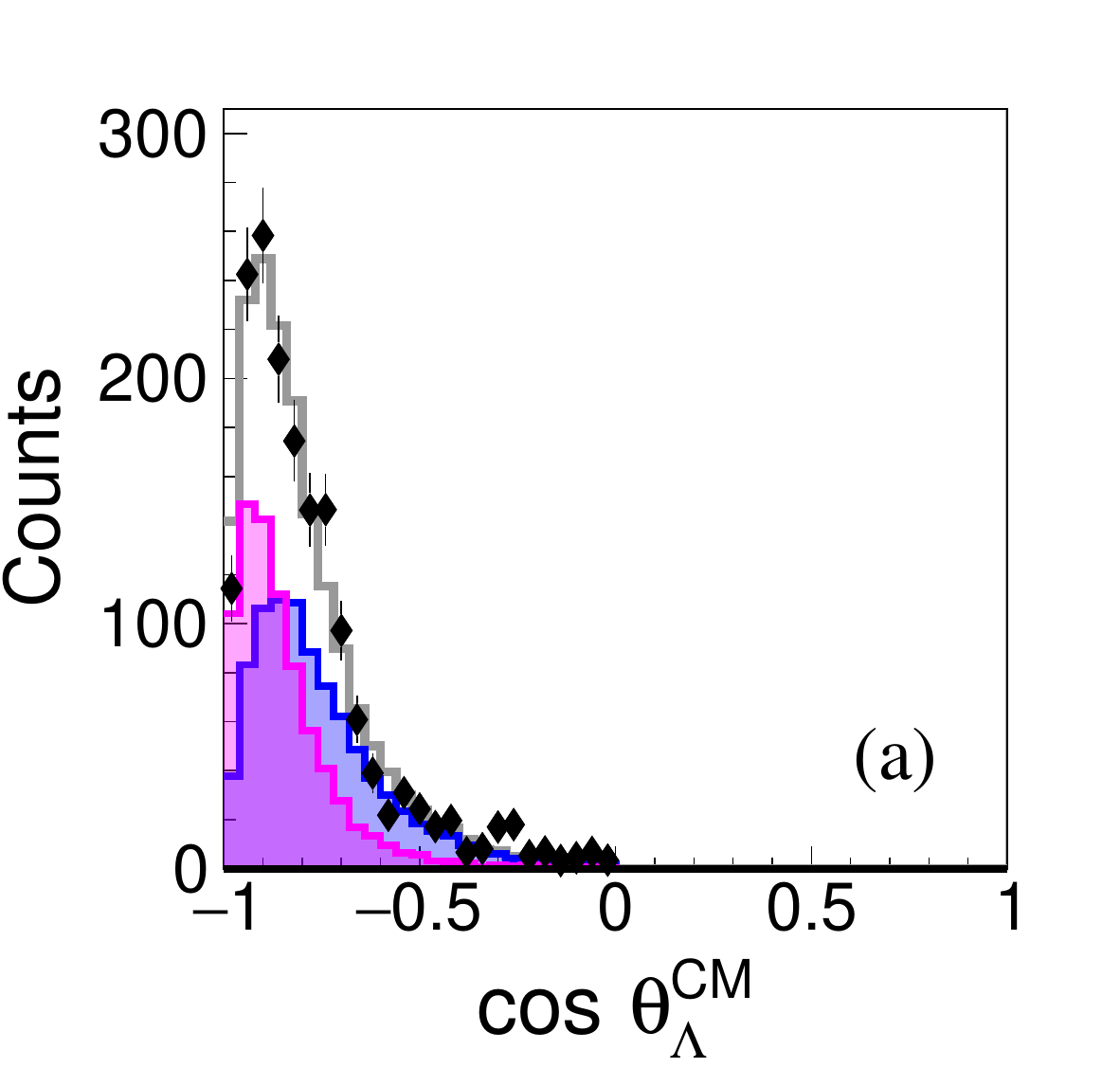}
	\end{minipage}
	\begin{minipage}[b]{\figwidth}
    		 \includegraphics[width=\textwidth]{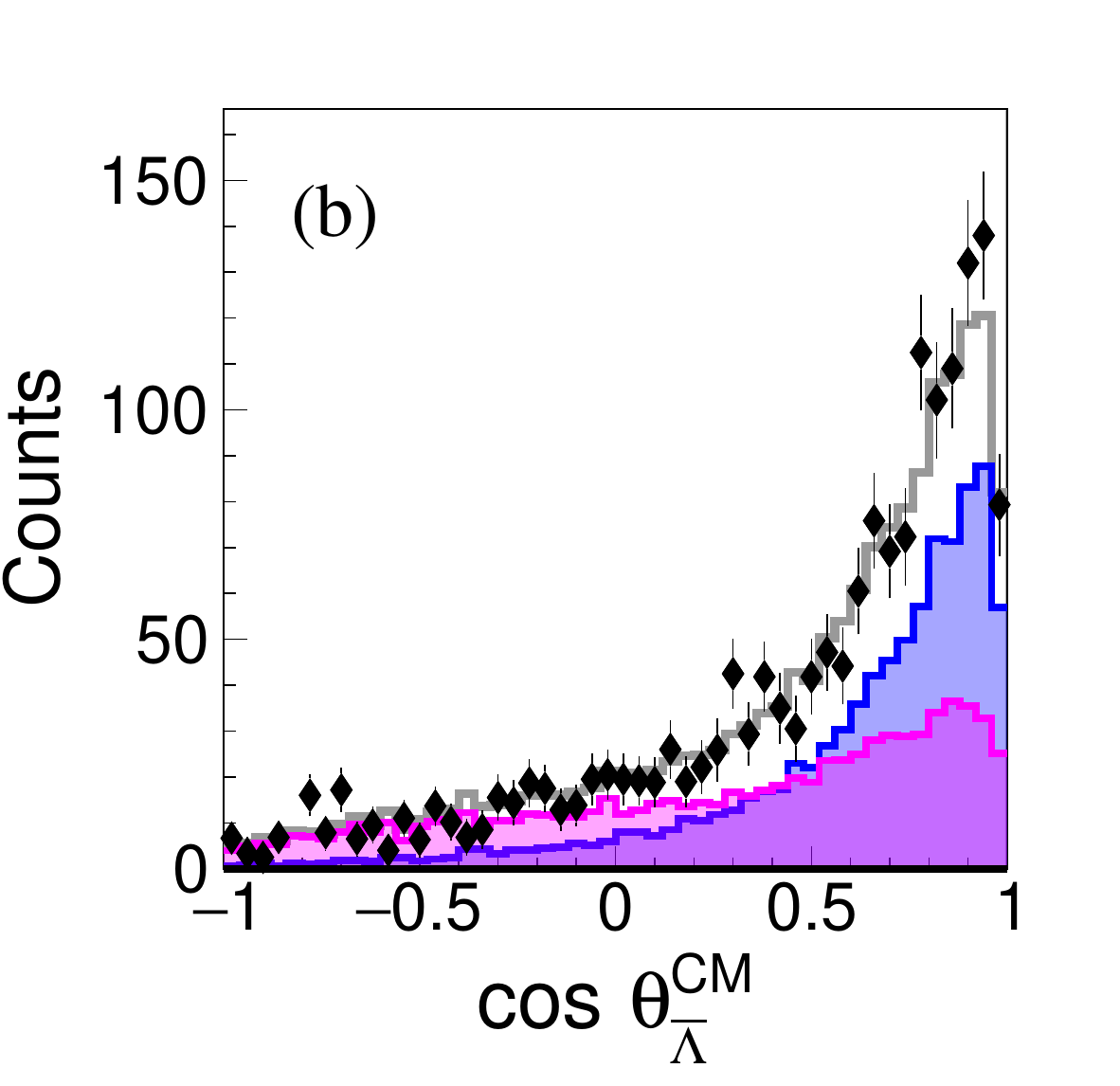}
	\end{minipage}
	\begin{minipage}[b]{\figwidth}
    		 \includegraphics[width=\textwidth]{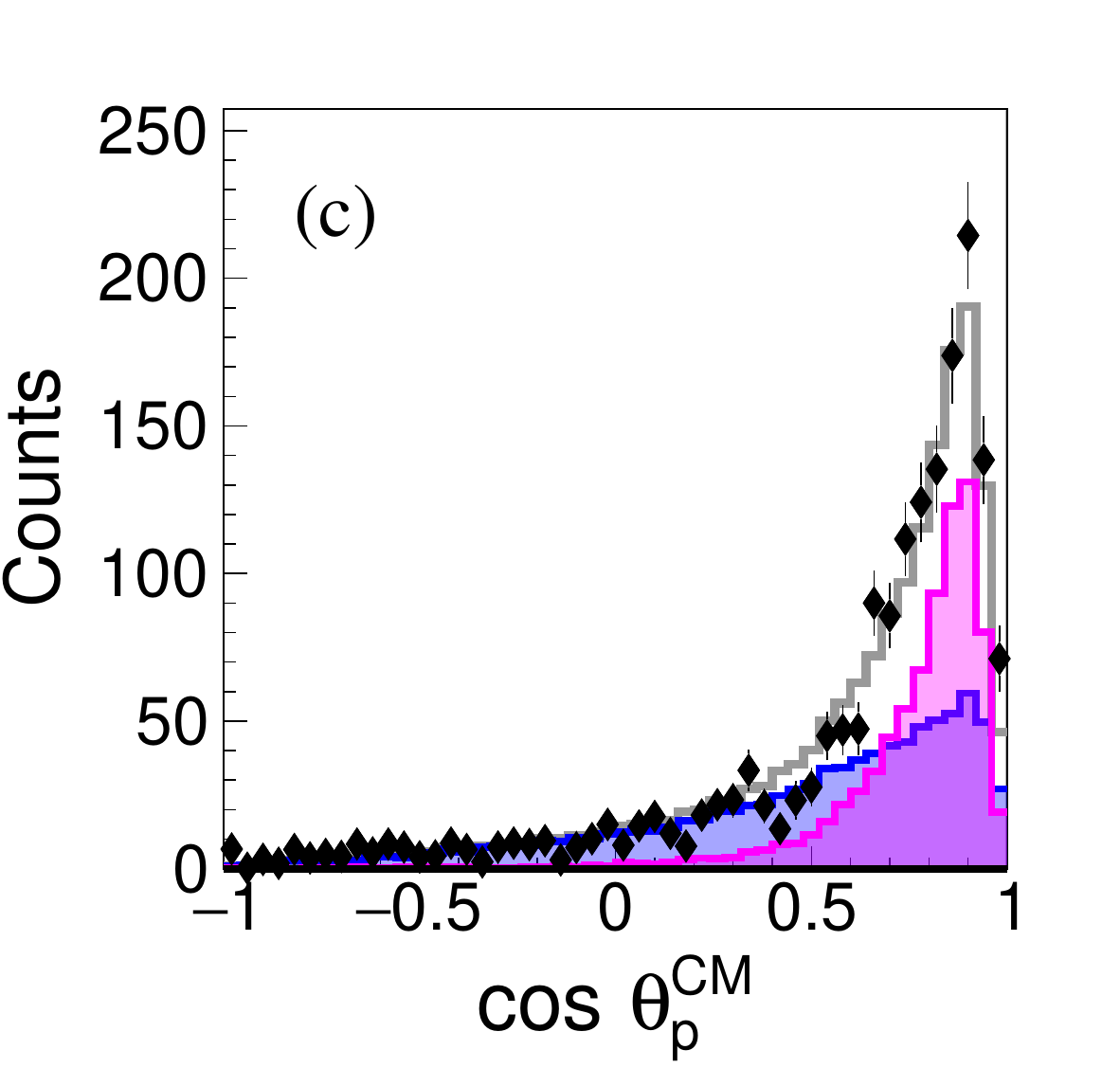}
	\end{minipage}
\caption{\label{fig:plambar_angular} Angular distributions of the three particles in the $\gamma p \rightarrow  \{ \bar{\Lambda} p \} \Lambda $ reaction channel in the overall CM system. The points and curves are organized as in Figs.~\ref{fig:ppbar_angular} and ~\ref{fig:lamlambar_angular},  but note that the roles of the proton and the $\Lambda$ are reversed when compared to Fig.~\ref{fig:lamlambar_angular}  in this reaction;  see text.}
\end{figure*}

Distinctly separate are the events with a backward recoiling  $\Lambda$ created in combination with a $p\antiLambda$ pair.  Figure~\ref{fig:plambar_angular} shows the angular distributions for events from $\gamma p \rightarrow  \{ p \antiLambda \}  \Lambda$, summed over the full range of beam energies and not acceptance corrected.  Again, this reaction has never been seen or studied in previous experiments. The $\Lambda$ takes the role of the recoiling backward particle, as seen in Fig.~\ref{fig:plambar_angular}(a).   This is possible when the exchanged meson is a kaon, as in Figs.~\ref{fig:feynman}(e, f).  Zero $\Lambda$'s appear in the forward hemisphere in Fig.~\ref{fig:plambar_angular}(a) due to the selection of the Van Hove angle, Sect.~\ref{sec:vanhove}.  In Fig.~\ref{fig:plambar_angular}(c), the produced protons are seen to peak strongly in the forward direction, consistent with being created as part of $p \antiL$  pairs via single exchange production.  The \antiLambdas distribution in Fig.~\ref{fig:plambar_angular}(b) again has a forward peak consistent with the single-exchange ansatz, but also a broad tail to large angles consistent with the double-exchange ansatz.  

\clearpage
\begin{widetext}
\subsection{Multiparticle correlations}
\label{sec:vanhove}
Three-body states can be categorized via the correlations among the final-state particles using a Van~Hove-style analysis~\cite{VanHove:1969xa} of the longitudinal phase space of the reactions. The momenta along the beam direction ($\hat{z}$) of the three particles are mapped onto a plane that correlates which of them go forward and which go backward in the overall CM frame.

\begin{figure*}[htpb]
  \centerline{\includegraphics[width=.8\textwidth]{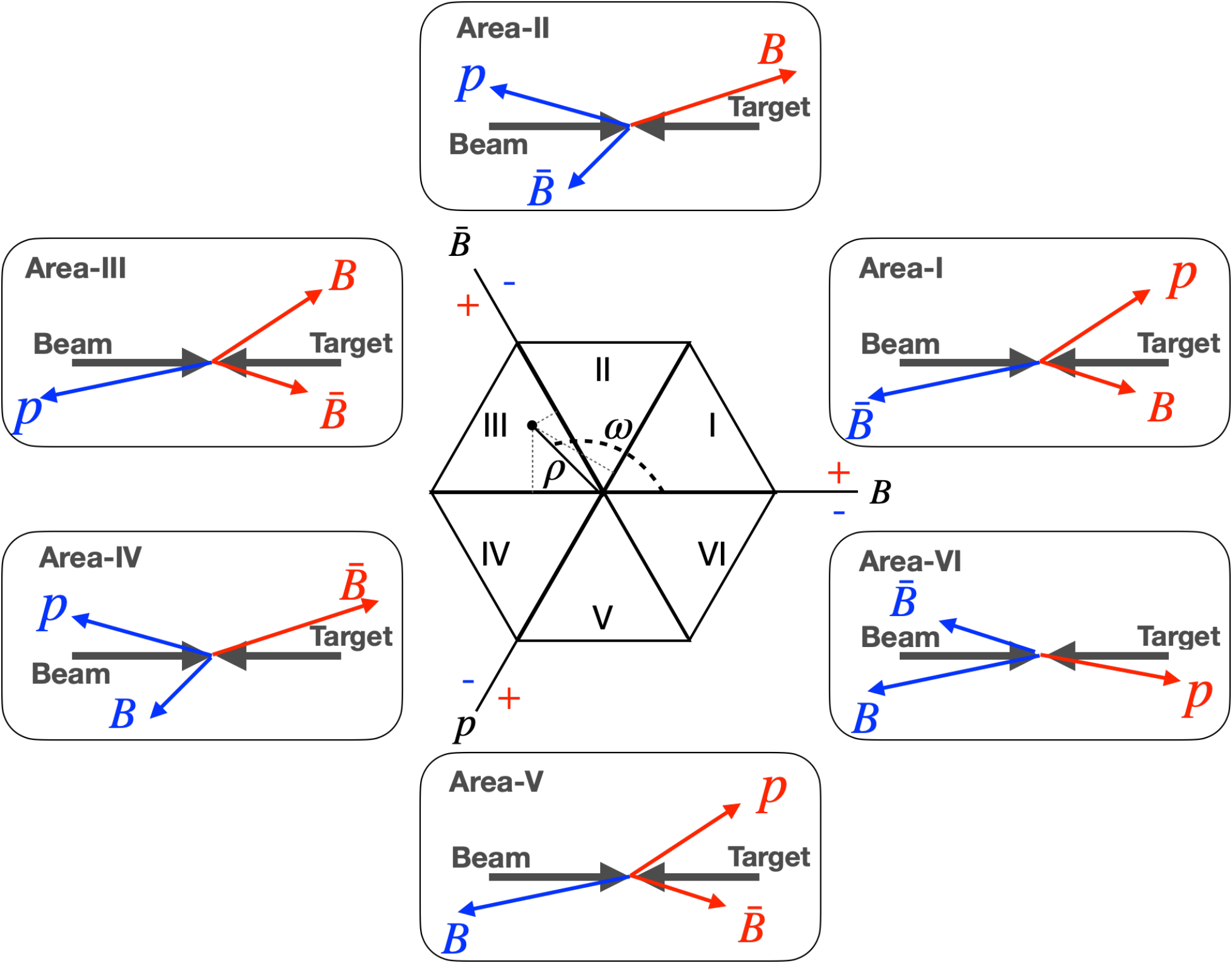}}
  \caption{Six sectors on a Van Hove diagram that differentiate three-body angular correlations in six separate kinematic regions in the CM frame, as depicted in the pictures located around the center and marked by Roman numerals.  See text for explanation. }
  \label{fig:vanhovediagram}
\end{figure*}

\begin{figure*}[htpb]
    \centering 
    \begin{minipage}{.245\textwidth}
        \centering
        \includegraphics[width=\linewidth]{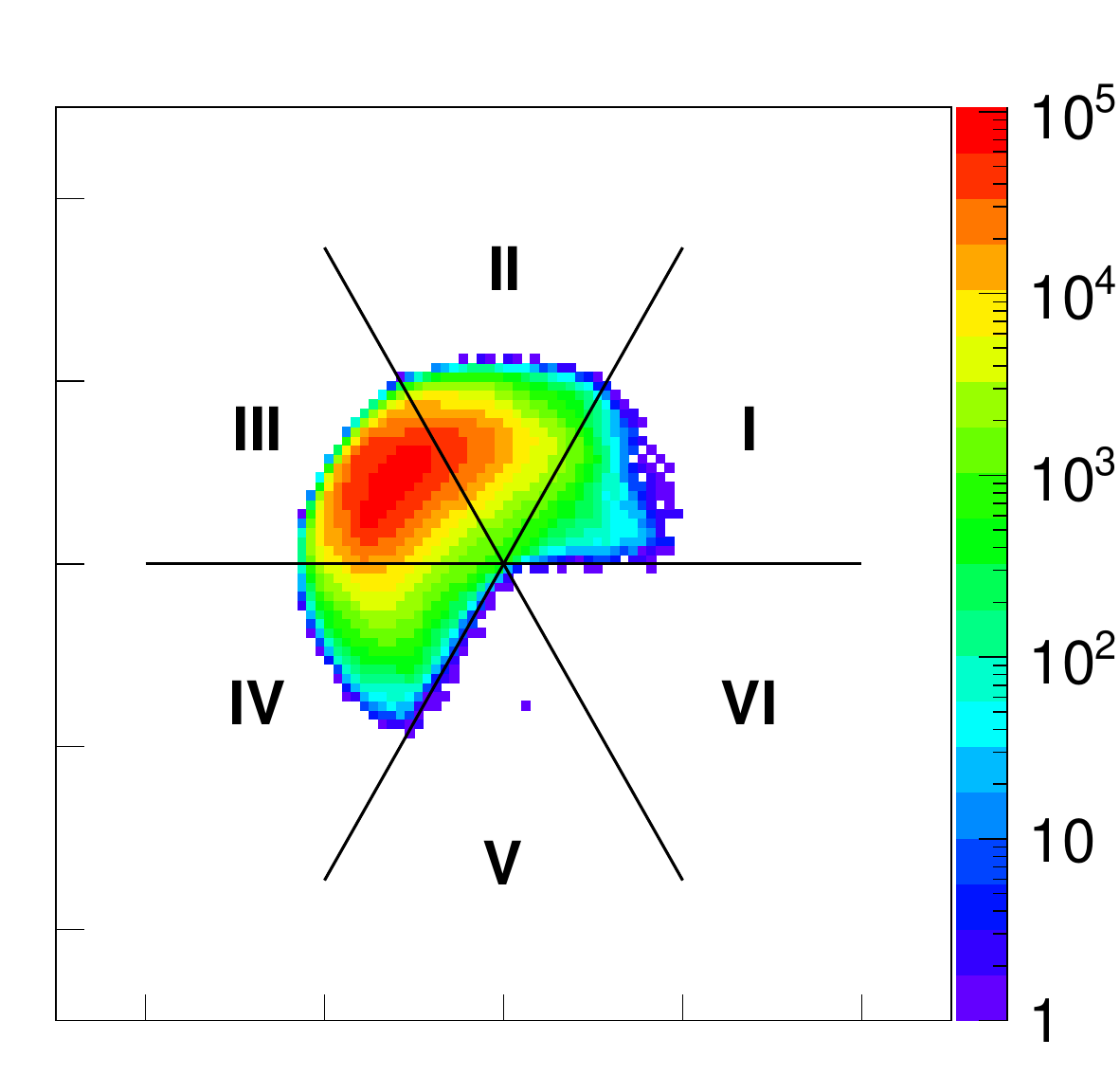}\\
        (a) GlueX: $\gamma p \rightarrow p\bar{p}p$
    \end{minipage}%
    \begin{minipage}{.245\textwidth}
        \centering
        \includegraphics[width=\linewidth]{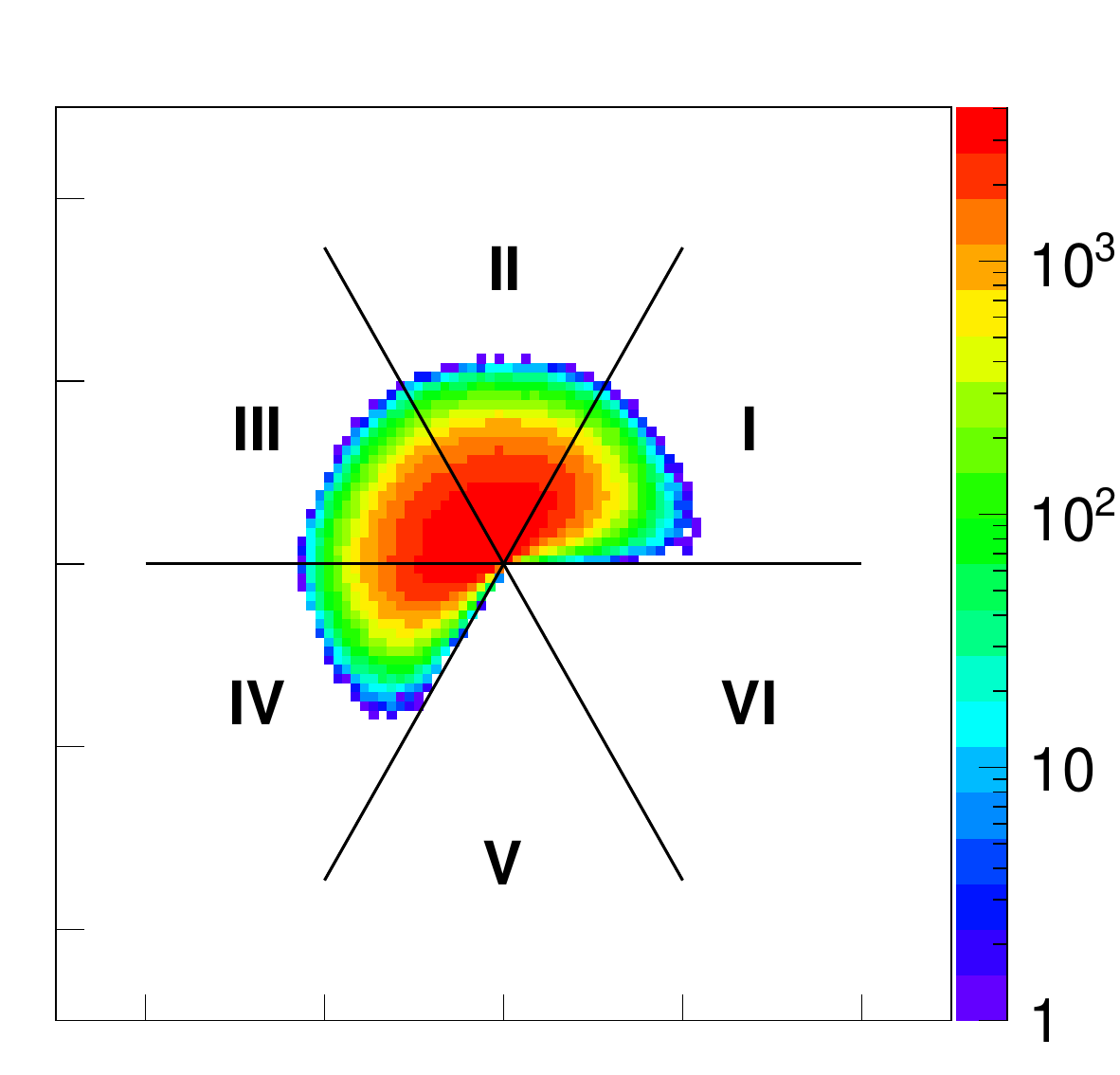}\\
        (b) MC: $p\bar{p}p$ phase space
    \end{minipage}%
    \begin{minipage}{.245\textwidth}
        \centering
        \includegraphics[width=\linewidth]{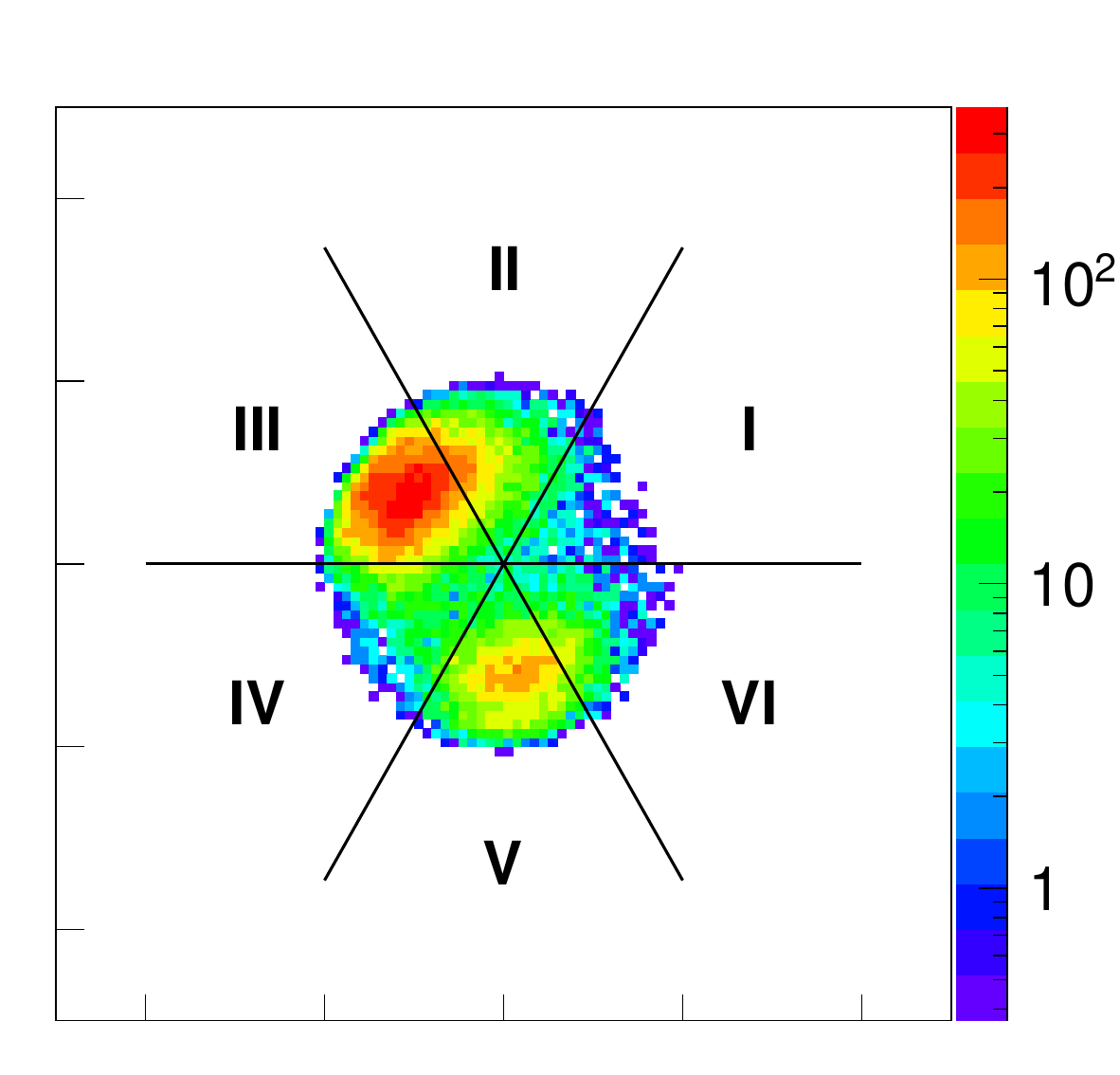}\\
        (c) GlueX: $\gamma p \rightarrow \Lambda \bar{\Lambda} p$
    \end{minipage}%
    \begin{minipage}{.245\textwidth}
        \centering
        \includegraphics[width=\linewidth]{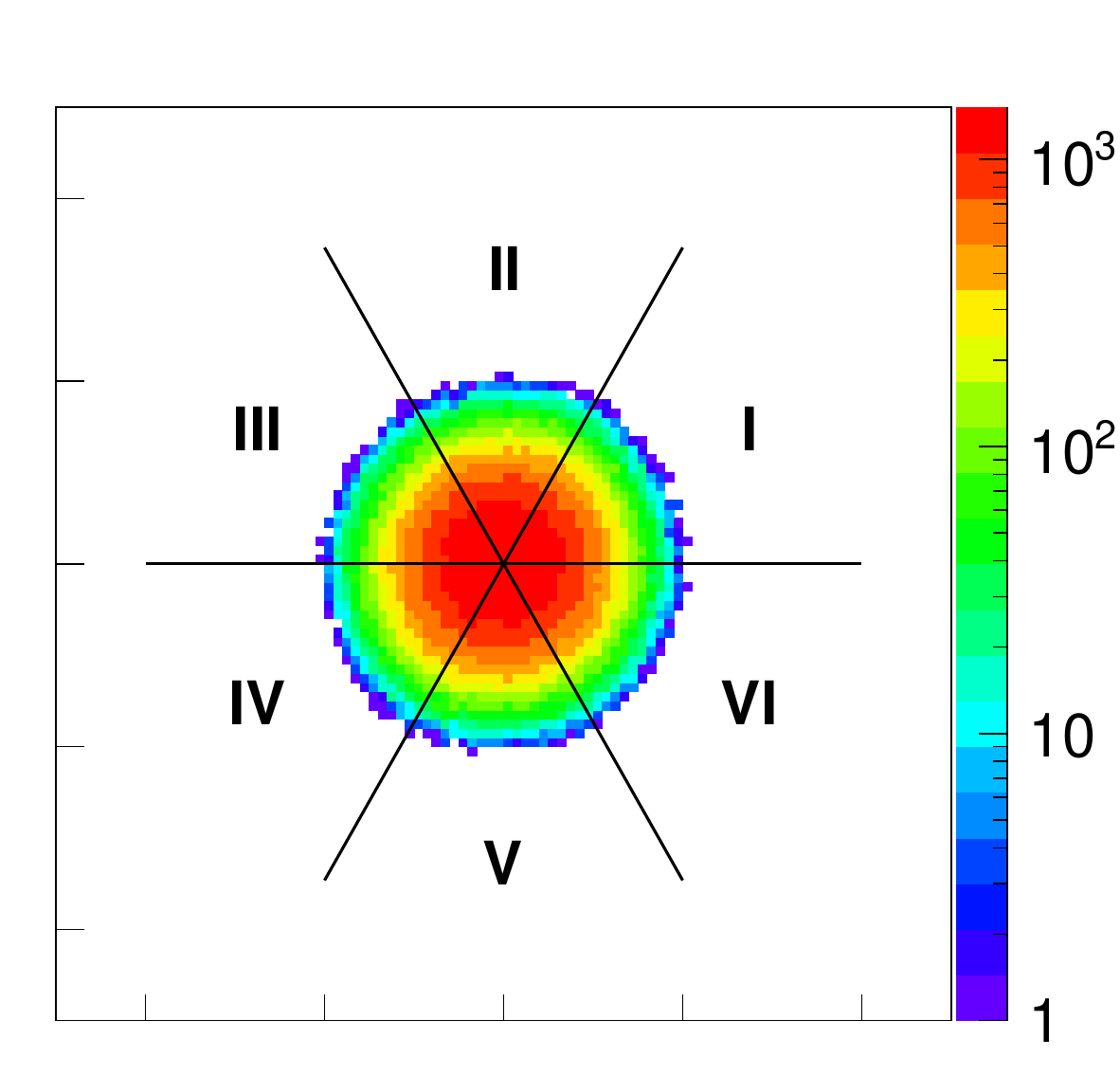}\\
        (d) MC: $\Lambda \bar{\Lambda} p$ phase space
    \end{minipage}
    \caption{Van Hove plots of the three-body final states: (a) Data from the $p\bar{p}p$ reaction, with Areas V and VI empty due to proton sorting. (b) Phase space MC distribution for the $p\bar{p}p$ reaction. (c) Data from the $\Lambda \bar{\Lambda} p$ reactions in which each particle is unique. Two clusters of data indicate the kinematic separation of $\Lambda \bar{\Lambda}$ and $p \bar{\Lambda}$ creation. (d) Phase space MC for the $\Lambda \bar{\Lambda} p$ reaction. The data shown in panels (a), (c) are not acceptance corrected.
    }
\label{fig:vanhovedata}
\end{figure*}
\end{widetext}
\clearpage

Figure~\ref{fig:vanhovediagram} illustrates the six combinations of proton, produced baryon and produced antibaryon that can each move either forward or backward in the CM frame.  Each event is represented as a point in the Van Hove plane, parameterized by polar angle $\omega$ and radius $\rho$.   
A brief overview of this categorization is included in Appendix~\ref{sec:vh}, defining the correlation variables $\omega$ and $\rho$.
Each particle ``$B$,'' ``$\antiB$,'' or ``$p$'' from the event that is denoted by a point on the ``+'' side of the indicated principal axis is moving forward in the CM frame, while for a point located on the ``$-$'' side of the same axis, the corresponding particle is moving backward.

Figure~\ref{fig:vanhovedata} shows the distribution of $p\antip p$ and $\Lambda\antiLambda p$ events summed over all photon beam energies for the real data [Figs.~\ref{fig:vanhovedata}(a) and ~\ref{fig:vanhovedata}(c)] and what is computed for a pure phase space distribution for these reactions [Figs.~\ref{fig:vanhovedata}(b) and ~\ref{fig:vanhovedata}(d)].  Events in ``Area~III,'' for example, correspond to the baryon and the antibaryon moving forward in the CM, with the proton moving backward.   Figure~\ref{fig:vanhovedata} (a) and (b) are not populated with events in Areas V and VI because the identical protons are sorted into ``more forward" for the produced baryon ``$B$" and ``more backward" for the recoil proton ``$p$.''

\begin{figure}[htpb]
    \centering
    \begin{minipage}[b]{.423\textwidth}
        \centering
        \includegraphics[width=\textwidth]{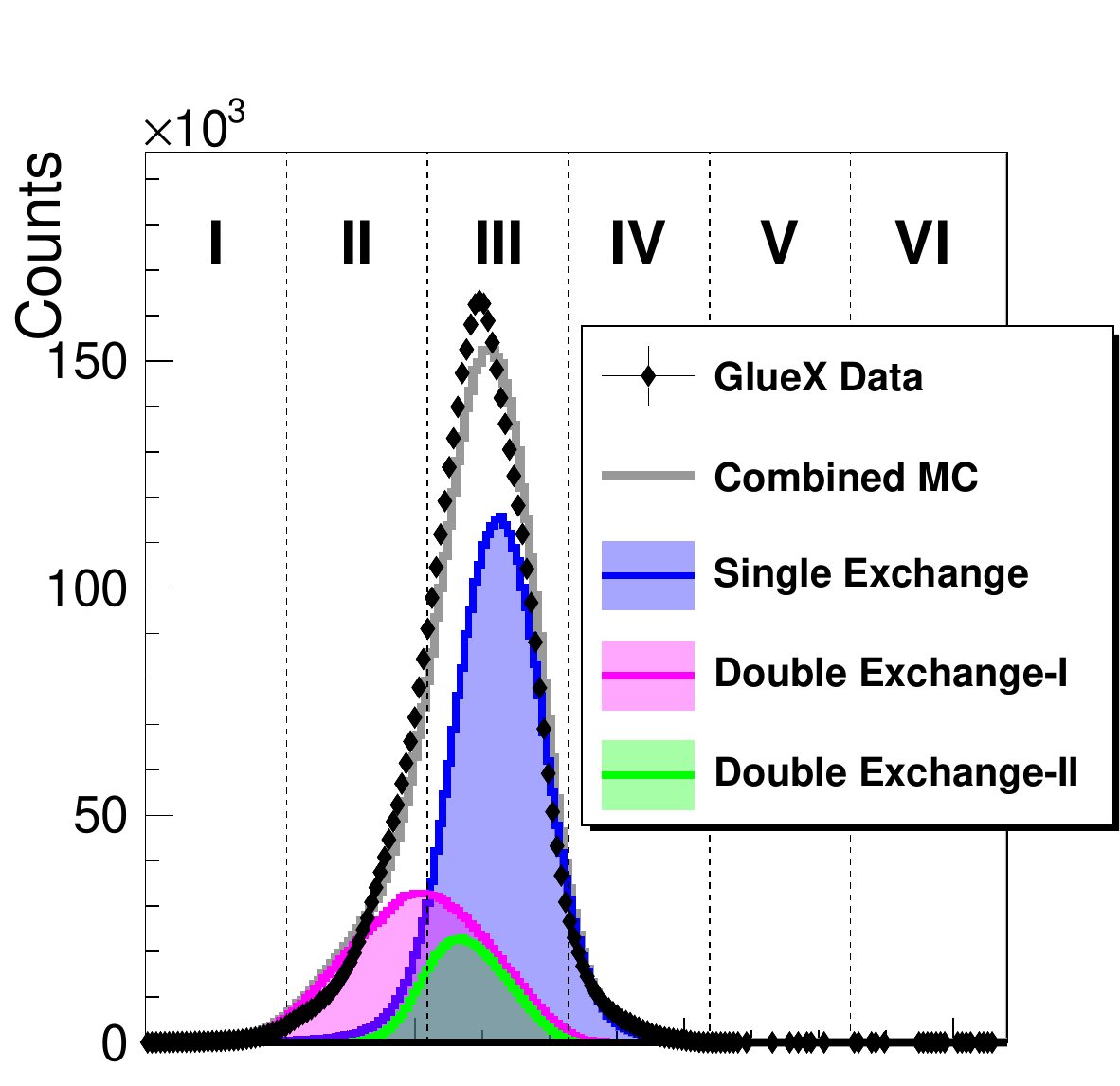}
    \end{minipage}
    \hfill
    \begin{minipage}[b]{.423\textwidth}
        \centering
        \includegraphics[width=\textwidth]{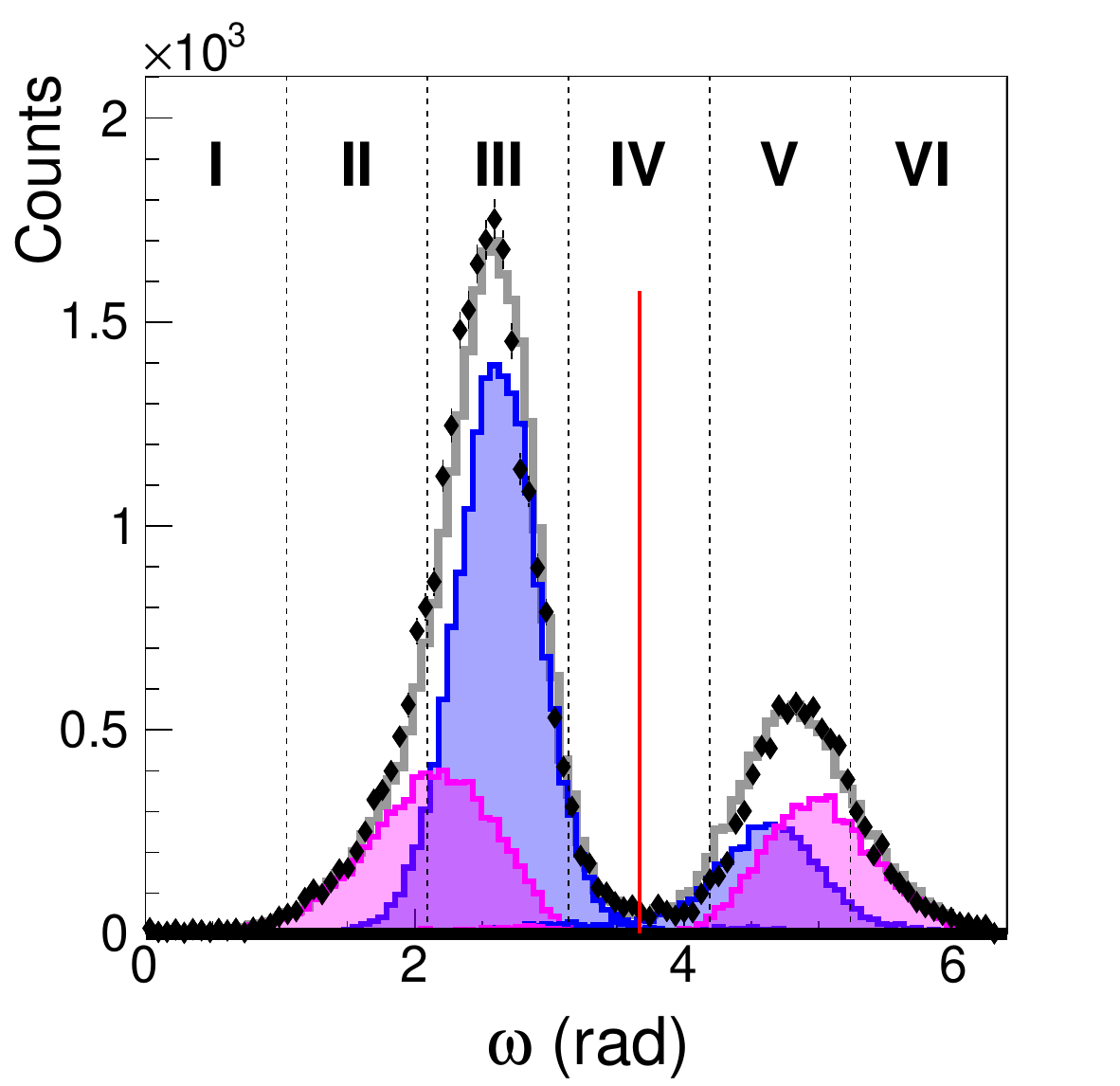}
    \end{minipage}
  \caption{Van Hove distributions of the three-particle final states as a function of the angular parameter, $\omega$, that sorts the events according to pair longitudinal momenta. (Top) Data from the $p \antip p$ reaction showing that most events are in Area~III, corresponding to single-exchange kinematics (blue in the model fit), but with significant overlap with events in the double-exchange-dominated region of Area~II (magenta and green in the model fit).  (Bottom)  Data from the $\Lambda \antiLambda p$ reaction showing the division between $\{ \Lambda \antiLambda \} p $ events mainly in Areas~II and~III, and $\{ p \antiLambda \} \Lambda $ events mainly in Areas~V and VI. 
  The cut separating $\LamantiLam$ and $\pantiL$ cases is indicated by the solid red line at $\omega = 3.67$.}
  \label{fig:vanhovedistributions}
\end{figure}

In contrast, the phase space panel Fig.~\ref{fig:vanhovedata}(d) for the $\Lambda\antiL p$ cases is uniformly populated in angle $\omega$, as expected.  Significantly, the data distribution for the $\Lambda\antiL p$ cases Fig.~\ref{fig:vanhovedata}(c) shows a separation of events where the two hyperons tend to travel together in the forward hemisphere in Area~III, but also where the proton and the $\antiL$ travel together predominantly in the forward hemisphere in Area~V. This was manifested in the single-particle angular distributions introduced in Sec.~\ref{sec:single_particle_correlations}.   For nonstrange particle exchange in the $t$ channel, the \LamantiLam pair will tend to be created in the forward hemisphere.  In the case of kaon exchange, the $p\antiL$ pair will tend to be created in the forward hemisphere, with a clear separation between the cases at \gx energies.  

The event distributions versus the Van Hove angle, $\omega$,  are shown in Fig.~\ref{fig:vanhovedistributions}.   
The model fit explained in Sec.~\ref{sec:reactionmodelfit} is shown as filled histograms, indicating how the single- and double-exchange domains overlap considerably, but both are needed to represent the data.  Double-exchange processes tend to populate the edge regions between sectors of this plane, especially Areas~II and III.   In the $p \antip p$ case, wherein the \pbar is found at larger angles, is shown at the top of Fig.~\ref{fig:vanhovedistributions}. The antibaryon asymmetry is reflected in excess events in Area~II compared to Area~IV.   
In the $\Lambda \antiLambda p$  case, wherein the \Lbar is produced at large angles, is shown at the bottom of Fig.~\ref{fig:vanhovedistributions}.  The antibaryon asymmetry is again seen in Area~II versus Area~IV, but also in Area~IV versus Area~VI, where the produced $p\antiLambda$ pair events preferentially have the antibaryon at large angles.  We effectively separated hyperon cases with a cut at $\omega = 3.67$ (or $210^\circ$), with \LamantiLam on the low side and \pantiL on the high side.  The effects of cross-contamination were small compared to statistical and other systematic uncertainties in this study.

\subsection{Invariant mass of pairs}
\label{sec:clustering}

\begin{figure}[htbp]
    \centering
    \includegraphics[width=\columnwidth]{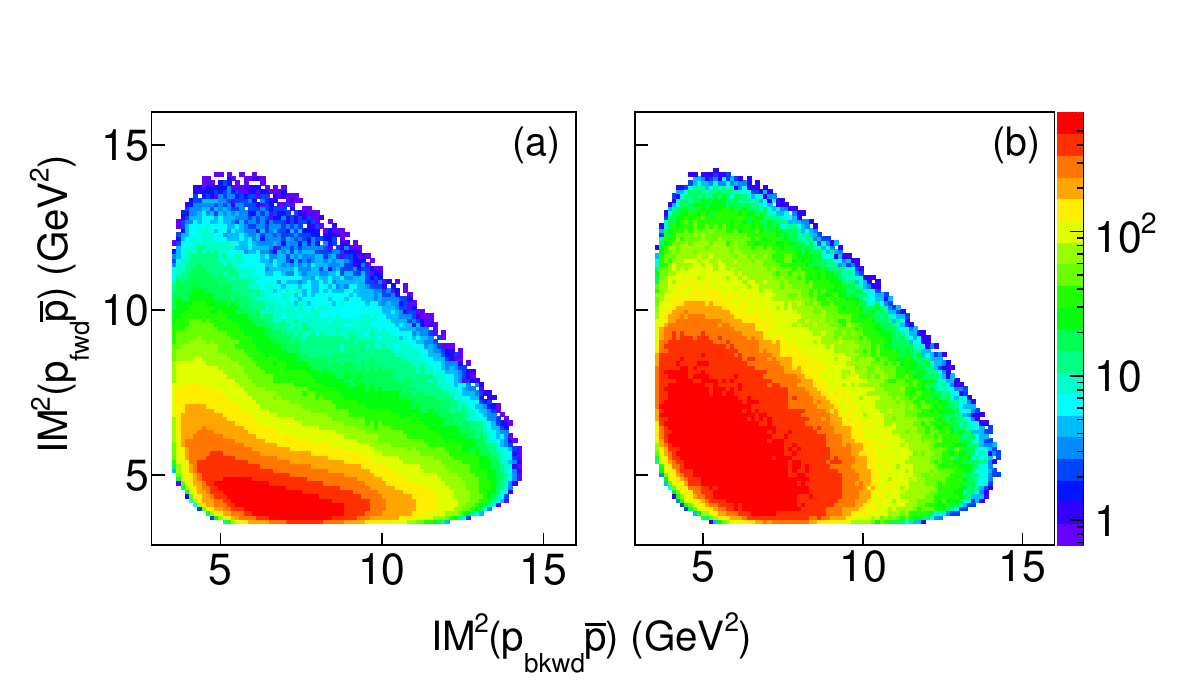}
    \caption{Dalitz plots for the $\gamma p \rightarrow \{p\bar{p}\} p$ reaction summed over the GlueX beam energy range. (a) \gx data, without acceptance correction. (b) three-body phase space MC. } 
    \label{fig:dalitz_ppbar}
\end{figure}

The invariant mass of the produced particle pairs was found to cluster toward low masses compared to phase space, as shown first in the Dalitz plot for $\ppbar$ pairs in Fig.~\ref{fig:dalitz_ppbar}.  We parameterize this as an attractive interaction independent of $t$ between the created baryons and antibaryons in the model introduced in Sec.~\ref{sec:reactionmodel}.  The asymmetric band in Fig.~\ref{fig:dalitz_ppbar}(a) is caused by the asymmetry between the $p$ and \antips distributions due to the double exchange contribution to the reaction.

\begin{figure}[htbp]
		\centering
		  \begin{minipage}[b]{.48\columnwidth}
          \centering
		    \includegraphics[width=\textwidth]{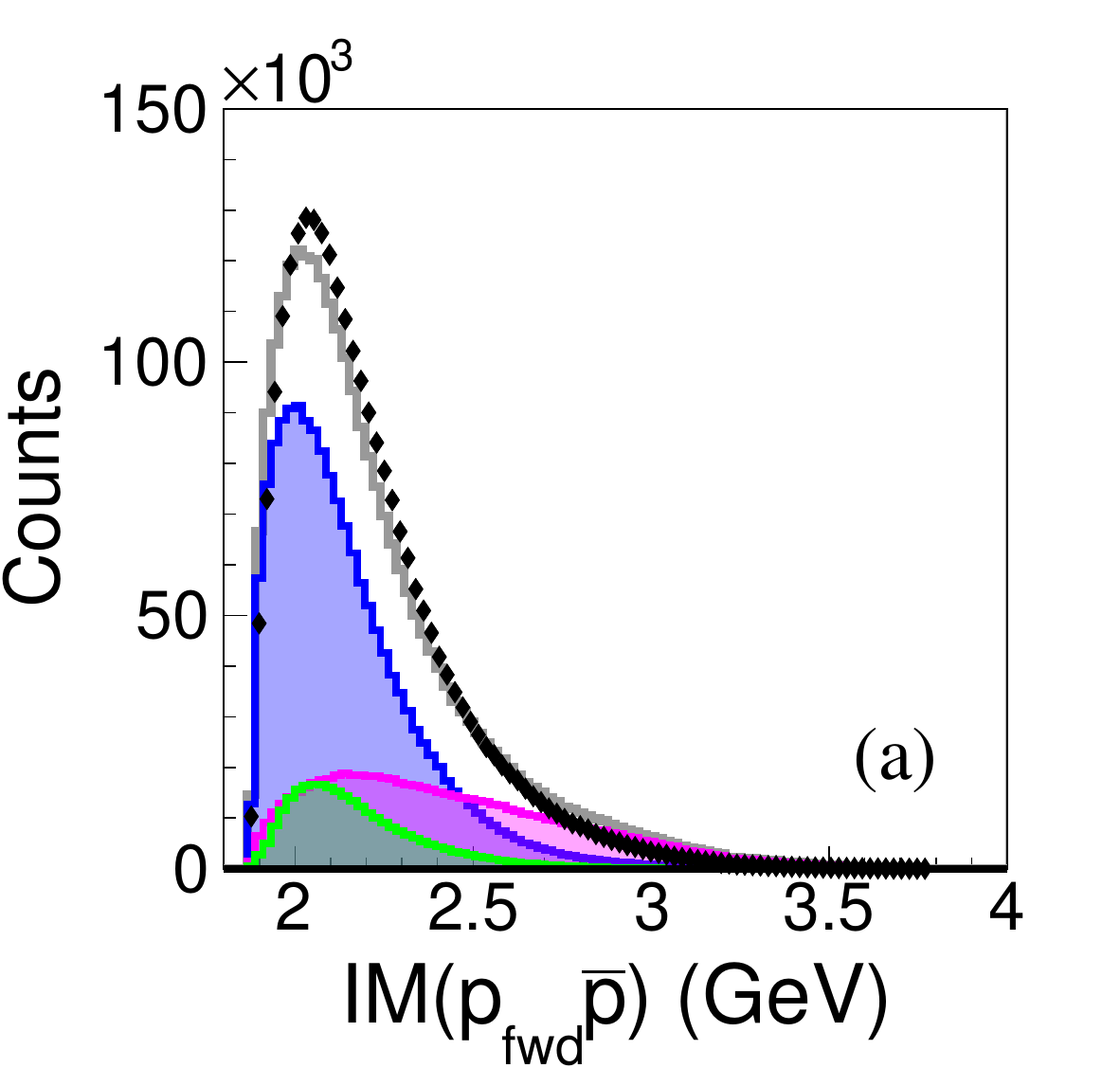}
		  \end{minipage}
		  \begin{minipage}[b]{.48\columnwidth}
          \centering
		    \includegraphics[width=\textwidth]{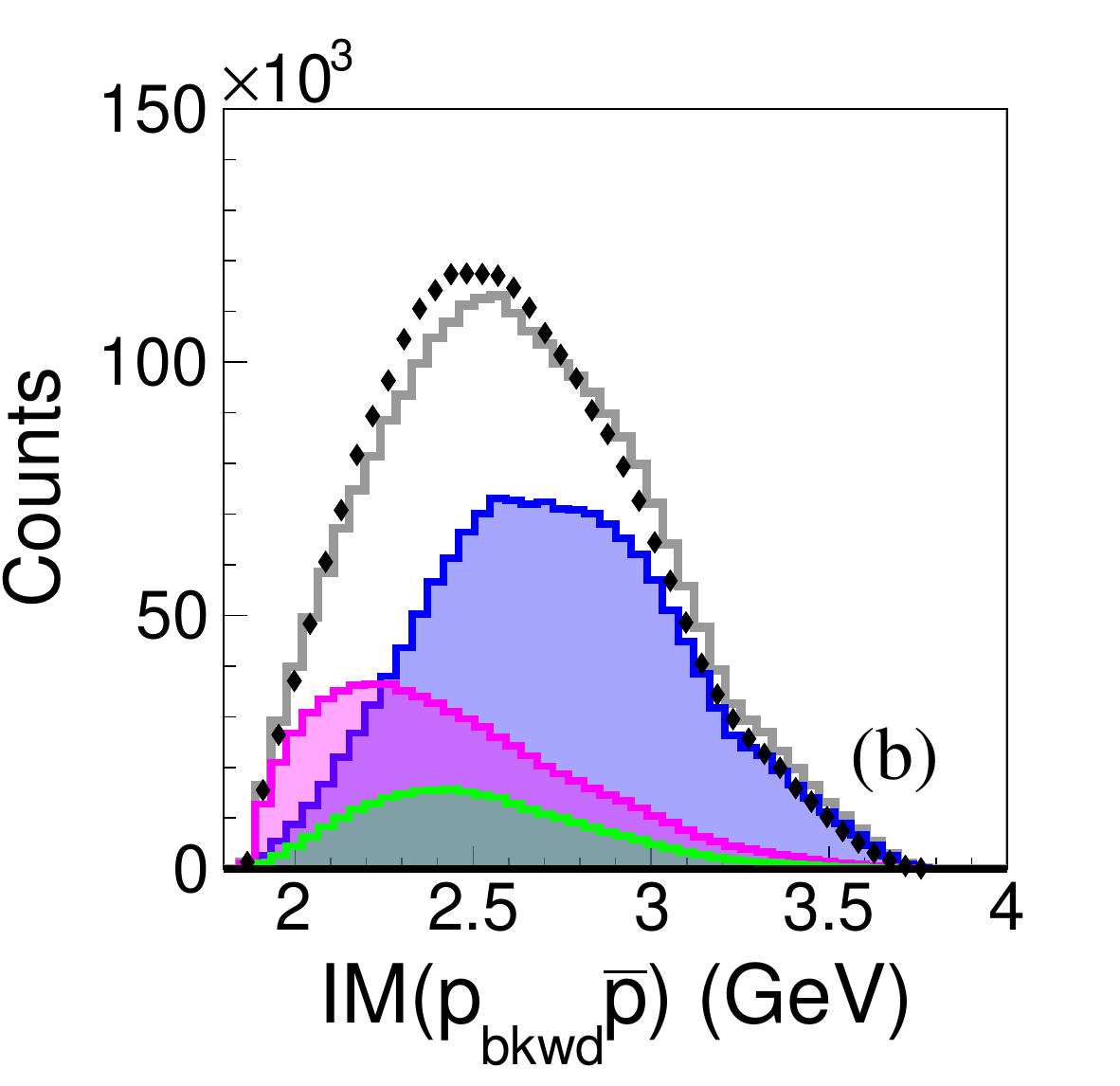}
		  \end{minipage}
\caption{\label{fig:IM_&_t_modelfit}
(a) Invariant mass of $p\antip$ pairs for the more forward proton (black points), with model fits for the single-exchange (blue) and double-exchange (magenta and green) processes, totaling to the histogram (grey). (b) Pairings of the \antips with the more backward $p$. The distributions are not acceptance corrected. }
\end{figure}

Figure~\ref{fig:IM_&_t_modelfit} shows the invariant mass distribution of the antiproton in both possible pairings: Fig.~\ref{fig:IM_&_t_modelfit}(a) with the more forward proton that is likely the created proton, and Fig.~\ref{fig:IM_&_t_modelfit}(b) with the more backward proton that is much more likely to be the recoil proton. The shaded regions are the result of the fitted model, in which the data in panel Fig.~\ref{fig:IM_&_t_modelfit}(a) were included in the fitting procedure, while the data in Fig.~\ref{fig:IM_&_t_modelfit}(b) was not. 

\begin{figure}[htpb]
     \centering
    \begin{minipage}{.49\columnwidth}
        \centering
        \includegraphics[width=\textwidth]{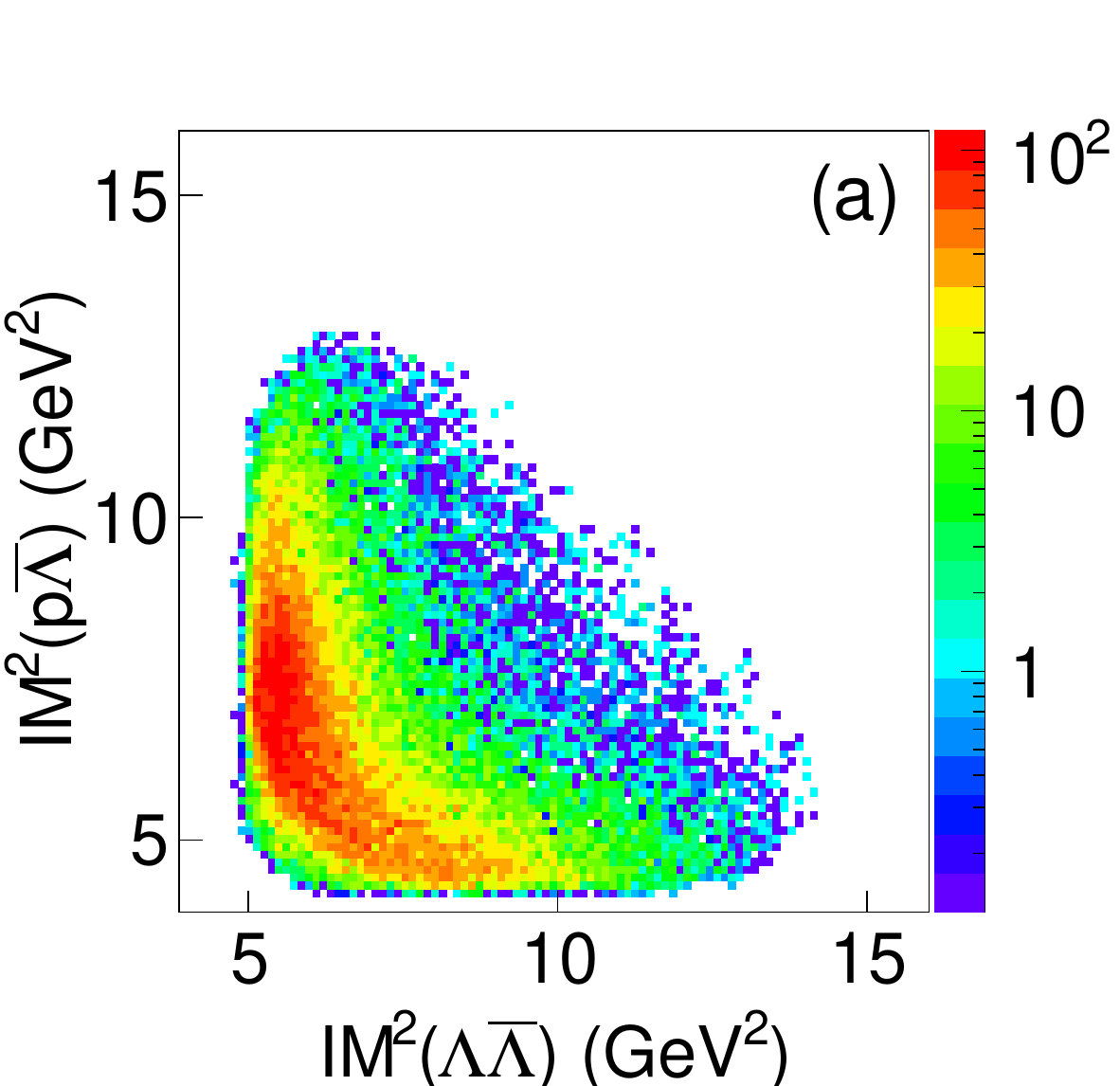}
    \end{minipage}%
    \begin{minipage}{.49\columnwidth}
        \centering
        \includegraphics[width=\textwidth]{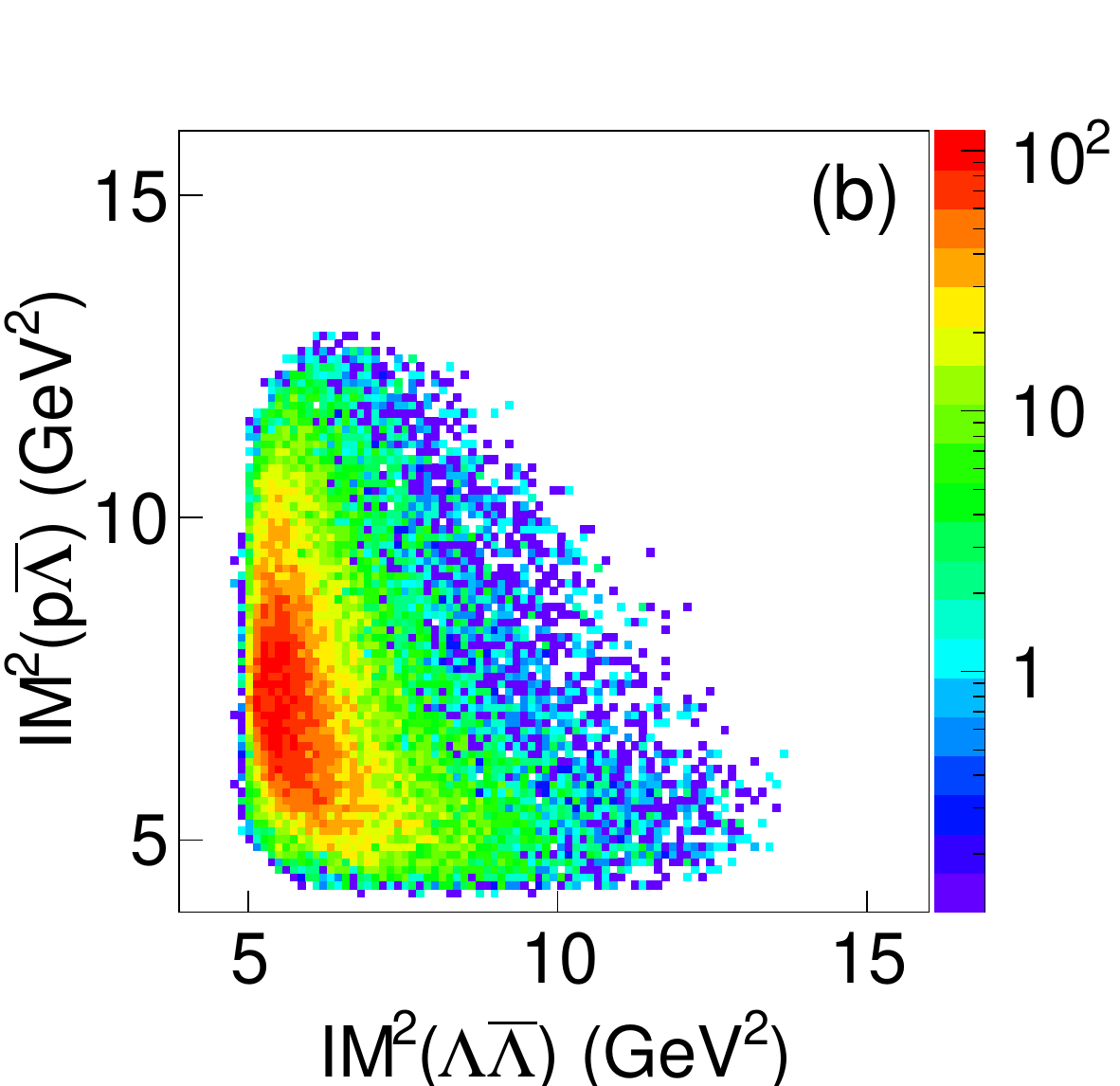}
    \end{minipage}
    \begin{minipage}{.49\columnwidth}
        \centering
        \includegraphics[width=\textwidth]{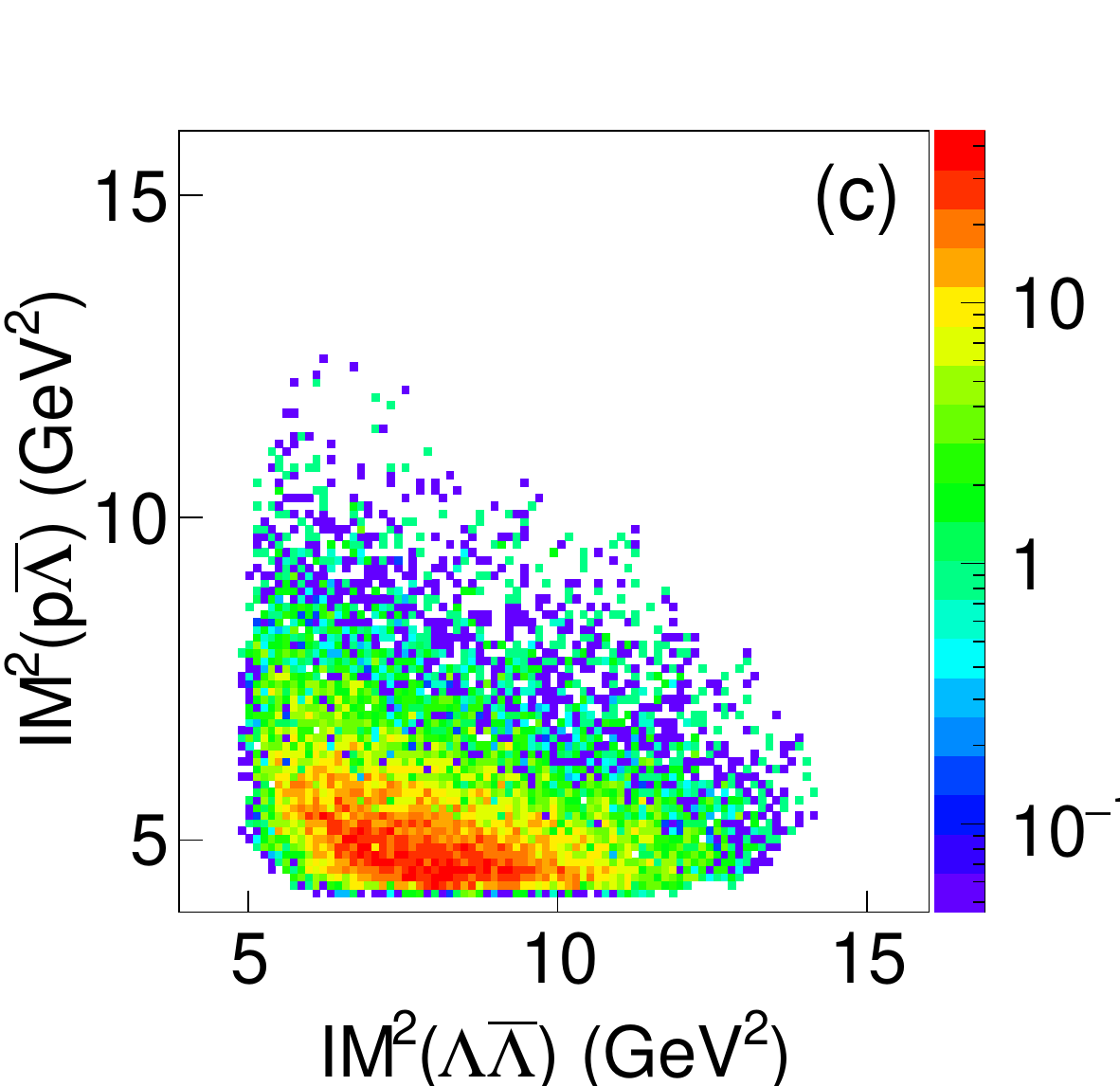}
    \end{minipage}%
    \begin{minipage}{.49\columnwidth}
        \centering
        \includegraphics[width=\textwidth]{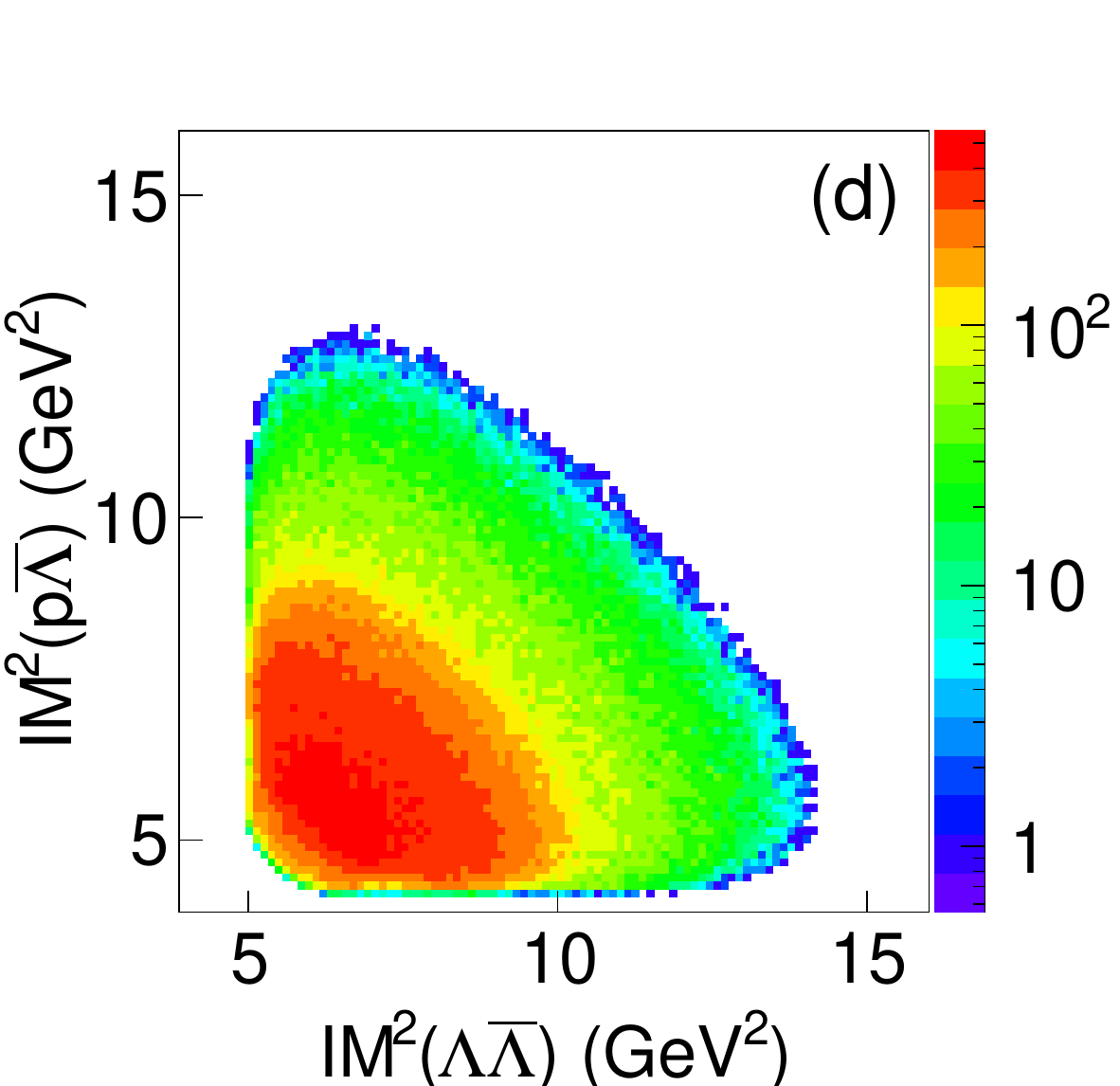}
    \end{minipage}
    \caption{\label{fig:dalitz_lamlambar}
        Dalitz plots for $\gamma p \rightarrow \Lambda\bar{\Lambda}p$ photoproduction for 
        (a) \gx data summed over the entire GlueX beam energy range,
        (b) data selected ($\omega<3.67$) for the case where $\lamlambar$ system recoils against $p$,
        (c) data selected ($\omega>3.67$) for the case where $\plambar$ system recoils against $\Lambda$ , and (d) three-body phase space MC.
        }  
\end{figure}

Analogous Dalitz plots for the hyperon channels are shown in Fig.~\ref{fig:dalitz_lamlambar}, where the two cases were separated using the Van Hove cut discussed in Sec.~\ref{sec:vanhove}.  The curving aspect of the intensity correlations again reflects the asymmetry in the two-body angular correlations observed in the CM frame.  It is again a signature of the double-exchange component of the production.

\begin{figure}[H]
		\centering
		  \begin{minipage}[b]{.48\columnwidth}
          \centering
		    \includegraphics[width=\textwidth]{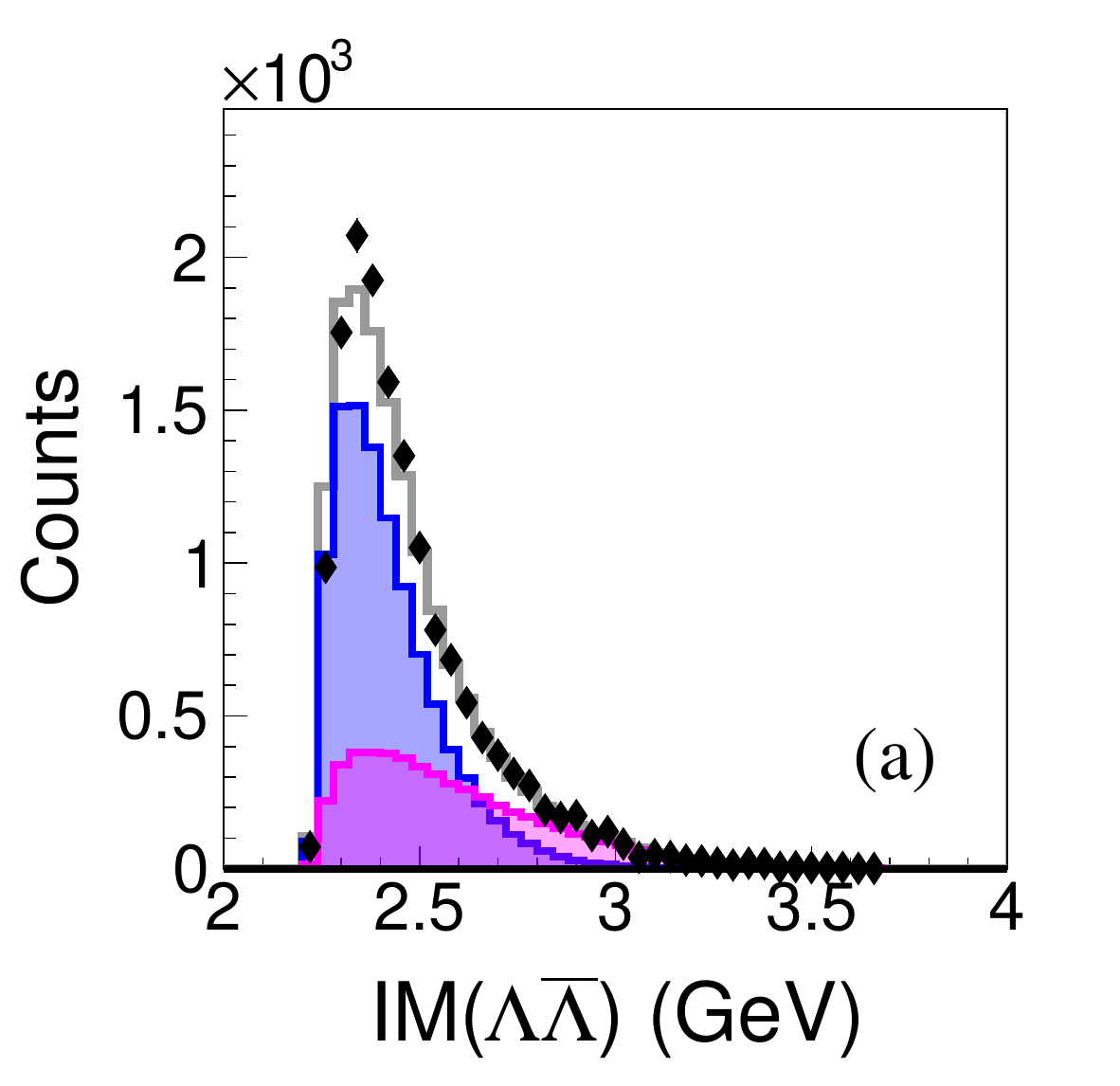}
		  \end{minipage}
		  \begin{minipage}[b]{.48\columnwidth}
          \centering
		    \includegraphics[width=\textwidth]{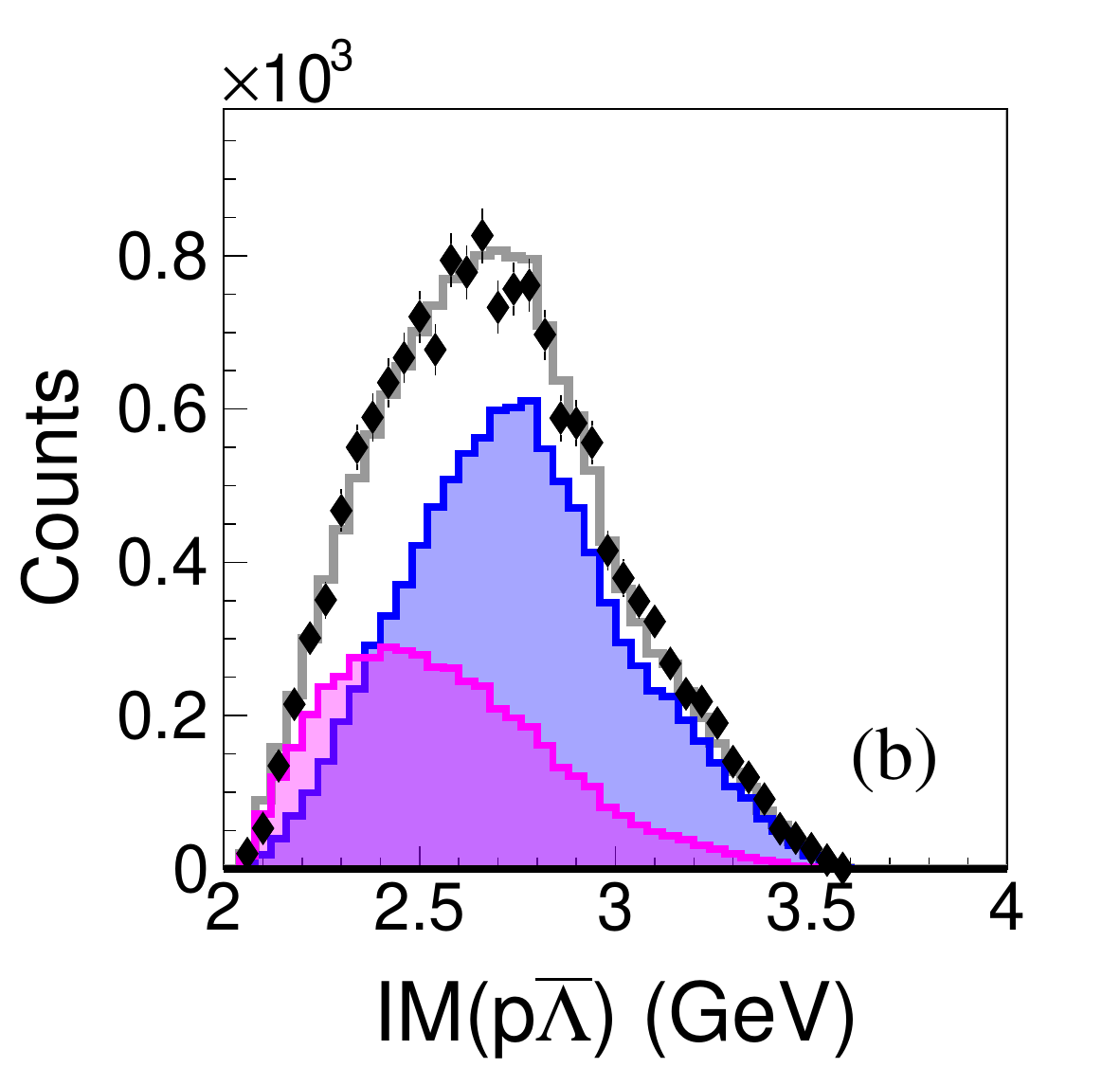}
		  \end{minipage}
\caption{\label{fig:lamlambar_IM_&_t_modelfit}
(a) Invariant mass of $\lamlambar$ pairs for forward-going hyperon pairs (black points), together with model fits for the single exchange (blue) and double exchange (magenta) processes.  (b) Pairings of the \antiLambda~ with the recoil proton.  The distributions are not acceptance corrected. }
\end{figure}

For the case of  $\Lambda\antiLambda$ production,  Fig.~\ref{fig:lamlambar_IM_&_t_modelfit} shows the invariant mass distribution of pairings, with Fig.~\ref{fig:lamlambar_IM_&_t_modelfit}(a) showing hyperons together going mostly forward, and Fig.~\ref{fig:lamlambar_IM_&_t_modelfit}(b) showing forward \antiLambdas with the recoil proton.  

\begin{figure}[H]
		\centering
            \begin{minipage}[b]{.48\columnwidth}
            \centering
		    \includegraphics[width=\textwidth]{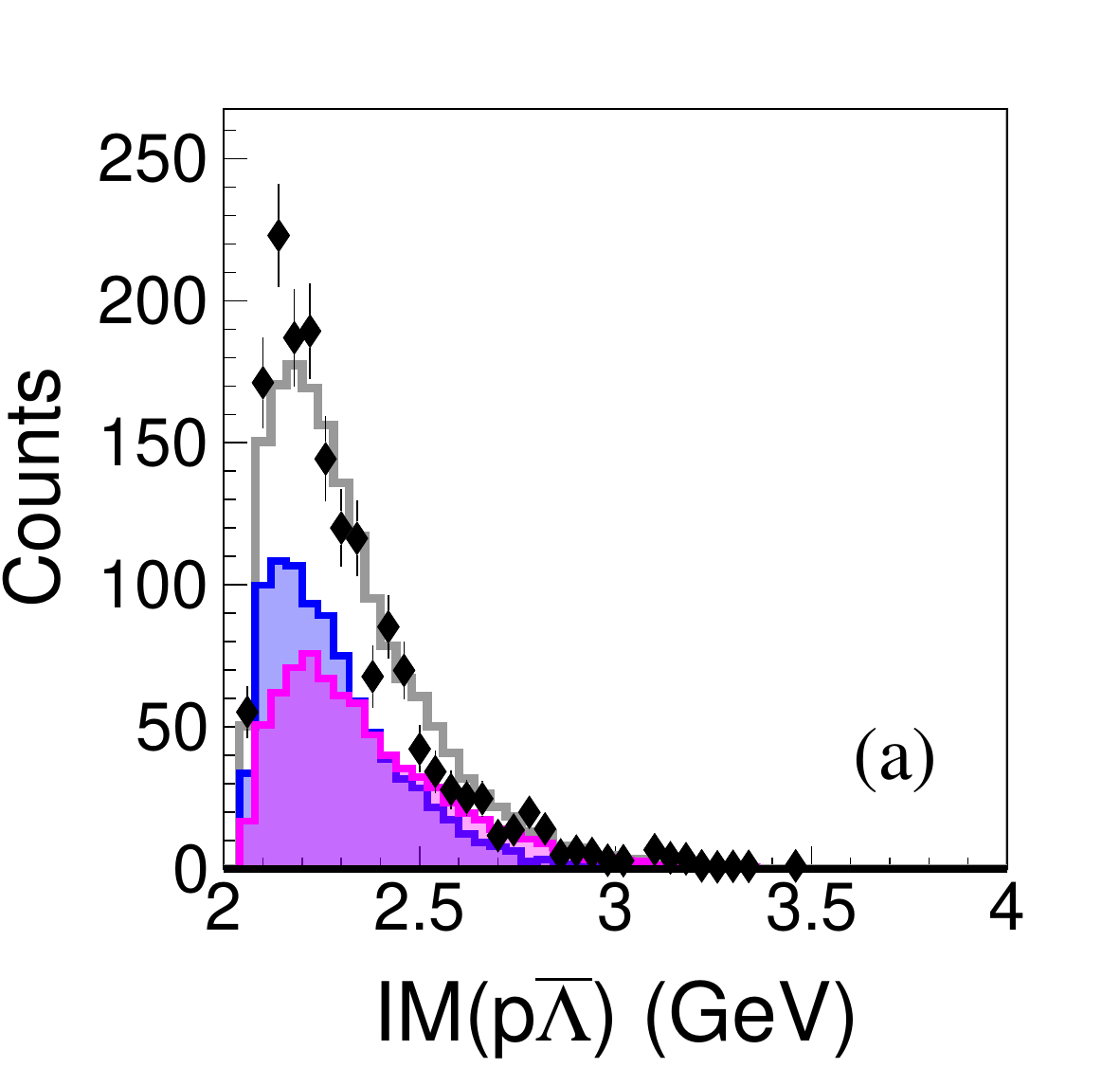}
		  \end{minipage}
		  \begin{minipage}[b]{.48\columnwidth}
          \centering
		    \includegraphics[width=\textwidth]{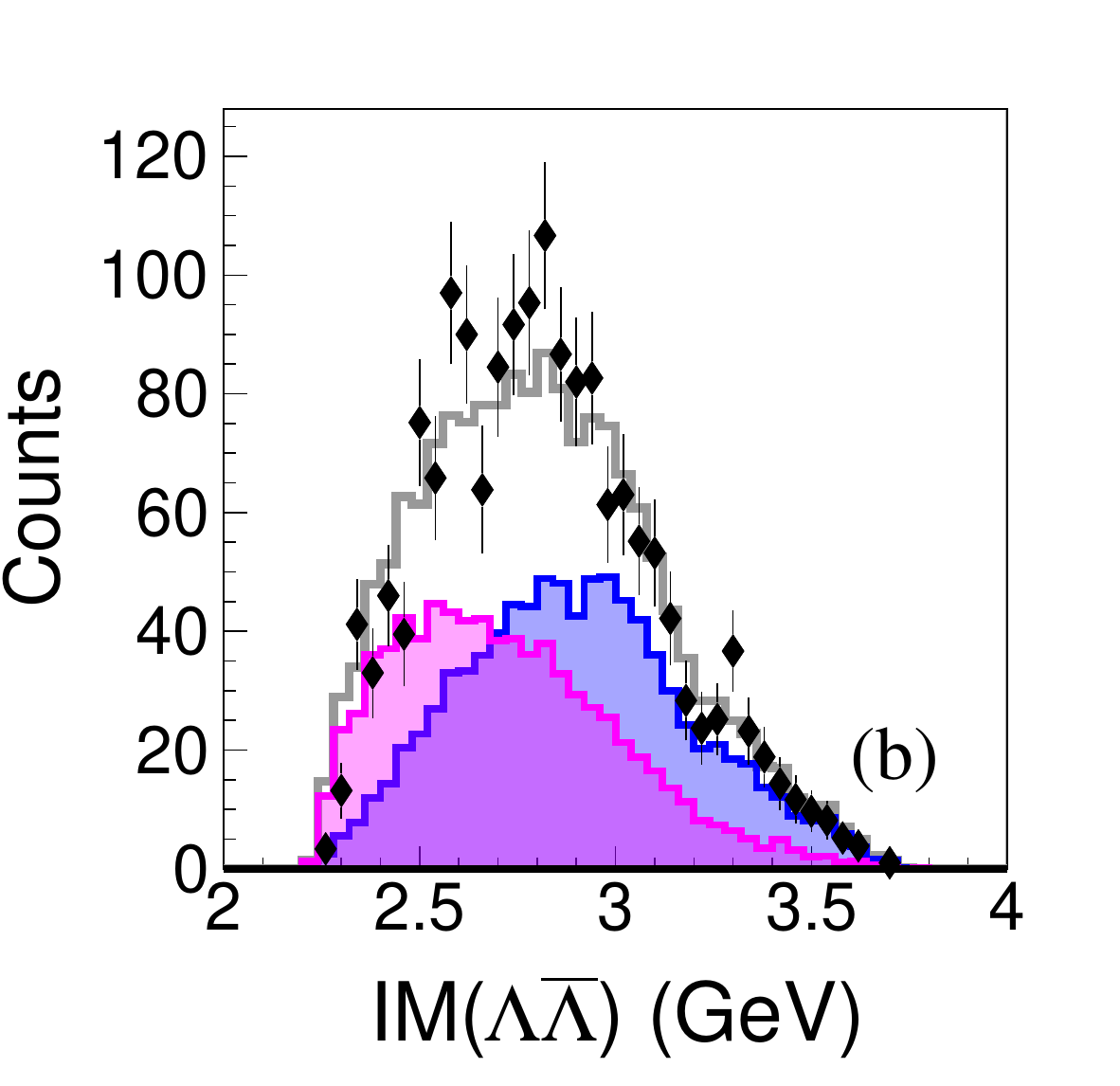}
		  \end{minipage}
\caption{\label{fig:plambar_IM_&_t_modelfit}(a) Invariant mass of $\plambar$ pairs for the more forward proton (black points), together with model fits for the single exchange (blue) and double exchange (magenta) processes.  (b) Pairings with the more backward proton.  The distributions are not acceptance corrected. }
\end{figure}

For the case of  $p\antiLambda$ production,  Fig.~\ref{fig:plambar_IM_&_t_modelfit} shows the invariant mass distribution of pairings, with Fig.~\ref{fig:plambar_IM_&_t_modelfit}(a) the produced proton with \antiLambdas together going mostly forward, and Fig.~\ref{fig:plambar_IM_&_t_modelfit}(b) the produced \antiLambdas with the recoil $\Lambda$.  The agreement of the fit of the model with the data in each of these cases is very good.

\section{Model Fitting}
\label{sec:reactionmodelfit}
     
Our reaction model was discussed in Sec.~\ref{sec:reactionmodel}, but we have not yet discussed the fitting procedure or the insights that could be gleaned from it.  A global likelihood function was defined to enable a maximum-likelihood estimation of the reaction model parameters on the basis of the experimental data.  The parameters were subsequently optimized to reproduce the observed kinematic distributions via Monte Carlo simulation.  The goal was to determine the reaction-dependent acceptances with a simple intensity model that introduces few parameters while remaining physically motivated.   No matrix elements from quantum amplitudes were computed.   
The components are added incoherently, so interference effects are not considered.   

We suppose future partial-wave analysis will incorporate quantum mechanical principles, but without clear resonant structures in the mass spectra, any amplitude-based analysis will need an ansatz consistent with this initial exploration of the reaction dynamics. 
    
For the statistically strongest channel, $p\antip$, a stochastic gradient descent (SGD) algorithm was developed to minimize the global negative-log-likelihood for the optimal set of parameters.  For the lower statistics \LamantiLam and \pantiL channels, a reduced parameter set was adapted from the \pantip case with iterative tuning without SGD.  As described in the following discussion, sets of parameters were obtained to accurately compute the acceptance of the \gx apparatus at all relevant beam energies and particle momenta.

\subsection{Parametric binned-likelihood method}
\label{sec:optimizing}
    

\input{p-anti-p_modeling_diagrams.tex}

For the baryon-antibaryon photoproduction processes reported here, the amplitudes illustrated in Figs.~\ref{fig:ppbar_combined_model}(a) and ~\ref{fig:ppbar_combined_model}(b) were evaluated phenomenologically as intensities given by Eqs.~\ref{eq:sR_modeling} and~\ref{eq:dR_modeling}.
Those intensities were incoherently summed as the components of the combined model as shown in Table~\ref{tab:all_reaction_parameters}, where all parameters and their descriptions are listed.  
    
For the $\gamma p \rightarrow p\ppbar$ process, the model components are labeled as $k=1,2,3$ for single-$t$-channel meson exchange ($k=1$), double-meson-baryon exchange ($k=2$), and the second instance of double meson-baryon exchange ($k=3$).
The nine parameters are: ``Single Exchange'' ($\theta_1 ... \theta_3$), ``Double Exchange I'' ($\theta_4 ... \theta_6$) , and ``Double Exchange II'' ($\theta_7 ... \theta_9$), respectively.  The second double-exchange parametrization was introduced as an {\it ad hoc} component to obtain a satisfactory fit to the data within the framework of our simple model.
With lower statistics available to fit the reactions $\Lambda\antiL$ and $p\antiL$, the first two components of the model with their six parameters, ``Single Exchange'' ($\theta_1 ... \theta_3$), ``Double Exchange I'' ($\theta_4 ... \theta_6$), were sufficient to describe the data distributions well.

The matching between data and Monte Carlo entails a binned likelihood \cite{barlow1993fitting} that employs Poisson statistics to incorporate the statistical uncertainties on both data and Monte Carlo distributions.
For an observable $\mathcal{X}$, we define bins as an array $\bm{x}$ of bin-edges indexed by $j$.  The joint likelihood function that matches the reconstructed Monte Carlo simulations with observed data is then the product of Poisson probability functions for each histogram bin: 
    \begin{align}
        \mathcal{L}_{\mathcal{X}}(\bm{x}; \bm{w},\bm{\theta}) = \prod_{j} \frac{e^{-M_j(\bm{x}; \bm{w}, \bm{\theta})} [M_j(\bm{x}; \bm{w}, \bm{\theta})]^{O_j}}{O_j!},
        \label{eq:likelihood}
    \end{align}
where $O_j$ is the observed or measured number of events at bin-$j$  for $\mathcal{X}$ in the data without acceptance correction.  $M_j$ represents the expected number of events in bin $j$, generated from the combined model with parameter set $\bm{\theta}$ and fractions $\bm{w}$ [see Eq.~\ref{eq:combined_histogram}], reconstructed using the reconstruction software and passing all event selection criteria as the observed data. The value of the likelihood function depends on $M_j$, which varies with different sets of model parameters. By minimizing the likelihood function, the optimal set of parameters that best describes the observable $\mathcal{X}$ is determined.

The combined histogram $\bm{M}(\bm{x}; \bm{w}, \bm{\theta})$ is constructed as a linear combination of the yields $\bm{N}^{(k)}(\bm{x};\bm{\theta})$ of all the components of the model, each scaled by a fraction $w_k$:
    \begin{equation}
        M_j(\bm{x};\bm{w},\bm{\theta}) 
        = 
        \sum_{k} w_k N_j^{(k)}(\bm{x};\bm{\theta}),
        \label{eq:combined_histogram}
    \end{equation}
where $j$ indexes the histogram bins, $k$ indexes the model components, and the fitted fractions sum to one, that is,  $\sum_k w_k = 1$ to preserve the overall normalization. 
    
For the $k^{th}$ component of the combined model, the binned-likelihood approach discretizes the continuous intensity function,
    $I_{\mathcal{X}}(x';\bm{\theta}): \mathbb{R} \to \mathbb{R}$, into an array of probabilities,
    \begin{align}
        p^{(k)}_{j}(\bm{x};\bm{\theta}) 
        =
        \frac{1}{\int I_{\mathcal{X}}(x';\bm{\theta}) dx'}
        \left(\int_{x_j}^{x_{j+1}} I_{\mathcal{X}}(x';\bm{\theta}) ~dx' \right), 
    \end{align}
each representing the observable $\mathcal{X}$'s probability density integrated over bin $j$, while ensuring $\sum_j p^{(k)}_{j} = 1$. The histogram representation of the expected yield for the model component-$k$ can be represented as: 
    \begin{equation}
         N_j^{(k)}(\bm{x}; \bm{\theta}) = \mathcal{N}_{\text{sample}} ~p^{(k)}_{j} (\bm{x}; \bm{\theta})  \epsilon^{(k)}_j(\bm{x}; \bm{\theta}), \label{eq:binned_hist_representation}
    \end{equation}
where $\mathcal{N}_{\text{sample}}$ is the sample size and $\epsilon^{(k)}_j$ is the joint acceptance and efficiency factor for the $j$-th bin.  In practice, $N_j^{(k)}$ is directly acquired by histogramming the reconstructed yields of the Monte Carlo simulation of each model component individually. 

    \begin{figure*}[tph]
		\centering
		  \begin{minipage}[b]{.3\textwidth}
		    \includegraphics[width=\textwidth]{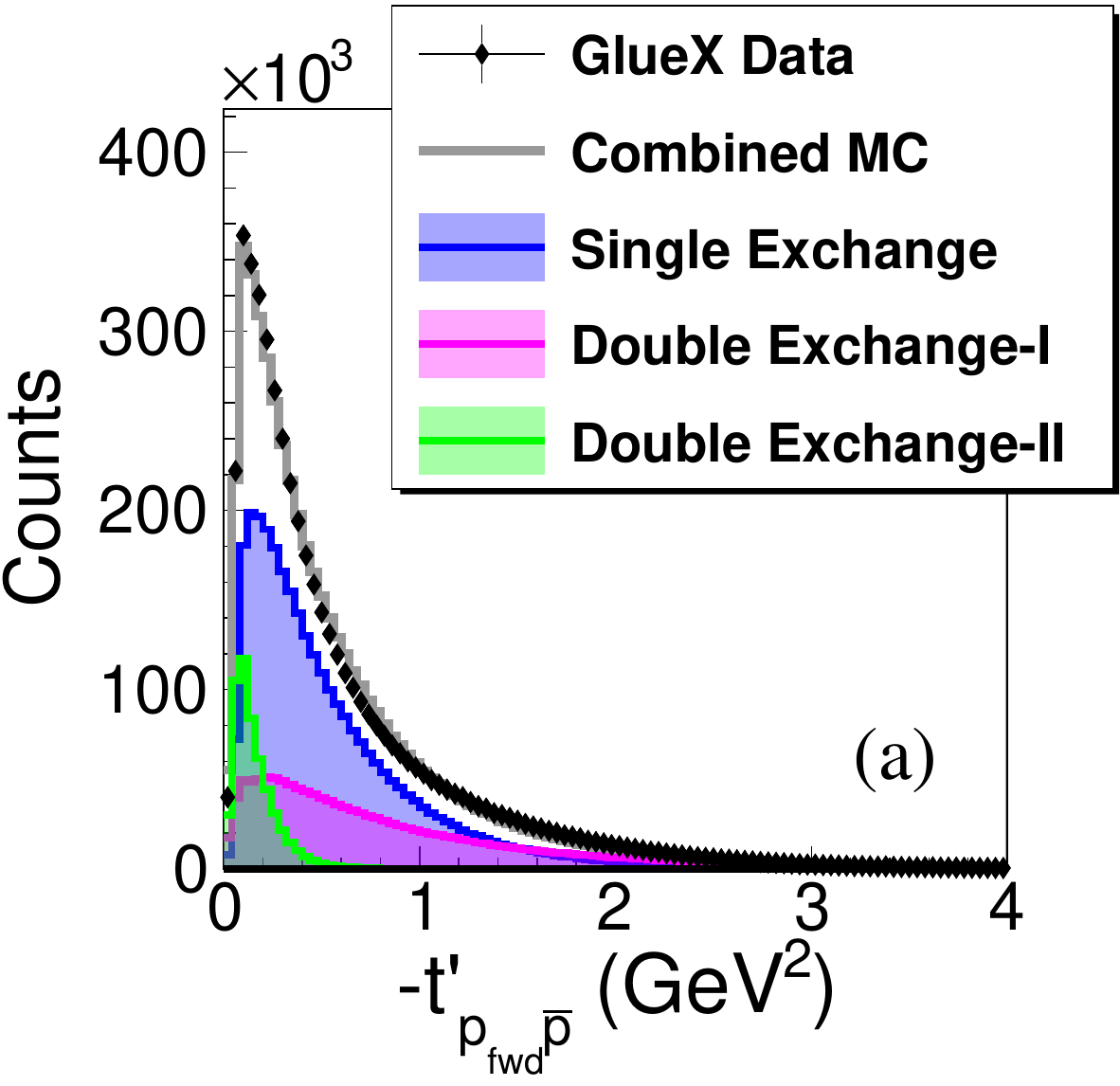}
		  \end{minipage}
		  \begin{minipage}[b]{.3\textwidth}
		    \includegraphics[width=\textwidth]{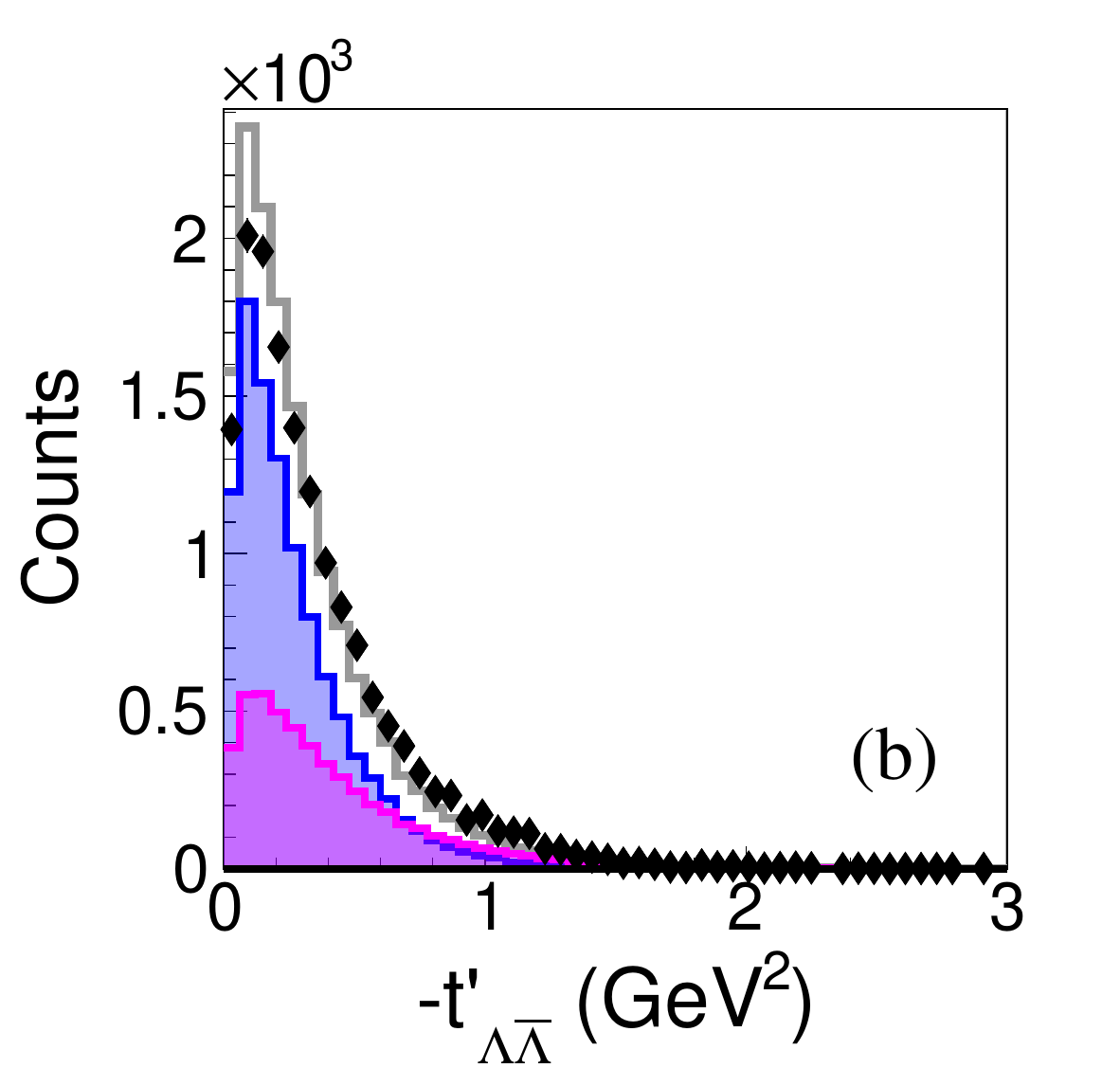}
            \end{minipage}
		  \begin{minipage}[b]{.3\textwidth}
		    \includegraphics[width=\textwidth]{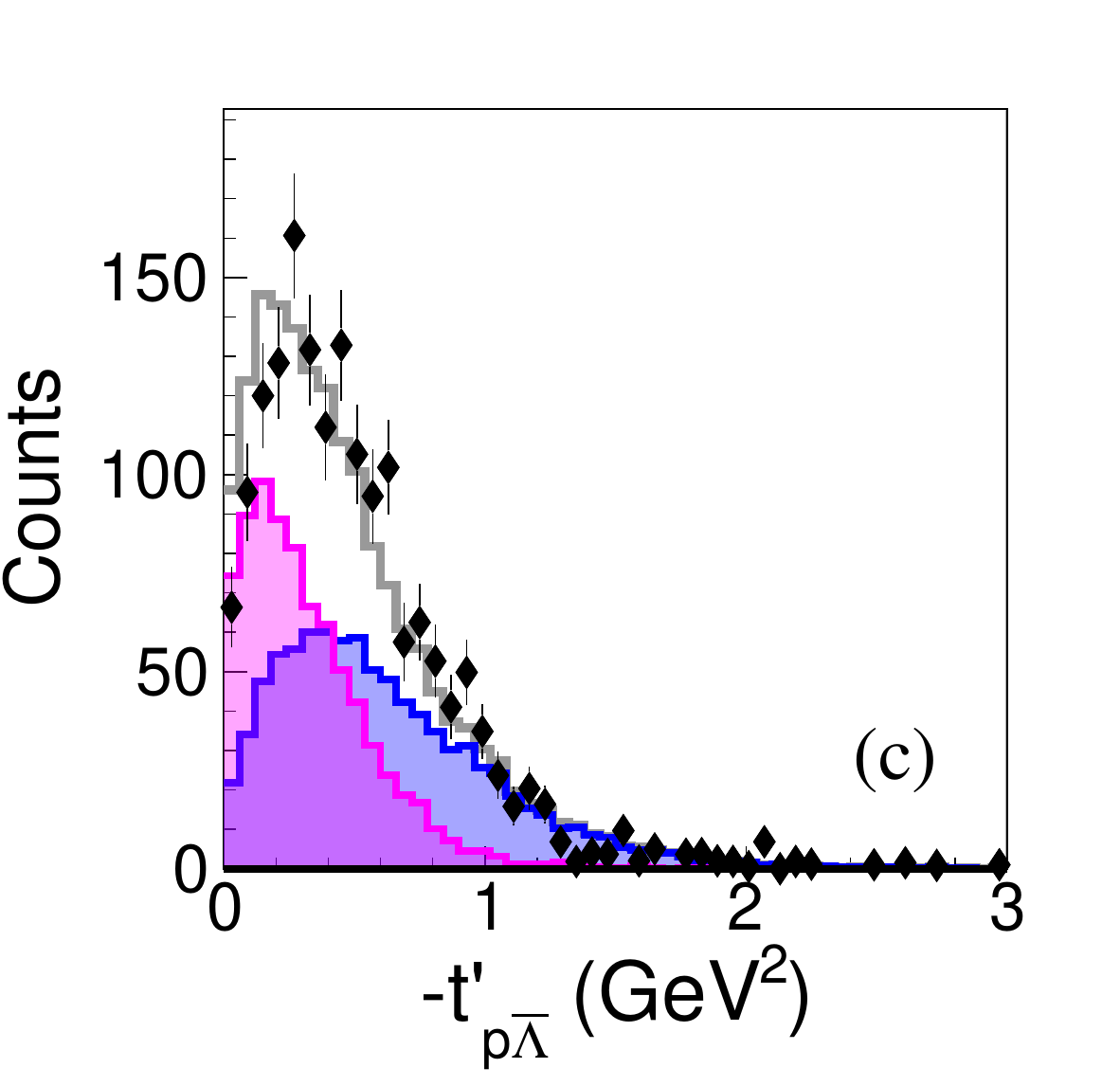}
		  \end{minipage}
    \caption{\label{fig:reduced_t_modelfit}
        Comparison between data and model distributions of the reduced four-momentum transfer, $t^{\prime}$, from beam photon to the target proton [Eq.~\ref{eq:reduced_t}], summed over all beam energies.
        (a) $p\antip$ pairs recoiling against $p$, 
        (b) $\lamlambar$ pairs recoiling against $p$,
        and (c) $\plambar$ pairs recoiling against $\Lambda$. 
        The model component colors are as in Fig.~\ref{fig:ppbar_angular}.
        These distributions are not acceptance corrected.}
    \end{figure*}

For the purpose of acceptance correction of the observed yields, $\epsilon^{(k)}_j$ was extracted from the ratio of accepted to generated MC events. 
The net acceptance can vary significantly between different model components due to the substantial differences in their characteristic kinematic distributions. For instance, the double-exchange events tend to have lower overall acceptance than the single-exchange events. The antibaryons move more backward in the CM frame, a region with different detector coverage in the \gx detector system.

To capture kinematic correlations between final-state particles, as discussed in Sec.~\ref{sec:data_analysis},  a global negative-log-likelihood function is defined through the sum of negative logarithm of Eq.~\ref{eq:likelihood},
    \begin{align}
        L_{\text{global}}(\bm{\theta}) &= 
        -\sum_{\mathcal{X}_i}\log \mathcal{L}_{\mathcal{X}_i}(\bm{\hat{w}}(\bm{\theta}), \bm{\theta})
        \label{eq:global_likelihood}
    \end{align}
serving as a measure of the similarity between data and reconstructed Monte Carlo simulations over a carefully chosen set of observables. 
The set, denoted as $\{\mathcal{X}_i\}$,
    \begin{align}
    \{\mathcal{X}_i\} = \{ \theta^{\text{CM}}_{B}, \theta^{\text{CM}}_{\bar{B}}, m_{B\bar{B}} \},\label{eq:observable_set}
    \end{align}
    consists of the CM frame polar angular distributions of the baryons and antibaryons, along with the invariant masses of all baryon-antibaryon pairs. 
    The local best-fitting fractions $\bm{\hat{w}}$ as a function of $\bm{\theta}$ are
    \begin{equation}
        \bm{\hat{w}}(\bm{\theta}) \equiv \mathop{\arg\min}_{\bm{w}} \left[ -\sum_{\mathcal{X}_i} \log \mathcal{L}_{\mathcal{X}_i}(\bm{w}, \bm{\theta}) \right],\label{eq:w_hat}
    \end{equation}
where the minimization is performed over $\bm{w}$ while keeping $\bm{\theta}$ fixed. This ensures that $\bm{\hat{w}}(\bm{\theta})$ is obtained deterministically as the conditional maximum likelihood estimate given a set of fixed parameters $\bm{\theta}$.  
    
The fitted fractions $\bm{\hat{w}}(\bm{\theta})$ have not yet been acceptance corrected. To remove acceptance bias, we define the acceptance-corrected fractions for each $B\overline{B}$ reaction channel as
    \begin{equation}
        \hat{w}^{\BBbar}_k(\bm{\theta}) \;=\; \left(\displaystyle \frac{\hat{w}_k(\bm{\theta})}{\bar{\epsilon}_k}\right)
        \Bigg/
        \left(\displaystyle \sum_{k'} \frac{\hat{w}_{k'}(\bm{\theta})}{\bar{\epsilon}_{k'}}\right),\label{eq:acceptance_corrected_w_hat}
    \end{equation}
    where 
    $
    \bar{\epsilon}_k = \sum_j p^{(k)}_j (\bm{x}; \bm{\theta}) \epsilon^{(k)}_j (\bm{x}; \bm{\theta})
    $
is the \textit{effective efficiency} for component $k$. The fractions sum up to one and represent the acceptance-corrected composition of each component.

\subsection{Optimizing model parameters}
\label{sec:optimizedparameters}

The combined model parameters $\bm{\theta}$ are effectively optimized by minimizing the global negative-log-likelihood $L_{\text{global}}$ [Eq.~\ref{eq:global_likelihood}] using any preferred optimization method:
   \begin{equation}
       \bm{\hat{\theta}} \equiv \mathop{\arg\min}_{\bm{\theta}} ~L_{\text{global}}(\bm{\theta}).
       \label{eq:optimal_parameters}
   \end{equation}
Table~\ref{tab:all_reaction_parameters} lists, for each baryon-antibaryon photoproduction reaction, the optimized parameters fitted to data covering the full beam energy range. The parameters corresponding to the individual components of the combined model are detailed in the respective rows.
    
The optimization of the parameters set $\bm{\hat{\theta}}$ for the \pantip dataset used an iterative numerical method based on ``online convex optimization'' \cite{flaxman2004online} employing a SGD method. The algorithm used to implement this method is described further in Appendix~\ref{appendix:sgd}. The global negative-log-likelihood surface in the parameter space is determined not only by the intensity of the three model components but is also significantly shaped by the detector acceptance embedded in the simulation and event reconstruction.  The method was found to be efficient in identifying the optimal set of the nine parameters in as few as 20 iterations.  
     
The accuracy of the optimized model is shown in comparison between the \pantip data and Monte Carlo simulation across a list of kinematic variables in Figs.~\ref{fig:ppbar_angular}, \ref{fig:IM_&_t_modelfit} and \ref{fig:reduced_t_modelfit}(a) where good agreement was achieved. The correlation among the three final-state longitudinal momenta, shown in Fig.~\ref{fig:vanhovedistributions} was also well approximated. A notable exception is the invariant mass of $p_{\text{bkwd}}\bar{p}$ [Fig.~\ref{fig:IM_&_t_modelfit}(b)], where the small discrepancy may stem from unmodeled effects, such as perhaps weak mass clustering within the double-exchange component.

For the lower statistics \LamantiLam and \pantiL channels, parameter optimization was performed using a grid search method, which minimized the same global negative-log-likelihood $L_{\text{global}}(\bm{\theta})$ without employing the SGD method. The comparison between the data and Monte Carlo simulations across various kinematic observables, as shown in Figs.~\ref{fig:lamlambar_angular}, 
    \ref{fig:plambar_angular}, 
    \ref{fig:vanhovedistributions}, \ref{fig:lamlambar_IM_&_t_modelfit}, \ref{fig:plambar_IM_&_t_modelfit}
and \ref{fig:reduced_t_modelfit}(b) and \ref{fig:reduced_t_modelfit}(c), indicates generally good agreement. 

As shown in the ``Fitted values'' columns of Table~\ref{tab:all_reaction_parameters}, the exponential mass clustering parameter $c_m$ (second row) is very similar across all reactions, at approximately 0.2~GeV.  For the single-exchange component of the nonstrange meson cases in Figs.~\ref{fig:feynman}(a) and~\ref{fig:feynman}(c), it appears that the slopes $\alpha'$ (first row) are compatible and similar, while for the strange meson exchange depicted in Fig.~\ref{fig:feynman}(e), the slope is smaller.  
    
For each of the $p\antip$, $\Lambda$\antiLambda,  and \pantiL reactions, the fitted fractions of each model components, $\bm{\hat{w}}$, obtained from the fit to the dataset spanning all beam energies, are summarized in Table~\ref{tab:all_reaction_fractions}. It appears that the single exchange component is dominant in both the $\pantip$ and $\LamantiLam$ reaction channels, but accounts for less than half of the $\pantiL$ channel.

    \begin{table}[htbp]
    \centering
    \caption{\label{tab:all_reaction_fractions}
     Fitted fractions of the combined model for different baryon-antibaryon photoproduction reactions. The values in the table have been acceptance corrected using Eq.~\ref{eq:acceptance_corrected_w_hat}.
     }
    \renewcommand{\arraystretch}{1.5}
    \begin{tabular}{l|c|c|c}
    \hline
    \textbf{Component} & \textbf{\boldmath$\mathbf{\hat{w}}^{\ppbar}$ (\%)} & \textbf{\boldmath$\mathbf{\hat{w}}^{\lamlambar}$ (\%)} & \textbf{\boldmath$\mathbf{\hat{w}}^{\plambar}$ (\%)} \\ \hline\hline
    $k=1$: Single Exchange    &$54.9 \pm 1.8$ & $61.3\pm2.1$ & $42.4\pm2.6$ \\ \hline
    $k=2$: Double Exchange-I  &$25.5 \pm 1.8$ & $38.7\pm2.1$ & $57.6\pm2.6$ \\ \hline
    $k=3$: Double Exchange-II &$19.6 \pm 1.8$ & -- & -- \\ \hline
    \end{tabular} 
    \end{table}
    \renewcommand{\arraystretch}{1.0}

We base the uncertainty estimations on the global negative-log-likelihood of the fits.  Confidence intervals are constructed by examining the probability of the log-likelihood  change when a model parameter varies. The probability is computed from the $\chi^2_{\nu}$ distribution in the Gaussian statistics limit. For the uncertainty on the optimal fractions $\bm{\hat{w}}$, as defined in Eq.~\ref{eq:acceptance_corrected_w_hat}, we perform a standard profile likelihood scan using a $1\sigma$ confidence interval. For the best parameters $\bm{\hat{\theta}}$, we employ a conditional likelihood parameter scan, in which each parameter is individually varied while all other parameters are fixed at their best-fit values. The confidence intervals in the best parameters are tailored to the curvature of each $\chi^2$/NDF curve, ensuring an accurate representation of the statistical uncertainty without imposing an arbitrary uniform selection.  The uncertainties we estimated are tabulated as parts of Tables~\ref{tab:all_reaction_parameters} and ~\ref{tab:all_reaction_fractions}. More information on the fitting and error estimation can be found in  Appendix~\ref{appendix:sgd}.
    
This approach is effective for the \pantip\ channel, where the bin-by-bin statistics are sufficient to approximate using Gaussian uncertainties. 
For the hyperon channels, the lower statistics and the alternative fitting method make the direct estimation of uncertainties less definite. The uncertainties for these parameters are possibly underestimated, but are assumed to be similar to those estimated for the \pantip\ channel given the similar kinematic distributions and systematic behaviors.  
    
\subsection{Energy-dependency of model parameters}
\label{sec:energy_dependence}
The robustness and stability of our phenomenological model can be tested by examining its sensitivity to photon beam energy. We expect the incident energy to affect the detailed nature of the interactions of the reaction components. The abundant statistics in the $\ppbar$ reaction channel allow us to partition the data further into seven bins of photon beam energy to investigate the evolution of the reaction parameters with energy. 
    
For fits restricted to specific beam energy ranges, the global negative-log-likelihood [Eq.~\ref{eq:global_likelihood}] was optimized following the SGD approach described in the previous section. The global negative-log-likelihood included the same distributions of observables as before [Eq.~\ref{eq:observable_set}].  The results revealed the model's robustness in reproducing the observed data across different beam energy ranges.

\begin{figure}[htpb]
  \includegraphics[width=.95\columnwidth]{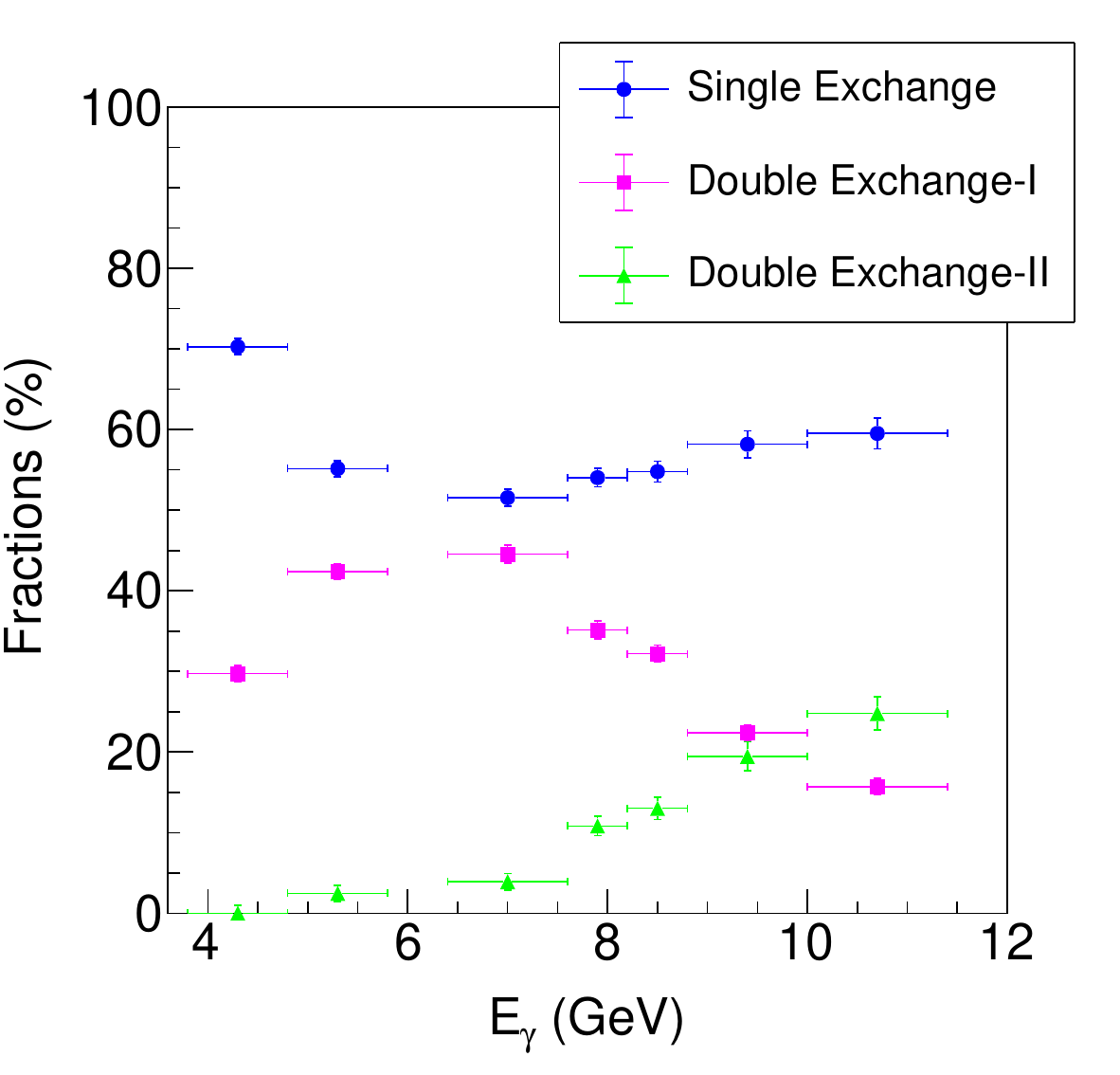}
  \caption{Proportions of the reaction mechanism components describing the \pantip photoproduction reaction as a function of beam energy: Single Exchange (blue), Double Exchange I (magenta), and Double Exchange II (green).
  }
  \label{fig:energydependentfractionspbarp}
\end{figure}

In the $\ppbar$ channel, the three reaction components in the combined model contribute with different strengths at different photon beam energies. This is shown in Fig.~\ref{fig:energydependentfractionspbarp}, where it is evident that the single exchange mechanism accounts for between 50\% and 70\% of the reaction.  The single exchange fraction varies somewhat with beam energy and seems to be smallest near 6.4 to 7.6 GeV, which is just where the total cross section is largest, as shown later.  The balance is composed of the two double-exchange mechanisms.  The fractions vary smoothly with beam energy,  showing the stability of the fitting algorithm but also hinting that there is some evolution of the reaction mechanisms with energy.  
    
\begin{figure*}[htpb]
  \centering
  \includegraphics[width=.90\textwidth] {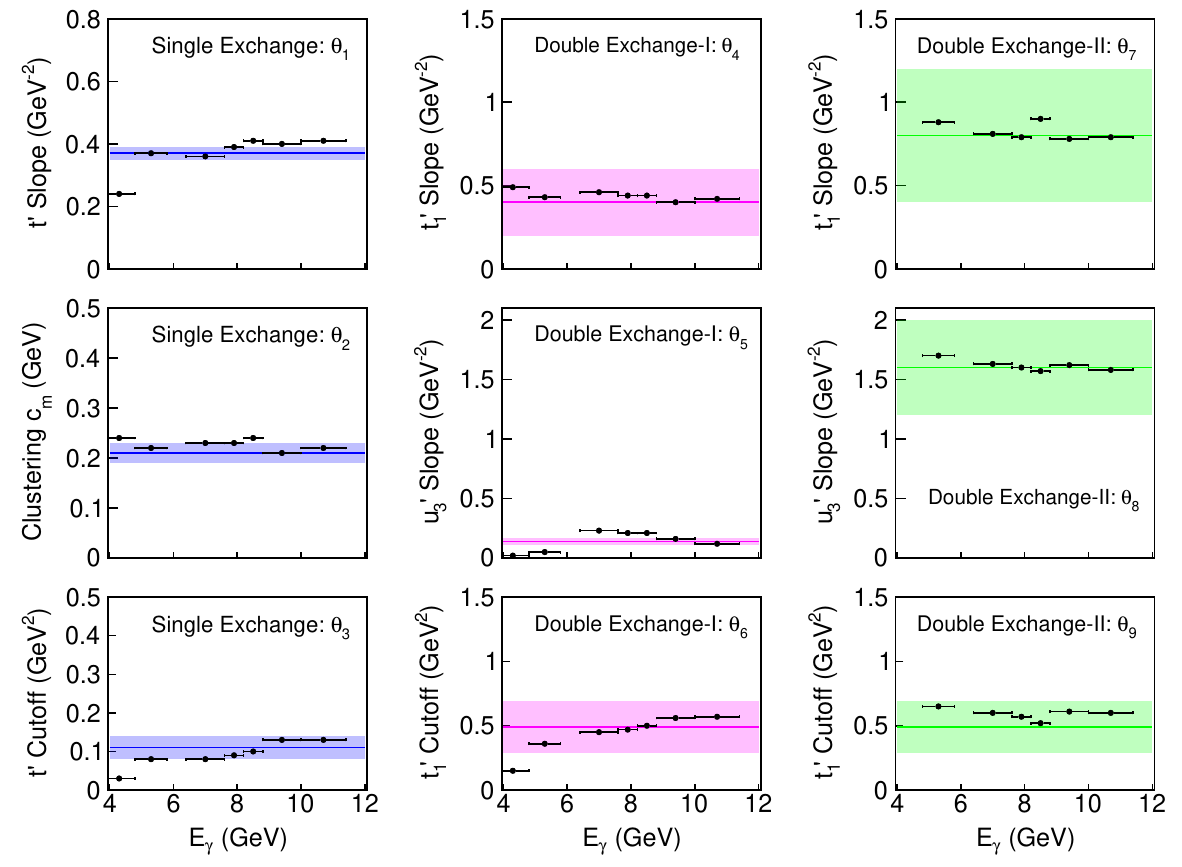}
  \caption{Beam energy dependence of the nine \pantip combined model parameters.  
  Left column: ($\theta_1$, $\theta_2$, $\theta_3$) for the Single Exchange component, depicted in black solid circle markers.
  Middle column: Double Exchange I ($\theta_4$, $\theta_5$, $\theta_6$). 
  Right column: Double Exchange II ($\theta_7$, $\theta_8$, $\theta_9$).  The colored bands summarize the global fit values and the estimated global fit uncertainties from Table~\ref{tab:all_reaction_parameters} for the whole GlueX energy range.} 
  \label{fig:modelparamspbarp}
\end{figure*}

As discussed previously, finding satisfactory fits to the \pantip data required including an {\it ad hoc}  second double exchange component, though not for the hyperon channels, due to lower statistics. It is possible that this rather smooth beam energy dependence is a sign that our method simply does not capture the quantum interference effects that must surely be present.  This remains an open question.

Figure~\ref{fig:modelparamspbarp} shows the results for the nine fitted reaction parameters that describe the \pantip photoproduction reaction on a set of seven beam energy bins.   Most of the parameters are quite stable across all beam energies.  The single-exchange $t$-slope $\theta_1$ (top left) stays nearly constant at about 0.4 GeV$^{-2}$ except at the lowest beam energy. The associated cut-off parameter $\theta_3$ (bottom left) increases smoothly to a maximum of about 0.12 GeV$^{-2}$, again with some deviation at the lowest beam energy.  

The exponential mass clustering parameter, $\theta_2$ ($c_m$ in Eq.~\ref{eq:clustering_modeling}), is largely independent of the beam energy at about 210 MeV, corresponding to an RMS relative momentum in the \pantip system of 0.46 GeV/c and a kinetic energy of 0.14 GeV.  This supports the hypothesis that it is a property of the pair system rather than the overall reaction dynamics.  

The double-exchange part of the fit is what enables the description of the broader angular distribution of antiprotons that is not mirrored by the protons. We find that the associated model parameters ($\theta_4$--$\theta_9$) show no apparent energy dependence except for the lowest beam energies, as shown in Figure~\ref{fig:modelparamspbarp}.

For the two hyperon production channels, event statistics were insufficient to compute beam-energy dependent model parameters.   A single set of parameters was extracted across all beam energies for each reaction channel.
A satisfactory description of the available data was achieved using a single double exchange component in the reaction mechanism for both hyperon creation cases. 

For the cross section measurements presented in the next section, a single optimized set of nine (or six) parameters was applied uniformly for each baryon-antibaryon photoproduction reaction in the Monte Carlo simulation. This ensured consistent modeling across the full energy range for each reaction, spanning up to 11.4~GeV. The optimized parameters and fractions used for these simulations are tabulated in Tables~\ref{tab:all_reaction_parameters} and \ref{tab:all_reaction_fractions}, and shown for the \pantip case in 
Fig.~\ref{fig:modelparamspbarp}.

\section{Results and Discussion}
\label{sec:resultsanddiscussion}

\subsection{Near-threshold mass spectrum of \texorpdfstring{$p\antip$}{Lg}}
\label{sec:thresholdstates}

\input{baryonium_table.tex}

Early interest in proton-antiproton photoproduction included the search for bound or unbound resonant states, summarized in Table~\ref{tab:earlymeasurements}.  (Cf.~Ref.~\cite{MONTANET1980201}).  A meson decaying to $p\antip$ could be visible, for example, in the threshold region.  Early measurements of this reaction with bubble chambers showed narrow accumulations with low-statistics in exclusive measurements.  A DESY experiment~\cite{Bodenkamp:1984dg, Bodenkamp:1981ui} reported possible narrow baryonium structures at 2.02 and 1.94 GeV with widths of about 29 MeV. Other experiments did not support these results and interest in the topic declined.   We revisit this type of search in the continuum region here. 

\begin{figure}[htpb]
  \centerline{\includegraphics[width=\columnwidth]{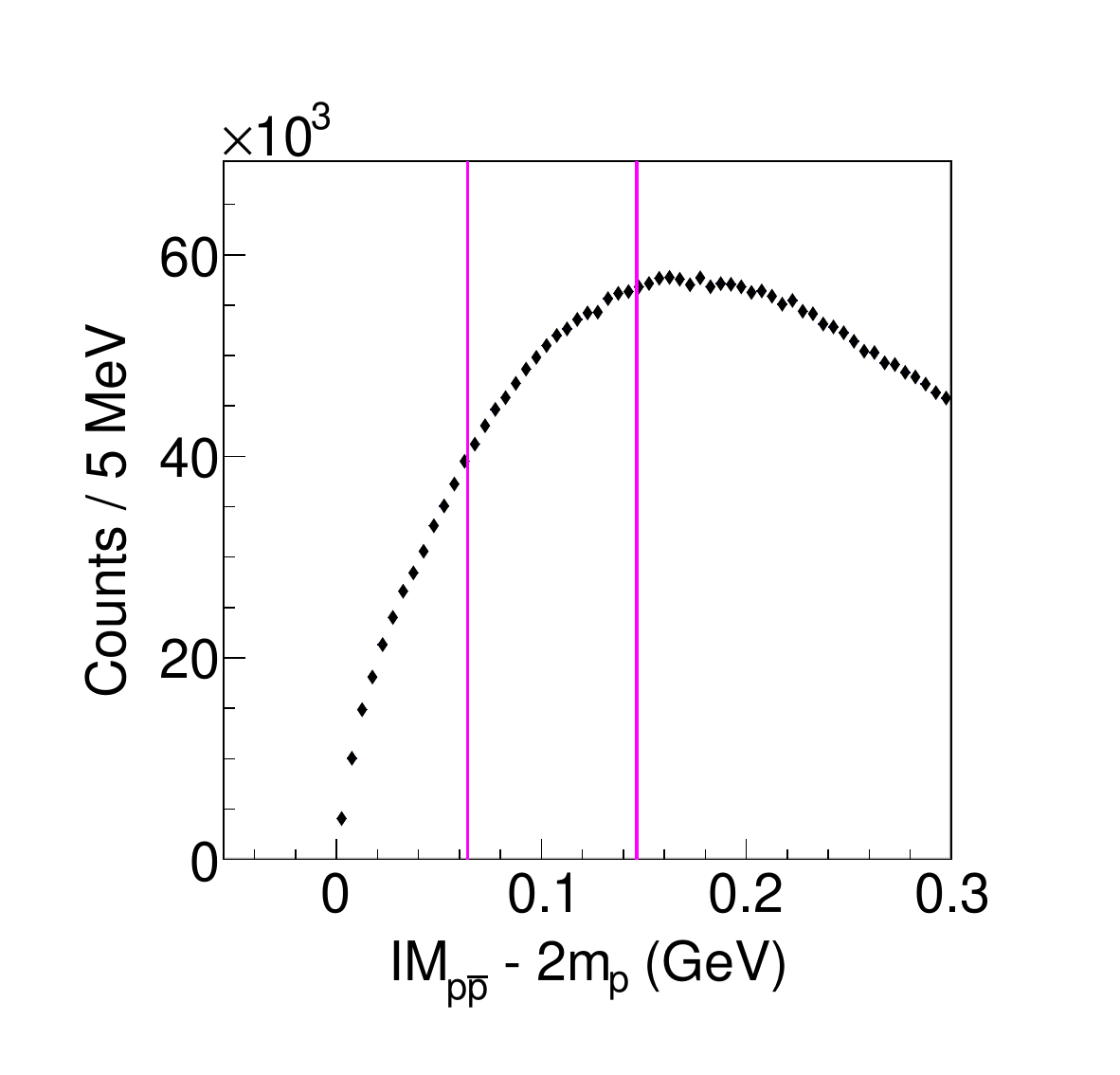}}
  \caption{Events per 5 MeV mass bin versus \pantip invariant mass for diffractively created pairs. The GlueX mass resolution for the forward-going $\ppbar$ system is 3 to 5~MeV in this region.  The threshold mass of $2m_p$ is subtracted.   The vertical magenta lines indicate where previous measurements (DESY~\cite{Bodenkamp:1984dg}) suggested that narrow peaks may occur.}
  \label{fig:pbarp_invariantmass}
\end{figure}

The threshold region $p\antip$ invariant mass for the present \gx dataset is shown in Fig.~\ref{fig:pbarp_invariantmass}.  Antiprotons were paired with the more forward of the two protons in the final state because these are more likely to be the proton created in the dominant single-exchange reaction mechanism.  
The plotted bin width is 5 MeV, which matches the \gx mass resolution in this mass region, estimated to be $3$ to $5$~MeV FWHM using the Monte Carlo simulation. The present data have at least $~10^3$ times the statistical power of the early measurements. Evidently, there are no narrow peaks visible here at these masses or elsewhere. 
A similar examination of the near-threshold behavior of the \LamantiLam and \pantiL reactions, though with much lesser statistics, also showed no narrow mass structures.  

%
%
\subsection{Total cross sections for \texorpdfstring{$p\antip$}{Lg}, \texorpdfstring{$\Lambda\antiLambda$}{Lg}, and \texorpdfstring{$p\antiLambda$}{Lg} photoproduction off the proton}
\label{sec:totalcrosssection}

\begin{figure*}[htpb]
  \centerline{\includegraphics[width=.995\textwidth]{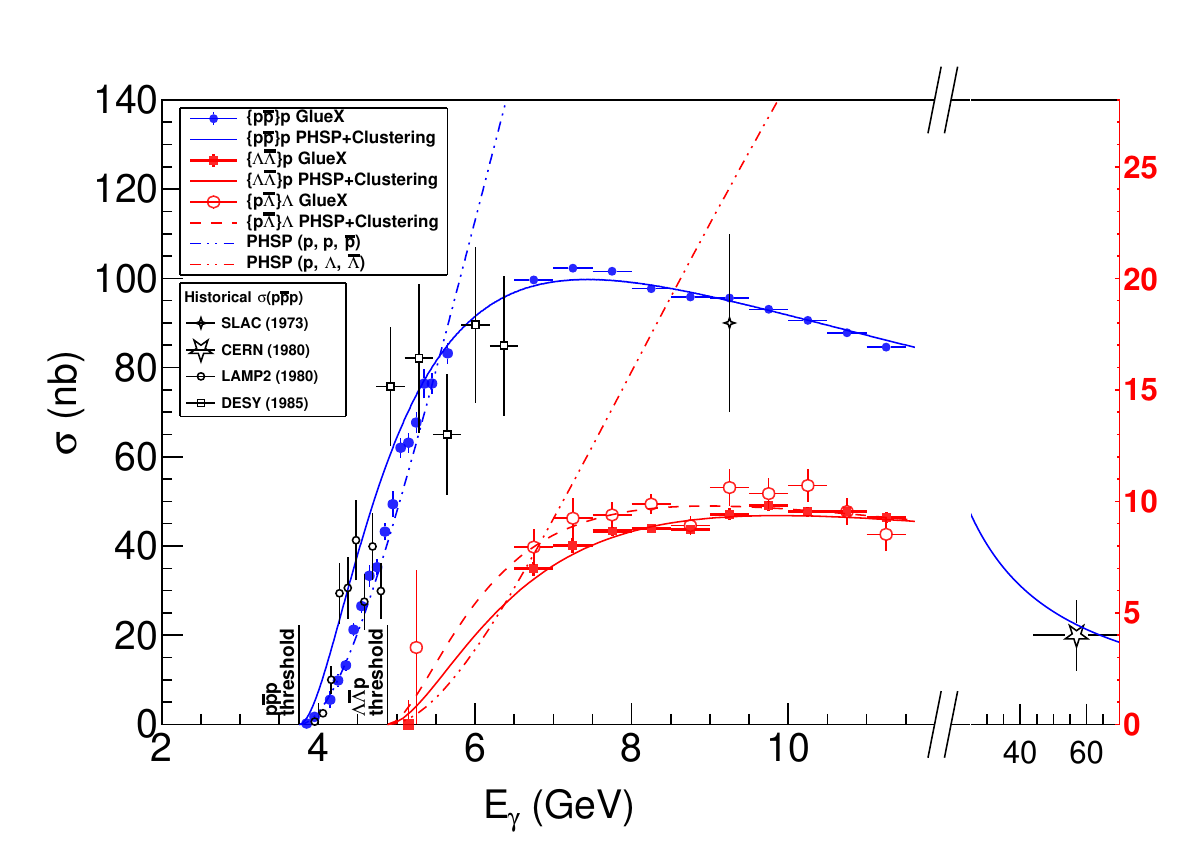}}
  \caption{Total photoproduction cross sections off the proton, as a function of beam energy, for the $\{\pantip\}$ (blue circles), $\{ \LamantiLam \}$(red squares), and $\{p \antiL\}$ (open red circles) systems.  The red axis on the right corresponds to the red data points and curves in the figure, while the black axis on the left corresponds to all other data points and curves.  The error bars show the statistical uncertainties only.  Comparison is made to a selection of previous $\{\pantip\}$ measurements from Daresbury (open circles)~\cite{Barber:1979ah},  DESY~\cite{Bodenkamp:1984dg} (open squares),  SLAC~\cite{Bingham:1973fu} (4-point star) and CERN~\cite{Aston:1980gri} (5-point star).  The curves represent the behavior of the total cross section modeled with the simplified picture [Eq.~\ref{eq:totalcrosssectionmodel}], for each reaction channel: \pantip (solid blue), \LamantiLam (solid red), and ${p \antiL}$ (dashed red). The blue model curve for the \pantip channel also includes an extrapolation to the high energy datum from CERN.  As a reference, the dotted-dashed curves illustrate integrated three-body phase space (Eq.~\ref{eq:phsp_sigma}) for the $pp\bar{p}$ (blue) and $p\Lambda\bar{\Lambda}$ (red) final states. 
  \label{fig:total_cross_section}
  }
\end{figure*}

Having established a reaction model that accurately describes the \gx data over the full energy range, simulations for each model component (\(k = 1, 2, 3\)) were performed using the optimized parameters given in Tables~\ref{tab:all_reaction_parameters} and \ref{tab:all_reaction_fractions}. 
This led to three Monte Carlo datasets for the \(\ppbar\) channel, with  \(O(10^7)\) events in each category: single exchange, and two double exchanges. Each dataset was processed through the full \gx simulation and reconstruction procedures, as well as the same selection criteria applied to the real data, to ensure a realistic reflection of the joint efficiency and acceptance.  Simulated datasets of similar size were generated for the \pantiL and \LamantiLam reactions for single and double exchange components.  
The overall acceptance for each reaction channel was computed in energy bins of 100~MeV from each reaction threshold up to approximately 5.8~GeV where the cross sections increase rapidly. For higher energies, from 6.5 to 11.4~GeV, 500~MeV bins were used.   The gap in beam energy coverage between about 5.8 and 6.5 GeV resulted from choices made in the \gx ``low-energy" and ``high-energy" measurement program.

The acceptance-corrected events in each photon energy bin were normalized to the integrated experiment luminosity.   The luminosity included the time-integrated beam photon flux and target thickness factors.   The total luminosity for all energies was $440~\text{nb}^{-1}$. 

The total cross section is plotted in Fig.~\ref{fig:total_cross_section} for the reactions $\{p\antip\}p$,  $\{\Lambda\antiLambda\}p$, and $\{{p\antiL}\}\Lambda$, and tabulated in the Appendix Tables~\ref{tab:totalcrosssections_pbarp}, \ref{tab:totalcrosssections_LambarLam}, and \ref{tab:totalcrosssections_pbarL}, respectively.  The error bars show only the statistical uncertainties.  The reaction-dependent systematic uncertainties are discussed in Sec.~\ref{sec:systematics}. 

\subsubsection{\texorpdfstring{$p\antip$}{Lg} total cross section}

The total cross section as a function of beam energy might na\"ively be expected to scale roughly as the available three-body final state phase space that grows rapidly with available energy, $\sqrt{s}$, as long as other final states do not drain flux from the particular channel of interest.   We can now show what is the actual case.

The $p\antip$ total cross section rises rapidly from threshold to a peak value of about 100~nb near $E_\gamma=8$~GeV, then starts a slow descent to the upper limit of the measurements at 11.4~GeV. These data span the largest beam energy range among all previous measurements.  The results are in good to excellent agreement with previous published work across smaller beam energy ranges
~\cite{Ballam:1971yd, Bingham:1973fu, Barber:1979ah, Aston:1980gri, Bodenkamp:1984dg}.  

A fair description of the total cross section result is achieved using our intensity-based model introduced in Sec.~\ref{sec:totalcrosssectiontheory}.  The total cross section is computed using the formulation for the dominant ($50\% - 60\%$) single-exchange process in Eq.~\ref{eq:totalcrosssectionmodel} using two free parameters:  an overall scaling factor $\mathcal{N}_{\BBbar}$ and a single mass clustering factor $c_{\text{total}}$, both fitted to the total cross section data points.  The total cross section model is independent of the slope parameter $\alpha^{\prime}$ as a consequence of the assumption made in the construction of $\norm{{\mathcal{M}_{fi}}}^2$. It is based on the observed $m_{12}$-dependence (Fig.~\ref{fig:q_pcm_vsIM}) for the single-exchange production picture but neglects the complications introduced by the sub-dominant ($40\% - 50\%$) double exchange contribution.  

The result for the $\ppbar$ channel is shown in Fig.~\ref{fig:total_cross_section} as the solid blue curve.   This demonstrates that the overall beam energy dependence can be reasonably related to the available phase space plus one additional parameter that describes what we call the mass clustering effect.  The extrapolation of the curve to about 60 GeV shows that it matches the single data point available at that higher energy.  In the threshold region of the $p\antip$ channel, it is evident that the cross section rises more slowly than in our simple two-parameter model. 
We note the likely presence of annihilation processes (e.g. Refs.~\cite{Haidenbauer:2022baz, Dai2017ont,  acharya2022investigating}) that may deplete the cross section where final state momenta are smaller than at higher energies.  Our simplified model for the total cross section completely neglects the double-exchange type of effects discussed previously.   They are evidently not apparent when looking at simply the total cross section parameterized with this minimal model.

There are other measurements at Jefferson Lab of the photoproduction of $p \antip$ pairs off the proton measured in a series of largely unpublished measurements from the CLAS Collaboration  ~\cite{bstokesPhD2006, wphelpsPhD2017} and~\cite{Phelps:2016huv} for laboratory photon beam energies up to 5.4 GeV. Consequently, we do not compare our results directly with those efforts.   

\subsubsection{\texorpdfstring{$\Lambda\antiLambda$}{Lg} and \texorpdfstring{$p\antiLambda$}{Lg} channels and strangeness suppression}
\label{suppression}

The total cross sections for $\Lambda \antiLambda$ and $p \antiLambda$ shown in Fig.~\ref{fig:total_cross_section} have not been previously reported.  It is notable that the two reactions have similar total cross sections within the precision of this measurement.   This means that the mechanisms of strange and nonstrange meson exchanges in Fig.~\ref{fig:feynman} (c) and (e) are equally significant.   The couplings at each of the internal lines for both diagrams have one strangeness- and one nonstrangeness-producing vertex, albeit in different places, so it is somewhat surprising that the cross sections are so similar.    The phase space and mass clustering leading to the overall shape of the total cross sections works well for both of these channels, as shown by the \LamantiLam (solid red) and \pantiL (dashed red) curves, as it did for the $p\antip$ case.   

We use these results to compare the intrinsic strength in these reaction channels of strange ($s$-quark) versus nonstrange light ($u$ and $d$-quark) final states.  This is of interest in hadronic production models using ``flux tube'' breaking approaches, such as the Lund model implementations of event generators including \texttt{JETSET} and \texttt{PYTHIA}.   (See Ref.~\cite{CLAS:2014gcd} for a compilation of relevant references.)  
Empirically, the Lund model requires the introduction of suppression factors for the production of heavier quark pairs:
\begin{equation}
u:d:s:c = 1:1:\frac{1}{3}:10^{-11},
\end{equation}
which have been examined in previous experiments, e.g., Refs~\cite{BOCQUET1996447, wroblewski1985strange, malhotra1983determination}.
Incorporated into the \texttt{PYTHIA} event generator,
the strangeness suppression factor $\lambda_s \approx 0.3$
has been shown to accurately describe the data from
electron-positron collision experiments \cite{PhysRevD.59.052001}
as well as from high-energy deep-inelastic scattering experiments
involving electrons and protons \cite{h12009strangeness}.

\begin{figure}[htpb]
  \centerline{\includegraphics[width=.995\columnwidth]{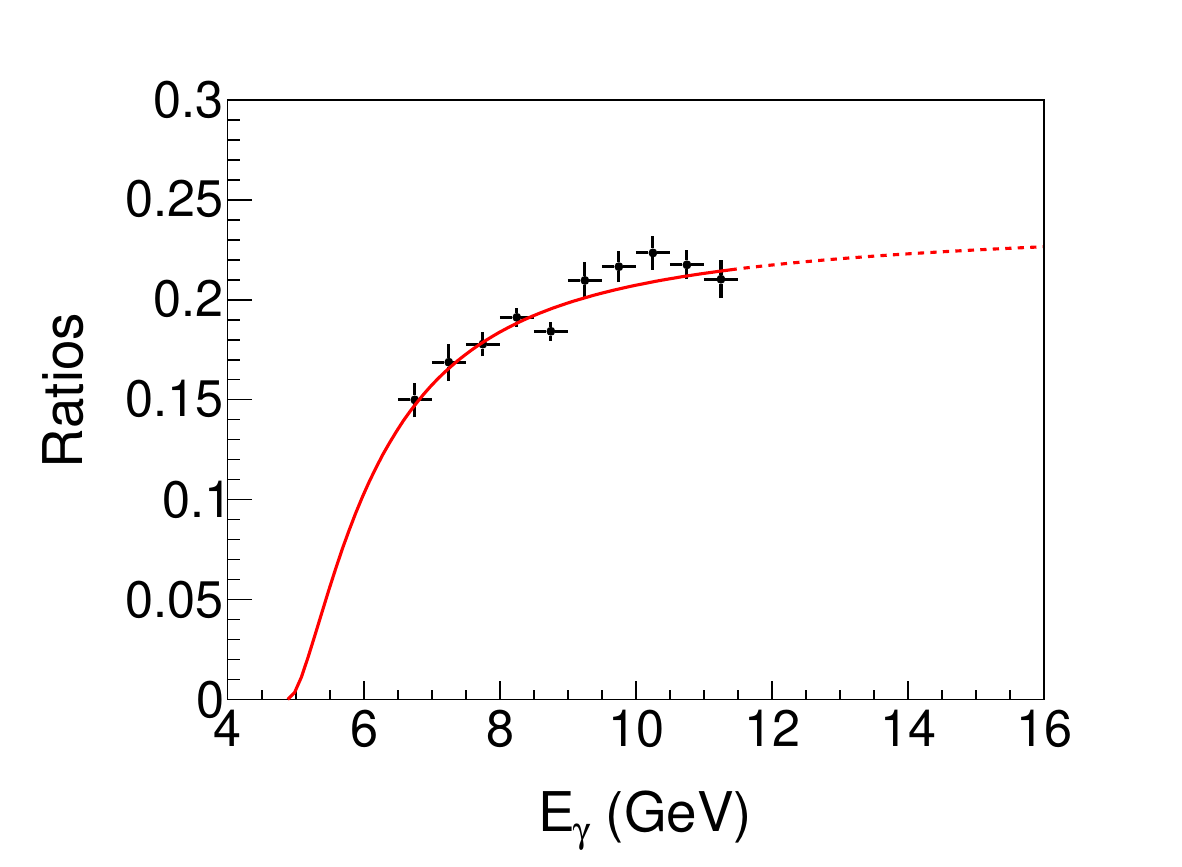}}
  \caption{Comparison of the total cross-section measurement ratios. Data points show the ratio between strange (sum of the $\lamlambar$ and $\plambar$ production) and nonstrange ($\ppbar$) production cross sections; the red line depicts the computed ratio using our total cross-section calculations.  The dotted line is an extrapolation to unmeasured energies. 
  \label{fig:total_cross_section_ratio}
  }
\end{figure}

The overall ratio of nonstrange to (singly) strange baryon-antibaryon production, in the asymptotically nearly flat part of the distributions (Fig.~\ref{fig:total_cross_section_ratio}), 
is estimated as 
\begin{equation}
    \sigma_{\gamma p \rightarrow \{\Lambda \overline{\Lambda}\} p + \{p \overline{\Lambda}\} \Lambda   } / \sigma_{\gamma p \rightarrow p \overline{p} p}  = 0.23.
\label{eq:suppressionratio}
\end{equation}
According to the extrapolation of our model of both total cross sections, this ratio appears to be stabilized at a value near one quarter. It includes the fact that there are {\it two} reactions wherein a single $s\antis$ quark pair was produced, and that they contribute about equally. Total cross-section data at higher beam energy would be needed to further confirm this observation.
We take this ratio to be an instance of ``strangeness suppression," where among the light unflavored quarks it is less probable that an $s\antis$ pair is created than either $u\antiu$ or $d\antid$ pairs.  In a simple factorization model~\cite{CLAS:2014gcd}, this ratio is a direct measure of probabilities for production out of the vacuum of either an $\antis s$ versus a $\antiu u$ quark pair:
\begin{equation}
    \frac{P(s\antis)} {P(u\antiu)} \approx 0.23.
\label{eq:sfraction}
\end{equation}

Closest to the present study was the work of Ref.~\cite{CLAS:2014gcd} in which exclusive electroproduction of $\Lambda K^+$ and $N\pi$ pairs was examined to estimate the ratio of strange to nonstrange quark pair production, $P(s \antis) / P(d  \antid)$.  Averaging their values based on differing assumptions about $P(u \antiu)/P(d \antid)$, a ratio which should be close to unity, we find their estimated value to be $P(s \antis)/P(d \antid) =0.21 \pm 0.03$.  Evidently, using the electromagnetic interaction as the initiator of strangeness production, our measurement finds compatible results.   

Our conclusion is that strangeness production is suppressed relative to nonstrange light quark production by about a factor of four. This observation is limited by the \gx beam energy range, and the fitted trend of the data suggests that at higher energy this suppression might be slightly less.  Future measurements at Jefferson Lab may test this possibility~\cite{JLab22GeV}.

\begin{figure*}[t]
	\centering
    \includegraphics[width=0.995\textwidth]{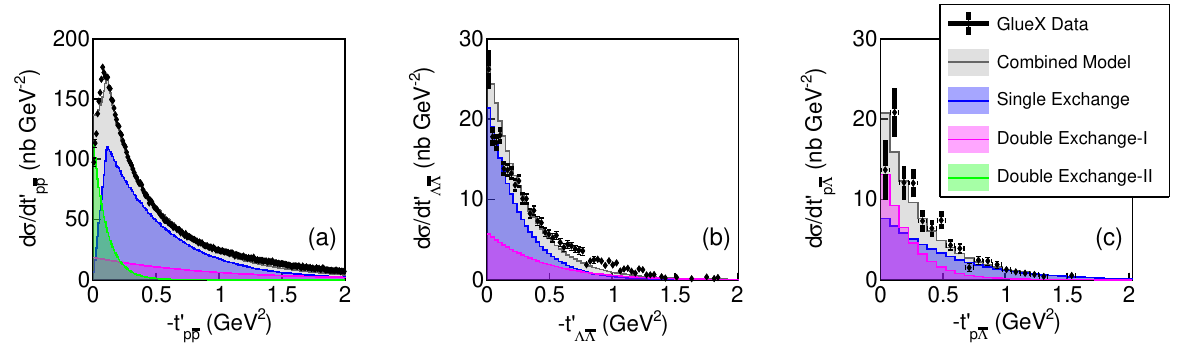} 
    \caption{\label{fig:differential_tprime}The differential cross section with respect to reduced four-momentum transfer $-t^\prime$ to the created pair: 
    (a) $\pbarp$,
    (b) $\lamlambar$, and
    (c) $\plambar$. The events are selected for the beam photon energy range $8.5-9.0$ GeV.
    Data points (black markers) are shown with statistical uncertainties only.
    The combined model (gray filled histograms) is normalized to the measured differential cross section. 
    Intensity distributions for each model component are depicted with the color scheme in the legend.
    Results for the full set of photon energy bins are presented in   Fig.~\ref{fig:difftprime}.
}
\end{figure*}
\subsection{Cross sections differential in reduced momentum transfer $t^\prime$}\label{sec:measured_dsigma_dtprime}

We have shown that the observed angular kinematic distributions and two-body correlations in these reactions call for substantial contributions from single- and double-channel $t$ exchanges.  It is now of interest to see how the full intensity-based model we have developed (Sec.~\ref{sec:fullreactionmodels}) can represent the $-t^\prime$ dependence [Eq.~\ref{eq:reduced_t}] of the data with acceptance correction. 

The measured differential cross sections in $t^{\prime}$ are displayed in Fig.~\ref{fig:differential_tprime} (black markers) for the beam energy bin of $8.5-9.0$~GeV, comparing the production of the (a) \pbarp, (b) \LambarLam, and (c) \pbarL ~systems.  The combined model curves are displayed with gray-filled histograms. The model components (colored filled histograms) were generated to cover the full energy range data and then projected onto the given beam energy range. Each component of the model was generated with the corresponding reaction mechanism described in Sec.~\ref{sec:fullreactionmodels} with the fit parameters summarized in Table~\ref{tab:all_reaction_parameters}, then incoherently summed with the fractions given in Table~\ref{tab:all_reaction_fractions}. The curves of the combined model are normalized in each case to the differential cross sections shown.  

The model we have developed provides a fair to good representation of the shapes for each of the reactions.   It is also qualitatively evident that omitting the double-exchange components in the model would have led to distributions not well represented by single individual exponential curves. 
We also note that the $t^\prime$-dependence of the \pantip reaction is broader than that of the two hyperon channels.
\begin{figure}[htpb]
    \centering
\includegraphics[width=0.975\columnwidth]{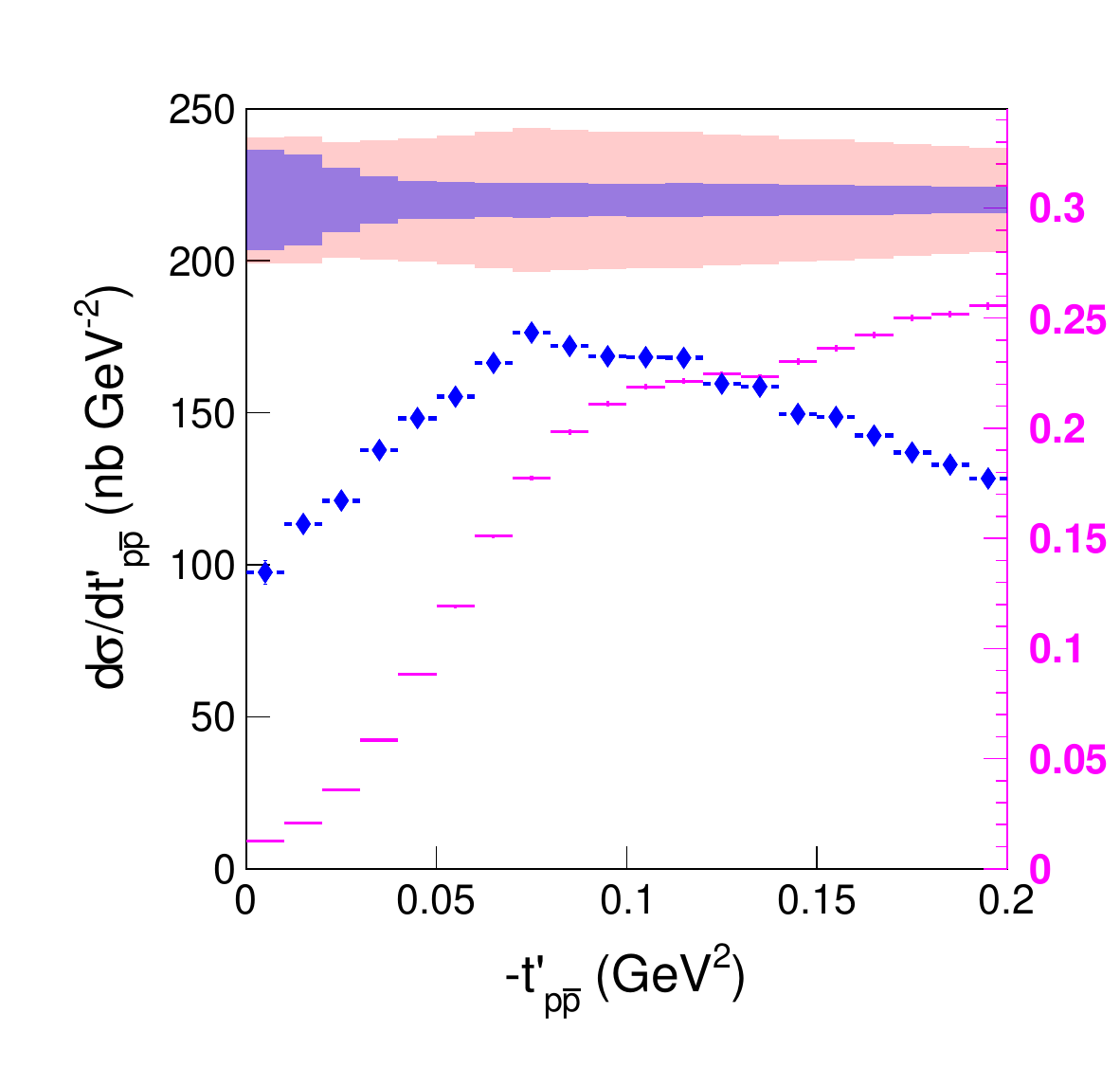} 
\caption{\label{fig:differential_tprime_zoomed} The differential cross section for the $\ppbar$ channel with respect to the reduced four-momentum transfer to the pair, emphasizing the low $t^\prime$ region for data in the slice of beam photon energy $8.5-9.0$~GeV with $0.01$ GeV$^2$ binning. The magenta points show the acceptance as a function of  $-t^\prime$, with the right-hand magenta axis.  The light blue band depicts the model-dependent systematic uncertainty and the pink band is the total systematic uncertainty of the cross section. 
}
\end{figure}

\begin{figure*}[t!]
	\centering
    \includegraphics[width=0.995\textwidth]{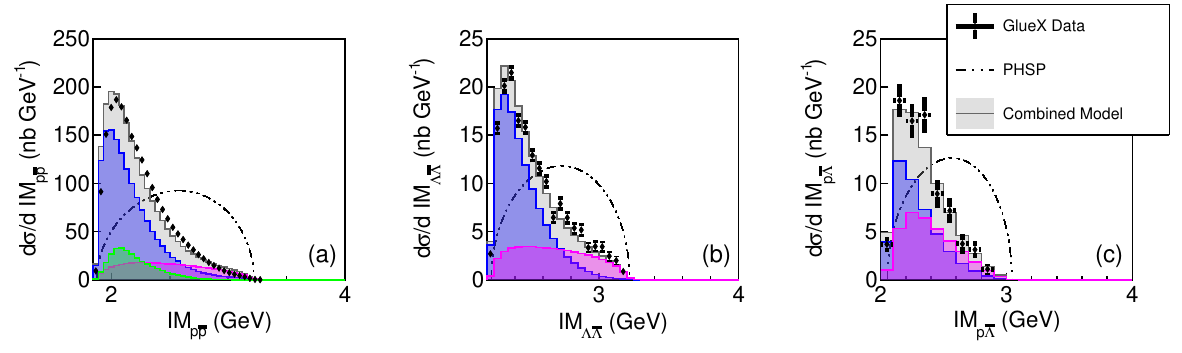} 
    \caption{\label{fig:differential_IM}The measured differential cross section with respect to the invariant mass of the baryon-antibaryon systems (a) $\pbarp$,
    (b) $\lamlambar$, and
    (c) $\plambar$. The events are selected for the beam photon energy range $8.5-9.0$ GeV.
    Data points (black markers) are shown with statistical uncertainties only.
    Phase space distributions (black dot-dashed curves) are normalized to the combined model (gray filled histograms) for comparison. 
    Model components are depicted with the same color scheme as in  Fig.~\ref{fig:differential_tprime}.
    Results for all photon energy bins are presented in Fig.~\ref{fig:diff_IM}.}
\end{figure*}

Figure~\ref{fig:differential_tprime_zoomed} highlights the low $-t^\prime$ behavior in the \pantip reaction.  It rises from its lowest value at $t^\prime = 0$ and then turns over near 0.1~GeV$^2$ to begin its exponential fall.  
The light blue band above the data points illustrates the estimated systematic uncertainty 
arising from the model dependence of the acceptance in the \gx detector and from the selection cut on low-momentum proton (antiproton) tracks. Our phenomenological approach was adapted to model this rising effect by introducing the ``$t$-cutoff" for this channel, as discussed earlier. 
Single and double exchange processes are modeled incoherently at all $-t^\prime$ values; in our model, we do not have an understanding of their possible coherent interplay.  Also, this effect of rising cross section at the smallest $t^\prime$ is not clearly visible in the other two channels, given limited statistics. 

\begin{figure}
    \centering
    \includegraphics[width=0.995\columnwidth]{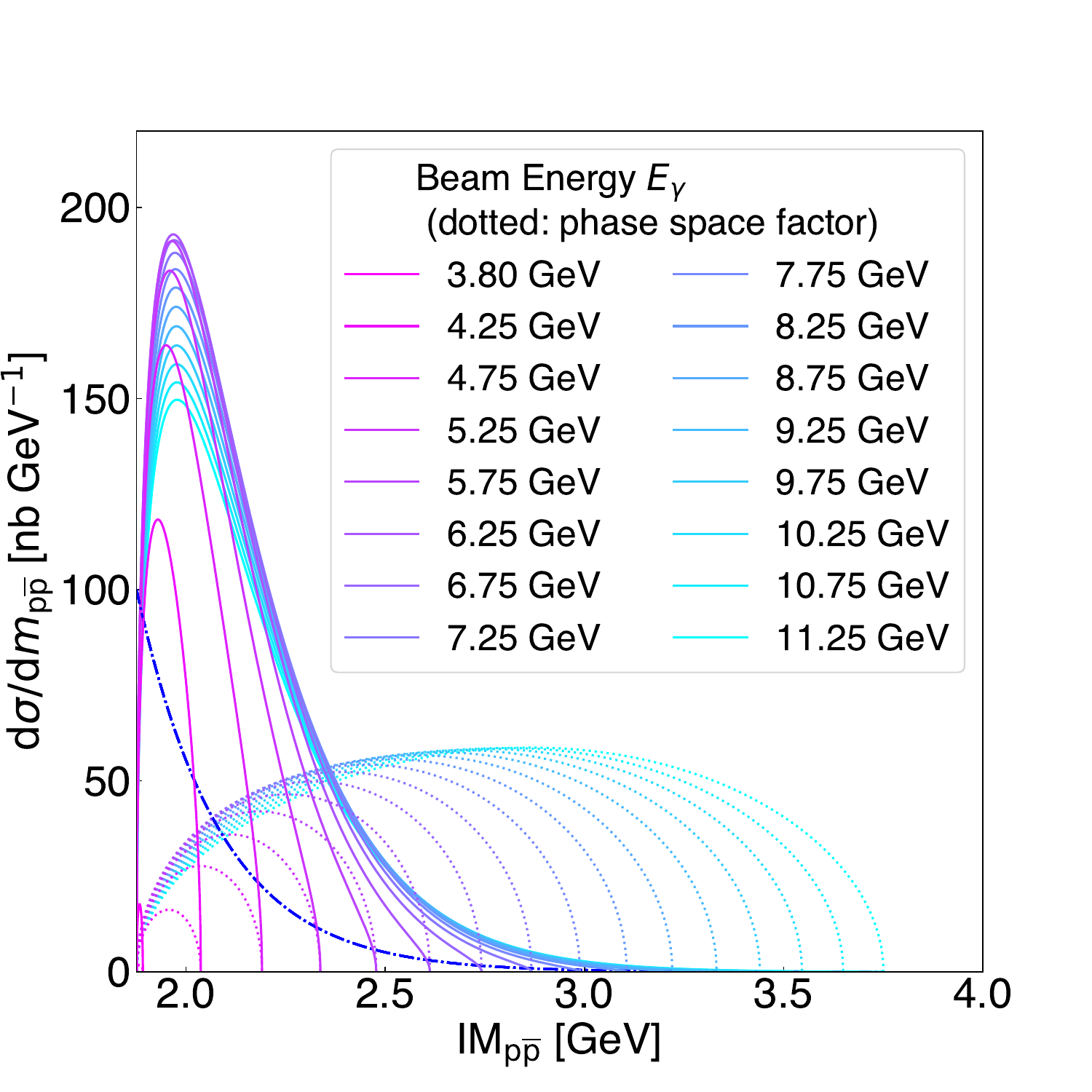}
    \caption{Components of the beam-energy-dependent clustering model of the $\ppbar$ mass distribution. A single exponential mass profile of the $\ppbar$ system before folding in three-body phase space is shown by the blue dash-dot curve. The phase space factors [Eq.~\ref{eq:phsp}] across beam energies are the light blue dotted lines. The resultant invariant mass distributions [Eq.~\ref{eq:dsigma_dm12}] at different beam energies are shown as solid lines of various shades. Compare to Fig.~\ref{fig:diff_IM}(a).} \label{fig:energy_dependent_mass_model}
\end{figure}

The complete set of data and combined models, for each reaction and across all beam energy bins, is displayed in Appendix~\ref{sec:difft_allE}, Fig.~\ref{fig:difftprime}.    One sees a consistent goodness of the fits to the model over most of the range of beam energies.   Close to the threshold, the model fits are only fair.

\subsection{Cross sections differential in mass of the created baryon-antibaryon pairs}
\label{sec:diffcrosssec}

The invariant mass of the produced baryon-antibaryon pairs serves as a key observable that characterizes the dynamics within the baryon-antibaryon systems. To examine this, the results were integrated over the complete range of the $-t$ momentum transfer to the produced pairs, and binned by beam photon energy, $E_{\gamma}$.   

The results for the beam energy range of $8.5$ to $9.0$~GeV are presented in Fig.~\ref{fig:differential_IM}, showing the (a) \(\pantip\)-system, (b) the \(\LamantiLam\)-system, and (c) the \(\pantiL\)-system.  Each model intensity component and their total are depicted as colored filled histograms. Their sums are normalized to the measured differential cross sections.    
The complete set of results for all photon energies is shown in Appendix~\ref{sec:difft_allE} in Fig.~\ref{fig:diff_IM}.  

Each channel exhibits a similar pattern of a fast rise from threshold with a tail out to the kinematic limit determined by the phase space available at each beam energy.  The three-body phase space distributions (Eq.~\ref{eq:phsp}) are included for reference, depicted as black dot-dashed curves,  with normalization chosen to equal the area under the total model curve.  For all of the reactions, the double-exchange components tend to populate the higher-mass end of the distributions.  This happens because the double-exchange pairs tend, on average, to be further apart in CM angle.  Hence, they tend to have higher invariant mass. 

The energy dependence of the phase space and the clustering model curves can be understood using Fig.~\ref{fig:energy_dependent_mass_model}, which shows the combined effect of mass clustering and phase space across photon beam energies. As discussed in Sec.~\ref{sec:reactionmodel}, the mass distributions are parameterized using an exponential clustering parameter \(c_m\), per Eq.~\ref{eq:clustering_modeling}. The shape of the phase space is governed by the interplay of the two-body momentum of the pair rest frame \(q^*\) and the two-body phase space breakup momentum \(p_f\) (Fig.~\ref{fig:q_pcm_vsIM}). 
The rapid rise from the mass threshold is driven by the exponential clustering effect combined with the opening of the three-body phase space, characterized by \(q^*\). 

Approaching the upper limit of the pair mass [Eq.~\ref{eq:m_bound}] for a given beam energy, the distribution is increasingly constrained by the available two-body phase space, determined by \(p_f\)  [Eq.~\ref{eq:p_f}], as shown in Fig.~\ref{fig:q_pcm_vsIM}. Consequently, the peaks in the model curves do not follow a strictly exponential behavior. 
Instead, the peak positions shift to higher mass as the beam energy increases. This behavior, influenced by the combined effects of three-body phase space and the exponential clustering parametrization, is reflected across increasing beam energies in both data and the model curves shown in Fig.~\ref{fig:diff_IM}(a).

\section{Systematic Uncertainties}
\label{sec:systematics}

The accuracy of each step in determining cross sections was investigated.   Table~\ref{tab:total_syst} lists the systematic uncertainties associated for baryon-antibaryon photoproduction cross sections for the $\pantip$, $\LamantiLam$, and $\pantiL$ channels.  Various sources of systematic uncertainty are enumerated, including kinematic fitting, data selection cuts, luminosity, tracking efficiency, model dependency, and variations seen across \gx run periods. 

\begin{table}[hbt]
  \centering
  \begin{tabular}{l|ccc}
    \toprule
    \textbf{Source} & $\ppbar$ & $\lamlambar$ & $\plambar$ \\
    \midrule
    Kinematic Fitting     & $8\%$  & $5\%$  & $2\%$ \\
    Data Selection        & $4\%$  & $10\%$ & $10\%$ \\
    Flux Normalization    & $5\%$  & $5\%$  & $5\%$ \\
    Tracking Efficiency   & $8\%$  & $13\%$  & $15\%$ \\
    Model Dependence      & $3\%$  & $3\%$  & $3\%$ \\
    Run Periods           & $1\%$  & $6\%$  & $10\%$ \\
    \midrule
    Total          		  & $13\%$  & $19\%$  & $22\%$ \\
    \bottomrule
  \end{tabular}
  \caption{Estimated range of global systematic uncertainties for each of the baryon-antibaryon photoproduction cross sections. }
  \label{tab:total_syst}
\end{table}

Kinematic fitting of exclusive final states was the main method for selecting  events in the data samples.  Confidence-level selections were varied on both real data and Monte Carlo samples to assess the sensitivity variations.  Mismatches between data and Monte Carlo modeling were seen for both more stringent cuts that affected the accuracy of detector modeling and for looser cuts that allowed nonexclusive background to affect the data samples.  For strangeness channels, the kinematic fits did not use a $\pi^-p$ or $\pi^+\pbar$ ~hyperon mass constraint to allow assessment of nonexclusive background contributions to the hyperon mass spectra.   The cut on the hyperon flight-significance parameter discussed earlier could thereby be inspected.   Again, this was done for both the real data and the Monte Carlo simulation data to assess the impact on the results.  The entries in Table~\ref{tab:total_syst} summarize these tests.

Data selection criteria were varied to test the sensitivity to the overall missing mass of the exclusive final states, the minimum acceptable momenta of each particle type, and the minimum accepted laboratory angle for each of the particle tracks.   These studies were carried out in each of the beam energy bins used in Sec.~\ref{sec:energy_dependence} to examine energy dependencies.  In the \pantip channel, these tests showed cross section variations at the few-percent level, with a  mild increase of about 5\% from the lowest to highest beam energies.   Variations in the hyperon channels were, within statistics, consistent with a 10\% range and are included in the final tabulation.   

The incident photon beam flux at \gx was measured and monitored by an $e^+ e^-$ pair spectrometer and calibrated against dedicated runs with an in-beam total absorption counter.   On the Monte Carlo side, the reconstruction of the photon-tagging spectrometer was modeled to estimate the effective efficiency of the tagging procedure.   
The uncertainty in the measurement of the photon flux was estimated to be $\pm5\%$ \cite{GlueX:2020idb}, and we take this value as the systematic uncertainty due to this quantity.

The tracking efficiency of charged particles in \gx was studied by reconstructing final states with one particle intentionally ``missing" and then testing how often that particle was reconstructed, in fact, in the data set~\cite{GlueX:2020idb}.  Again, this was done for both real data and Monte Carlo events, and tabulations of the estimated mismatch between the two were developed.  For pions and protons, across various angular and momentum ranges, maps of this two-dimensional mismatch were used to estimate the reconstruction accuracy for the particular final states in this study.   This was done using the observed momentum distributions of the particles in the experiment.    We made the assumption that these systematic uncertainties may be combined as being uncorrelated within any single event.  Table~\ref{tab:total_syst} reflects the fact that the hyperon channels each contain two more charged particles than the \pantip channel, leading to a greater systematic uncertainty.  

The reaction model used in this article for the $\gamma p \rightarrow \{p \antip \} p$ reaction contains nine parameters, as discussed in Sec.~\ref{sec:reactionmodel}.   For each parameter, an estimated uncertainty was included in Table~\ref{tab:all_reaction_parameters}.  We investigated what additional systematic uncertainties exist in the final measured cross sections because of possible variations in the parameters of the model, since the model itself is used to compute the acceptances of the reaction.  
When each parameter was varied independently, it was possible to estimate the range of variation of the results in the recomputed cross sections as a function of beam energy, $-t$, and \pantip invariant mass.  Assuming independence between these estimations, the results were aggregated using a quadrature sum. For the combined model parameters of the $\ppbar$ reaction, the overall systematic uncertainty remained consistently in the range of 3\% throughout the entire beam energy range, which supports the model not introducing significant larger biases into the measurement from inaccurately fitted parameters. Due to statistical limitations of data in the hyperon channels, a similar factor-by-factor study could not be performed, so we allowed that their uncertainties are the same, as indicated in Table~\ref{tab:total_syst}.

In addition to these uncertainties, there are systematic effects arising from the fitted fractions used to incoherently combine the three model components —  single exchange and two double exchanges — each of which exhibits distinct net acceptance because of their distinct kinematic distributions.  The fitted fractions, which varied within the uncertainties estimated (as shown in Table~\ref{tab:all_reaction_fractions}),  could impact edge regions such as the low $t^{\prime}$ region of the differential cross section measurement, illustrated by the blue band in Fig.~\ref{fig:differential_tprime_zoomed}. 
For systematic estimation of uncertainty of all $\ppbar$ cross sections in other kinematic regions, the results appeared to be less sensitive ($< 1\%$) to uncertainties in the fit fraction compared to the dominant effects of acceptance differences between different run periods. Hence, we allow the systematics on model fitting fractions to be already accounted for within the model-parameter-dependent and run-period-dependent systematics.

The results reported here came from three independent \gx data-taking periods separated by several months, each with somewhat different calibration and reconstruction details.   Comparing the results from these run periods, after all individual corrections were complete, allowed another measure of the repeatability of the measurements and assessment of the systematic uncertainties.  The $\ppbar$ results were reproduced with a possible variation at the $\sim1\%$ level, while the hyperon channels were not quite as good, with variations across beam energy estimated at $\sim 6\%$ in the $\lamlambar$ channel and $\sim10\%$ in the $\plambar$ channel.  These are reflected in Table~\ref{tab:total_syst}.

Further tests compared small discrepancies between results acceptance corrected with individual model component choices.   For the $\pbarp$ channel, the run-period dependent systematic uncertainties were estimated to be $4\%$ for the total cross section corrected by acceptance modeled using only the single-exchange component, but $2\%$ when using only Double Exchange-I, and $10\%$ when using only Double Exchange-II.  Combining the first two components reduced the uncertainties to $3\%$, and to $1\%$ when all three components of the model were used together. This means that the comprehensive model combination yielded the most consistent acceptance correction for the total cross sections. 

To summarize, the total systematic uncertainties aggregated from all significant sources, the $\Lambda\overline{\Lambda}$ and $p\bar{\Lambda}$ channels exhibit higher systematic uncertainties at $19\%$ and $22\%$, largely driven by data selection choices, tracking efficiency, and variations between run periods.  The $\pbarp$ channel presents a smaller systematic uncertainty at $13\%$. Having found no evidence to the contrary,  we took these uncertainties to be uncorrelated and combined each independent contribution in quadrature.   As implied in Table~\ref{tab:total_syst}, these uncertainties may be considered global in all beam energies and invariant masses.

\section{Further Discussion of the Findings}
\label{sec:furtherdiscussion}

\subsection{Cross sections}
\label{sec:crosssectiontheory}

There was a pioneering theoretical calculation~\cite{Gutsche:2017xtm, Gutsche:2016wix} for both \pantip and \LamantiLam  processes at beam energy of 9.0 GeV, performed with the goal of exploring the possible coupling of exotic states such as wide sub-threshold glueballs decaying to baryon-antibaryon final states.  A set of vector/axial-vector meson exchanges was considered, with possible couplings estimated from other reactions.   We do not show the comparison on graphs of our $-t^\prime$ dependence because the predictions peaked near 20 nb/GeV$^{-2}$ rather than the observed 150~nb/GeV$^{-2}$ for the \pantip case, and 1~nb/GeV$^{-2}$ rather than the observed 20~nb/GeV$^{-2}$ for the \LamantiLam case, thus missing the bulk of the reaction mechanism.   Revised theoretical calculations of the reaction mechanism are needed. 

\subsection{Structure near threshold}
\label{sec:subthreshold}

The BES-II Collaboration presented evidence of a threshold enhancement in \pantip mass in radiative $J/\psi$ decays, $J\psi \rightarrow \gamma p \pbar$~\cite{BESIII:PhysRevLett.91.022001, BESIII:2010vwa}.  A Breit-Wigner line shape was fitted together with phase space factors and an estimation of a background distribution that resulted in a sub-threshold centroid of about $M_{\pbarp} = 1861$~MeV, and a fitted width was given as an upper limit $\Gamma < 38$~MeV.   This structure could possibly correspond to a baryonium state with 15 MeV binding energy, or it could correspond to effects of FSI between near-threshold particles~\cite{Zou:2003zn,Kang:2015yka}.   Multiple effects could be in play.  

It is of interest to see whether such a structure is manifested in photoproduction.  With three massive particles in the final state, the phase space at the threshold goes to zero steeply, as seen, for example, in Fig.~\ref{fig:pbarp_invariantmass}, where no additional selection of the reaction products was made. A simple examination of the threshold is not fruitful in the same way as the radiative decay measurements.
However, three-proton kinematics does have a more favorable region when the \pantip produced pair momentum, $q^*$, goes to zero while the recoil momentum between the pair and the recoiling target proton, $p_{f}$, is maximized, as shown in Fig.~\ref{fig:q_pcm_vsIM}.   At \gx this study is ongoing.

In the present study, evidence for an general  ``enhancement"  consists of the evidence for a broadly attractive interaction between the produced proton and antiproton.   It was shown that the model calls for an exponential mass clustering effect with a scale of $\sim 210$~MeV that is largely beam energy independent.   This is very much broader than the effect seen at BES, so it is unlikely that these phenomena are related.   

\subsection{Differences in baryon versus antibaryon angular distributions}
\label{sec:asymmetry}

The asymmetry between the baryon and the antibaryon angular distributions is perhaps the most difficult observation to understand in this study.   The created antibaryons have CM angular distributions that reach all angles, unlike the created baryons, which generally are ``forward peaked". The single-exchange picture treats the created particles symmetrically, and this does not adequately describe the data, as was shown in Figs.~\ref{fig:ppbar_angular} to \ref{fig:plambar_angular}.   Our double exchange mechanism has only the antibaryon at the ``middle" vertex and fits the relevant parameters in a way that results in the observed distributions.   But at the outset, the diagrams in Fig.~\ref{fig:feynman} suggest that there is no reason to exclude diagrams with the created particles interchanged.  

An alternative reaction picture that is not mediated by an exchanged $t$-channel meson might be considered.  A direct quark knockout by the photon of a valence $u$ or $d$ quark from the target proton may qualitatively favor a produced baryon preferentially moving in the forward direction.   That is, the subsequent hadronization of a fast forward-going struck quark must generate additional quark pairs.   The created quarks combine with the fast forward struck quark, while the created antiquarks remain behind.   Hence, the created antibaryons may be less kinematically constrained to move in the forward CM direction.   
Sea quarks in the target proton are largely symmetric between quark and antiquarks and occupy the low end of the Bjorken $x$ range.  The hypothesis suggested here assumes that the incoming photon interacts primarily with the valence quarks, as might be expected.  

A similar phenomenological explanation was posited in a study of hyperon pair production in diffractive $pp$ collisions at the ISR~\cite{SMITH1985267}. It was concluded that partonic production processes could be the cause for produced antibaryon distributions that were not mirrored by those of the produced baryons.

\section{Conclusions}
\label{sec:conclusions}

We have presented the first comprehensive survey of the photoproduction of baryon-antibaryon pairs off a proton target, including the $p\antip$, $\Lambda \antiLambda$, and \pantiL channels.  
Using the large acceptance of the \gx detector, the momentum and angular correlations of the three-body final states were delineated and a phenomenological model was developed that successfully reproduced the observed distributions.   With this model, the differential and total cross sections (Sec.~\ref{sec:resultsanddiscussion}) were obtained, and several conclusions about the underlying mechanism were drawn. 

The reactions were described within a simple model based on non-interfering intensities of single- and double-$t$ channel exchanges. 
Although the fitted slopes and cutoff values did not arise from a full amplitude-level theory, they provide useful benchmarks for future theoretical work. 
For single-exchange intensities, in addition to the $t$-slope parameters, the description required a mass clustering parameter that is nearly independent of the beam energy and reaction channel.  We interpret this as evidence for an attractive interaction of the created baryon-antibaryon pairs.   
It was also observed that the $t^\prime$-dependence of the $\pantip$ channel decreases at small values of $-\tprime$.  This effect was taken into account by introducing an {\it ad hoc} cutoff parameter.  
To fully describe the data, a second set of double-exchange parameters was necessary to describe the $\pantip$ data, but not the $\LamantiLam$ or $\pantiL$ data.   This may indicate the limitations of our non-interfering intensity-based model for the $\pantip$ reaction. 
The model parameters are compiled in Table~\ref{tab:all_reaction_parameters}. 

The extracted cross sections also revealed that the ratio of singly-strange to nonstrange pair photoproduction near 10 GeV beam energy, well above threshold, is small and compatible with related measurements in meson electroproduction. 

Perhaps the most striking open question raised by this study is the broader angular distributions of produced antibaryons compared to baryons.   This asymmetry was built into our model by introducing a double-exchange component, with the antibaryons appearing only at the middle vertex.  We hypothesized that the wide-angle production of antibaryons could be due to incoming photons preferentially interacting with valence quarks, rather than sea antiquarks, favoring forward-going created baryons over antibaryons.   

Selected numerical results for the data are tabulated in Appendix~\ref{appendix:numericalresults}.  Complete numerical results for the cross sections can be obtained from the contact authors or from~\cite{dataarchive}.

Other channels of interest in the same energy range would include $n\antineutron$,  $\Sigma\antiSigma$, and $\Xi\antiCascade$. However, \gx has limited sensitivity to neutrons, and while other hyperon reactions were observed, the event statistics were insufficient for a detailed analysis.  

Beyond the results presented in this article, the same analysis exploited the photon beam linear polarization of the Hall D / \gx system to measure the beam spin asymmetry (``$\Sigma$") of the reactions.  In addition, the produced hyperons were found to be polarized, as measured through their parity-violating weak decays.  Their individual and spin-pair correlations were investigated.  These studies may be the subject of a future article.

\begin{acknowledgments}

We acknowledge the outstanding efforts of the staff of the Accelerator and the Physics Divisions at Jefferson Lab that made the experiment possible. This work was supported in part by the U.S. Department of Energy, the U.S. National Science Foundation, the German Research Foundation, GSI Helmholtzzentrum f\"ur Schwerionenforschung GmbH, the Natural Sciences and Engineering Research Council of Canada, the Russian Foundation for Basic Research, the UK Science and Technology Facilities Council, the National Natural Science Foundation of China and the China Scholarship Council. This material is based upon work supported by the U.S. Department of Energy, Office of Science, Office of Nuclear Physics under Contract No. DE-AC05-06OR23177.
This research used resources of the National Energy Research Scientific Computing Center (NERSC), a U.S. Department of Energy Office of Science User Facility operated under Contract No. DE-AC02-05CH11231. This work used the Extreme Science and Engineering Discovery Environment (XSEDE), which is supported by National Science Foundation Grant No. ACI-1548562. Specifically, it used the Bridges system, which is supported by NSF Award No. ACI-1445606, at the Pittsburgh Supercomputing Center (PSC).
Carnegie Mellon University researchers were supported by DOE Grant No. DE-FG02-87ER40315.  We acknowledge the help of Carnegie Mellon University undergraduate students Samuel Dai, Viren Bajaj, Brandon Neway, Lewis Liu, and graduate students Byron Daniel and Willow Hagen who ably assisted in the early exploratory phases of this analysis.

\end{acknowledgments}


\appendix
\section{Kinematics}
\label{appendix:kinematics}

For a photoproduction experiment at \gx with a fixed proton target ($m_N$), consider a two-body ($\gamma + p$) initial state characterized by energy $\sqrt{s}$, and a quasi-two-body final state characterized by a produced pair ($m_{12}$) consisting of a baryon ($m_1 = m_{B}$) and antibaryon ($m_2 = m_{\bar{B}}$) characterized by energy $\sqrt{s_{12}}$ together with a recoiling baryon ($m_3 = m_p \text{ or } m_{\Lambda}$). 
The CM frame two-body initial and final state momenta, $p_i, p_f$, and the additional decay momentum $q^*$ in the  $m_{12}$ rest frame can be calculated using
    \begin{align} 
        p_i &= \frac{1}{2\sqrt{s}}\lambda^{1/2}(s, 0, m^2_N),\label{eq:p_i}\\  
        p_f &= \frac{1}{2\sqrt{s}}\lambda^{1/2}(s, s_{12}, m^2_3),\label{eq:p_f}\\  
        q^* &= \frac{1}{2\sqrt{s_{12}}}\lambda^{1/2}(s_{12}, m_1^2, m_2^2). \label{eq:q}
    \end{align}
    Here $m_N$ is the mass of the nucleon (proton), and $\lambda(\alpha, \beta, \gamma) = \alpha^2 +\beta^2+\gamma^2-2\alpha\beta-2\alpha\gamma-2\beta\gamma$ is the K\"{a}ll\'{e}n function, commonly used in calculations of the two-body decay kinematics~\cite{ParticleDataGroup:2024cfk}.
    
    The allowed physical region is constrained by both $m_{12}$ and  $t^{\prime}$.
    The lower and upper bounds for the pair's invariant mass $m_{12}$ for fixed total energy $\sqrt{s}$ are given by
    \begin{align}
        m_{12}^{\min} &= m_1 + m_2, \label{eq:m_lowerbound} \\
        m_{12}^{\max} &= \sqrt{s} - m_3\label{eq:m_bound}.
    \end{align}
    Given a produced pair, we have the upper and lower bounds on $t^{\prime}$ as function of $m_{12}$:
    \begin{align}
        t^{\prime}_+(m_{12}) &= 0,\\
        t^{\prime}_-(m_{12}) &= -4p_ip_f. \label{eq:tprime_bound}
    \end{align}
    For given $t^{\prime}$ in its allowed range $[t^{\prime}_-, 0]$, the local upper bound for \( m_{12} \) can also be solved from \ref{eq:tprime_bound} as a function of $t^{\prime}$:
    \begin{align}
        &m_{12}^{+}(t^{\prime}) = \bigg[ (s + m_3^2) 
        - \Big(4s m_3^2 + \frac{s^2 \cdot (t^\prime)^2}{(s - m_N^2)^2} \Big)^{\frac{1}{2}} \bigg]^{\frac{1}{2}},\label{eq:m12plus}
    \end{align}
    which will be used for all integrations in the full range of $m_{12}$ to derive the singly differential cross section in $t^{\prime}$.
    The maximum upper bound on the mass given in \ref{eq:m_bound} is achieved when $t^{\prime} = t^{\prime}_+ = 0$.

        For pure phase space, we take the invariant amplitude matrix element squared $\|\mathcal{M}_{fi}(t^{\prime}, m_{12})\|^2$ as unity in Eq.~\ref{eq:dsigma_dtprimedm}:
        \begin{align}
            \frac{d^2\sigma_{\text{Phsp}}}{dt^{\prime}dm_{12}}
            =\frac{\pi}{(4\pi)^5}\frac{1}{s}\frac{q^*}{p_i^2} \int d\Omega^*
            =\frac{1}{(4\pi)^3}\frac{q^*}{4 p_i^2 s}. \label{eq:phsp_dsigma_dtprimedm}
        \end{align}
    
        To give quantitative descriptions of the three-body phase space kinematics, we derive here the cross section differential with respect to $m_{12}$ and $t^{\prime}$ within their bounds given in the previous section.
        
        For the phase space differential in $m_{12}$, the integration over the full range of $t^{\prime}$ is straightforward because the integrand \ref{eq:phsp_dsigma_dtprimedm} is flat with respect to $t^{\prime}$:
        \begin{align}
            \frac{d\sigma_{\text{Phsp}}}{dm_{12}}
            &= \frac{1}{(4\pi)^3}\frac{q^*}{4 p_i^2 s}
            \int^{t^{\prime}_+}_{t^{\prime}_-(m_{12})} dt^{\prime} 
            \\
            &= \frac{1}{(4\pi)^3}\left(\frac{p_f q^*}{p_i s}\right)\label{eq:phsp_dsigma_dm12},
        \end{align}
        where we recognize the parenthetic factor as the analytic expression of the \textit{three-body phase space kinematic factor}, $\text{Phsp}(m_{12})$ shown in Eq.~\ref{eq:phsp}. Note there is a multiplicative difference of $\int^{t^{\prime}_+}_{t^{\prime}_-} dt^{\prime} = 4p_ip_f$ between Eqs.~\ref{eq:phsp_dsigma_dtprimedm} and ~\ref{eq:phsp_dsigma_dm12}, even though both differential functions are independent of the variable $t^{\prime}$. This is because the kinematic boundary for the $(t^{\prime}, m_{12})$ space is constrained due to energy-momentum conservation, as characterized by the bounds given in Eq.~\ref{eq:tprime_bound}.

         For the phase space differential in $t^{\prime}$, since the upper bound of mass is a function of $t^{\prime}$ we keep the integral form for convenience and simplicity, where the upper bound of the integral is given in Eq.~\ref{eq:m12plus}:
        \begin{align}
            \frac{d\sigma_{\text{Phsp}}}{dt^{\prime}}  
            =  
            \frac{1}{(4\pi)^3} 
            \left(
            \frac{1}{4 p_i^2 s}\int^{m_{12}^+(t^{\prime})}_{m_{12}^{\min}} q^* dm_{12} 
            \right)\label{eq:phsp_dsigma_dtprime}.
        \end{align}
        Similar to the $\text{Phsp}(m_{12})$ expression, the parenthetic factor here describes the three-body phase space as a function of  $t^{\prime}$ and has the same integrated area under the curve as $\text{Phsp}(m_{12})$.

        One may choose to further integrate either Eq.~\ref{eq:phsp_dsigma_dm12} over $t^{\prime}$, or integrate Eq.~\ref{eq:phsp_dsigma_dtprime} over $m_{12}$ to arrive at the same total cross section. For simplicity, we keep the integral form here:

        \begin{align}
            \sigma_{\text{Phsp}}(s) &= \frac{1}{(4\pi)^3}
            \int^{m_{12}^{\max}}_{m_{12}^{\min}}
            \text{Phsp}(m_{12})
            dm_{12}.
            \label{eq:phsp_sigma}
        \end{align}
        
        Note that this formula differs depending on the choices of the produced baryon--antibaryon pair and its recoil particle, as the corresponding kinematic thresholds [Eqs.~\ref{eq:m_lowerbound} and ~\ref{eq:m_bound}] vary accordingly. However, the general line shapes are very similar but significantly differ from our parameterized total cross section model [Eq.~\ref{eq:totalcrosssectionmodel}] which incorporates the mass clustering model [Eq.~\ref{eq:clustering_modeling}], as shown in Fig.~\ref{fig:total_cross_section}.
        
\section{Van Hove diagrams}
\label{sec:vh}
Taking the CM $\hat{z}$ components of the baryon, antibaryon, and proton momenta as  $p_z^{B}$, $p_z^{\scalebox{.6}{$\scriptscriptstyle\antiB$}}$, $p_z^{p}$, it can be shown~\cite{VanHove:1969xa} that one can define polar quantities 
\begin{eqnarray}
    \omega = \tan^{-1}\left[ - \sqrt{3} \frac{p_z^{B}}{p_z^{B} + 2 p_z^{\scalebox{.6}{$\scriptscriptstyle\antiB$}}} \right] + \pi ,\\
\label{omega}        
    \rho =  \sqrt{\frac{3}{2}} \sqrt{ [p_z^{B}]^2 +  [p_z^{\scalebox{.6}{$\scriptscriptstyle\antiB$}}]^2 + [p_z^{p}]^2 }  .
\label{rho}       
\end{eqnarray}
Alternatively, one can write 
\begin{align}
    p_z^{B} &= \rho \sin \omega ,\nonumber \\
    p_z^{\scalebox{.6}{$\scriptscriptstyle\antiB$}} &= \rho\sin (\omega - \frac{2}{3}\pi) ,\nonumber  \\
    p_z^{p} &= \rho\sin (\omega - \frac{4}{3}\pi) .\label{eq:vanhove_formula}
\end{align}

\noindent The radial parameter, $\rho$, is related to the energy of the system in the longitudinal direction, while the angular parameter, $\omega$, encodes the relative directions and the correlations of the particles in the CM inertial frame.  
This procedure defines six wedge-shaped sectors corresponding to the six combinations of particle directions, as presented in Fig.~\ref{fig:vanhovediagram}.
Equation~\ref{eq:vanhove_formula} shows that there is a phase of $\pi/3$ radians between the three $(\rho,\omega)$ representations of the longitudinal momenta. In the Van Hove diagram with three principal axes separated by $120^\circ$, this means that the perpendicular projections of any point on the three principal axes encode the positive/negative longitudinal momenta of the particle represented by each axis.

\section{Monte Carlo event generation}
\label{appendix:optimization}
When generating Monte Carlo events for a reaction of the single-exchange type, events were initially drawn from three-body phase space. These events were rotated in the overall CM frame to achieve the desired angular distribution of the produced baryon-antibaryon pair.  That is, a created pair was rotated in 3-momentum space in such a way that the reduced invariant four-momentum transfer, $-\tprime$,  had the desired distribution at the upper vertex [cf. Sec.~\ref{sec:t_dependence} and ~Fig.~\ref{fig:feynman}(a)].  This rotation did not alter the invariant mass of the produced pair.  The events were then further selected by a rejection method to satisfy the exponential invariant mass distribution described by Eq.~\ref{eq:clustering_modeling}. These steps were sufficient to model events in the single-exchange picture.   

For generating events in the double-exchange category, a sampling method with additional steps was needed.   The $-t_1^\prime$ distribution of the single baryon at the top vertex was selected using the same rotation method described above. The desired $-u_3^\prime$ distribution at the bottom vertex could not then be set with another three-momentum space rotation of the whole event.  Instead, the event rejection method was used to discard events not consistent with the intended $-u_3^\prime$ distribution at the bottom vertex. The slope parameter at the bottom vertex was biased by the nonflat parent distribution for the latter rejection selection method.   Our method of regenerating the complete Monte Carlo data set at each step of the fitting procedure led to good overall fits, but skews the physical interpretation of the parameter $\theta_8$ in Table~\ref{tab:all_reaction_parameters}.  Although not a bias-free determination of the lower $t$-slope in the double-exchange picture, the observed data is reproduced accurately.

\section{Model optimization}
\label{appendix:sgd}
    \subsection{Modeling with embedded reconstruction efficiency}
    The global negative-log-likelihood function, defined in Eq.~\ref{eq:global_likelihood}, is a quantitative measure of the extent to which the data can be approximated by the reconstructed Monte Carlo simulations.
    The likelihood incorporates an efficiency factor, $\epsilon(\bm{x}; \bm{\theta})$, embedded in the binned histogram representation of the observable $\bm{x}$ [Eq.~\ref{eq:binned_hist_representation}]. This factor represents the probability that all final-state particles in an event will be detected by the experimental setup, successfully reconstructed, and that their kinematics satisfy all subsequent event selection criteria. 
    The function $\epsilon(\bm{x}; \bm{\theta})$ also depends on the reaction model and its parameters $\bm{\theta}$ because its value is a convolution of the global efficiency and the local kinematic distribution $\mathcal{X}(\bm{x}';\bm{\theta})$ determined by the model parameters:
    \begin{align}
        \epsilon(\bm{x}; \bm{\theta}) = \int g(\bm{x}') \mathcal{X}(\bm{x}'; \bm{\theta})~ d\bm{x}'.
    \end{align}

    The global efficiency $g(\bm{x}')$ is a $n$-dimensional mapping, $g: \mathbb{R}^n \to \mathbb{R}^1$, and the space $\{\bm{x}'\}\in\mathbb{R}^n$ is spanned by variables such as the momentum and angle of the reconstructed particles.

\subsection{Online optimization focusing on local efficiency mapping}
Instead of attempting to map the entire global efficiency $g(\bm{x}')$ as a fixed function and optimize the likelihood function across the entire parameter space (which is referred to as ``offline optimization'' in this context), we reformulate the problem as an ``online optimization" \cite{flaxman2004online}: minimizing a sequence of likelihood functions, 
$\mathcal{L}_1, \mathcal{L}_2, \mathcal{L}_3, \dots$, where each iteration involves a slightly different local efficiency determined by a small batch of Monte Carlo simulation directly. The local efficiency varies due to the changing kinematic distributions as the parameters are updated iteratively, resulting in a distinct (but smoothly varying)
local likelihood function at each step.  This approach, as further discussed below, allows for fast and efficient sampling of the high-dimensional parameter space, facilitating the optimization process in a computationally tractable manner.
      
The minimization uses a stochastic gradient descent-based algorithm, Algorithm~\ref{alg:sgd_improved}, to make decisions about the parameters to update for the subsequent iteration until reaching a minimum.  In each iteration, the algorithm stochastically approximates the gradient of the local likelihood function by generating small Monte Carlo samples with randomly chosen model parameters from a defined local region of the parameter space. These samples undergo the full Monte Carlo simulation, event reconstruction, and selection processes as the real data to ensure consistency. The region of parameter perturbation is kept sufficiently small so that the local efficiency can be assumed to be constant. As a result, the likelihood \(\mathcal{L}_t\) is effectively decoupled from the efficiency factor, allowing the estimated gradient to align more closely with the direction in which the model parameters can be improved. Compared to an ``offline'' optimization algorithm, this method enables efficient exploration of the parameter space by allowing the computation to autonomously decide how to generate Monte Carlo data, effectively adapting the simulation process to the evolving parametric landscape, shaped by both the computed intensities of the reaction model and the reconstruction efficiency.
    
        \par\medskip
           \refstepcounter{algo}
            \noindent\textbf{Algorithm \thealgo.}
            \label{alg:sgd_improved}
            An iterative algorithm that optimizes the model parameters by matching simulated data to experimental observations. The algorithm repeatedly generates Monte Carlo samples incorporating realistic detector effects and event selection criteria, computes a global negative-log-likelihood quantifying model-data agreement, and updates parameters using stochastically approximated gradient.
            
        \smallskip
        \begin{algorithmic}[1]
            \State \textbf{Initialization:}
            \State \quad Select start point $\bm{\theta_0} \in \mathcal{D}$ arbitrarily
            \State \quad Choose step size $\eta > 0$
            \State \quad Choose search radius $r$
            \State \quad Choose convergence criteria $\delta > 0$
            \State \quad Set $t = 0$
            \While{$t \leq t_{\text{max}}$ \textbf{and} $\|\bm{\theta_{t+1}} - \bm{\theta_t}\| < \delta$}
                \State Sample $N$ random unit vectors $\{\bm{v}_i\}_{i=1}^N$ in a $d$-dimensional hypersphere with radius $r$ such that
                    $|\sum_i \bm{v}_i| \leq 10^{-4}$
                \State Generate $N$ MC samples for the combined model, each containing $S$ events, for the testing parameters $(\bm{\theta_t} + r \bm{v}_i)$, where $i = 1, \ldots, N$. 
                \State Simulate MC events with realistic detector acceptance, apply event reconstruction/selection with the same criteria as applied to the observed data.
                \State For each set of testing parameters $(\bm{\theta_t} + r \bm{v}_i)$, match Monte Carlo Simulation to data distributions and compute the corresponding global negative-log-likelihood $L_{global}(\bm{\theta_t} + r \bm{v}_i)$ as defined in Eq.~\ref{eq:global_likelihood}.
                \State Evaluate effective gradient using the relation (Eq.~\ref{eq:grad})
                $$\nabla \tilde{f}(\bm{\theta}_t) = \mathbb{E}_{\bm{v}} \left[ \frac{d}{r} L_{global}(\bm{\theta}+r\bm{v}) \right]$$
                \State $t \leftarrow t + 1$ and update new parameter as: 
                $$\bm{\theta}_{t+1} \leftarrow \bm{\theta}_t - \eta \nabla \tilde{f}(\bm{\theta}_t)$$
            \EndWhile
            \State Declare convergence at $t=t_f$ and report $\bm{\hat{\theta}} = \bm{\theta}_{t_f}$.
        \end{algorithmic}
        \vspace{0.8\baselineskip}

    \subsection{Stochastic gradient descent}

    Conventional gradient-based optimization techniques can be very effective with direct access to the gradient value.
    However, since the global negative-log-likelihood defined in Eq.~\ref{eq:global_likelihood} does not have any analytic form, direct computation of its gradient is infeasible.  Instead, we use a stochastic approximation to
    the gradient \cite{flaxman2004online} which can handle noisy sampling processes where only small Monte Carlo samples are iteratively generated and reconstructed. In typical applications of SGD in deep learning,  gradients are directly computed from small data batches via back-propagation. The gradients in the present context are indirectly approximated through function evaluations, providing only a ``noisy'' estimation of the gradient.
    The algorithm implemented for this optimization procedure is described in Algorithm~\ref{alg:sgd_improved}.

    The SGD method can be understood by beginning with the basic gradient descent (GD) method. GD is a standard iterative numerical optimization method used to find the local minimum of a convex function $f(\bm{\theta}): \mathbb{R}^d \rightarrow \mathbb{R}$, where $\bm{\theta}$ represents a single point in a $d$-dimensional parameter space. At iteration-$(t+1)$ the parameter $\bm{\theta}$ is updated following the steepest descent direction, using the update rule with step size $\eta$:
    \begin{align}
        \bm{\theta}_{t+1} := \bm{\theta}_t -\eta \nabla f(\bm{\theta}_t).
    \end{align}
    Given convexity and L-smoothness in the parameter space, it is mathematically guaranteed \cite{boyd2004convex} 
    that GD converges to the $\epsilon$-neighborhood of the global minimum within a finite 
    number of iterations.
    
    When the closed form of the function $f(\bm{\theta})$ is  not available,  the gradient $\nabla f(\bm{\theta})$ is not accessible.
    One can then estimate the effective gradient from values of $f(\bm{\theta})$ directly.
    Specifically, in a $d$-dimensional parameter space, one assumes that function $f(\bm{\theta})$ is smooth within a hypersphere of radius $r$.
    The effective gradient \cite{flaxman2004online}, $\nabla \tilde{f}(\bm{\theta})$, at point $\bm{\theta}$ in $d$-dimension parameter space is then defined as:
    \begin{align}
        \nabla \tilde{f}(\bm{\theta}) \equiv
        \mathbb{E}_{\bm{v}} \left[ \frac{d}{r} f(\bm{\theta}+r\bm{v}) \right] , \label{eq:grad}
    \end{align}
    where the function
    $\tilde{f}(\bm{x}) = \int_{\|\bm{v}\|_2 \leq 1} f(\bm{x} + r\bm{v}) \, d\bm{v}$
    represents an averaged form of $f$ over the unit sphere, effectively smoothing $f$ within local neighborhood. 
    From the Divergence Theorem, the gradient of $\tilde{f}$, as an stochastic approximation of $\nabla f$, can be computed using $\mathbb{E}_{\bm{v}}[\cdot]$, which represents the expectation of $N$ randomly sampled unit vectors $\bm{v}_i$, weighted by the values of the functions evaluated at each point, $\bm{\theta} + r\bm{v}_i$.

    In our optimization problem of the set of nine parameters for the $\ppbar$ reaction, we found $N$ between $10$ to $15$ to be sufficient for our purpose. As shown in step $\#8$, it is crucial to ensure that the sum of $N$ random unit vectors remains unbiased to improve stability, particularly when the gradient is weak. The step size is fixed at $\eta = 0.5$ through all iterations.  The search radius is determined by the resolution of the model parameters allowed by the data statistics. In practice, we typically set the search radius to $r = 0.05$. 
    For each Monte Carlo sample generated with a test parameter set, a sub-sample size of $S_k = 2\times10^4$ was found to be sufficient for each individual model component.

\section{Cross sections differential in momentum transfer and invariant mass}\label{sec:difft_allE} 
\begin{sidewaysfigure}[htbp]
  \centering
  \makebox[\textheight][c]{ 
    \begin{minipage}[b]{0.37\textheight}
      \includegraphics[width=\linewidth]{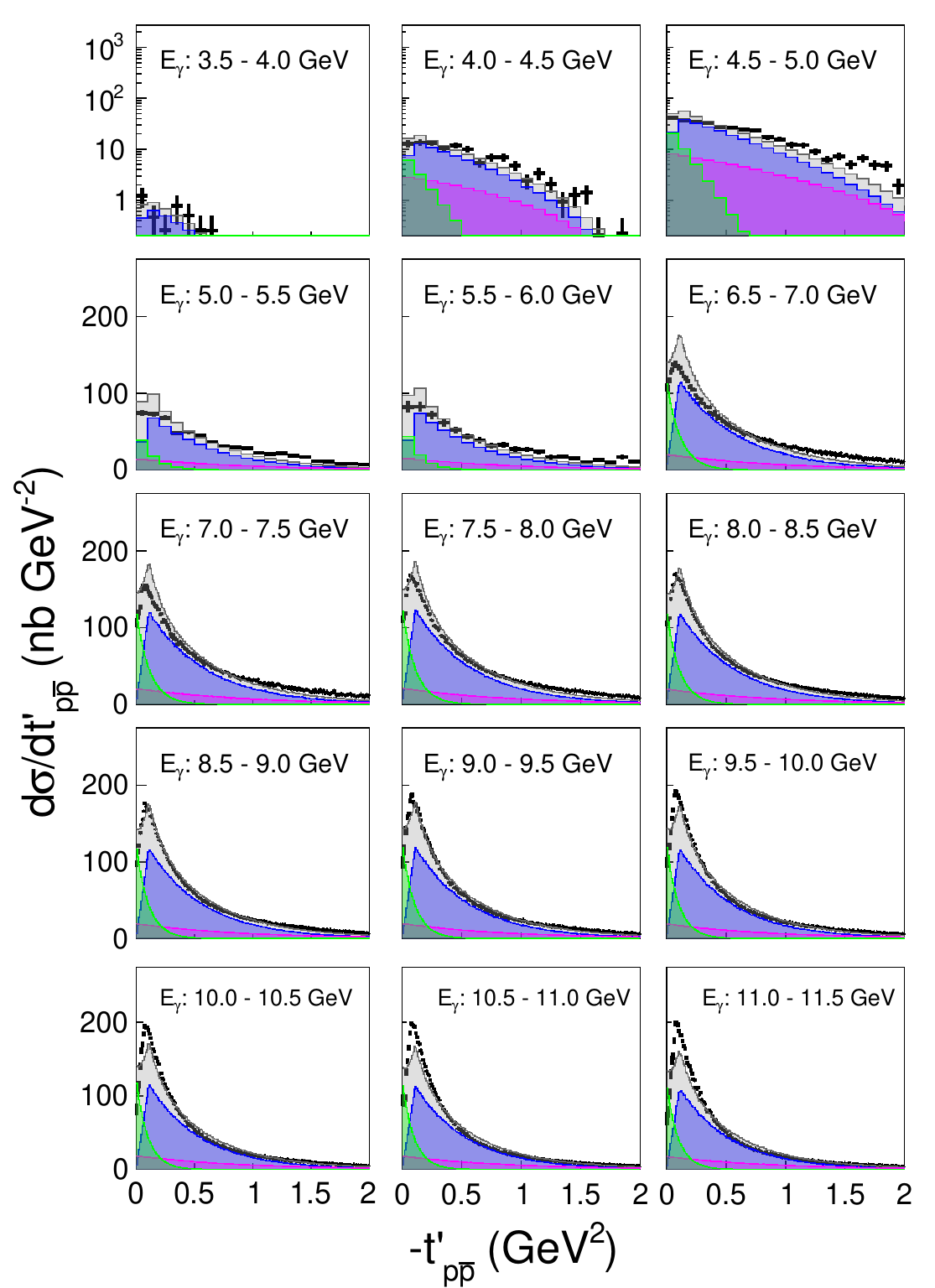}\\
      (a)
    \end{minipage}
    \hspace{0.02\textheight}
    \begin{minipage}[b]{0.30\textheight}
      \includegraphics[width=\linewidth]{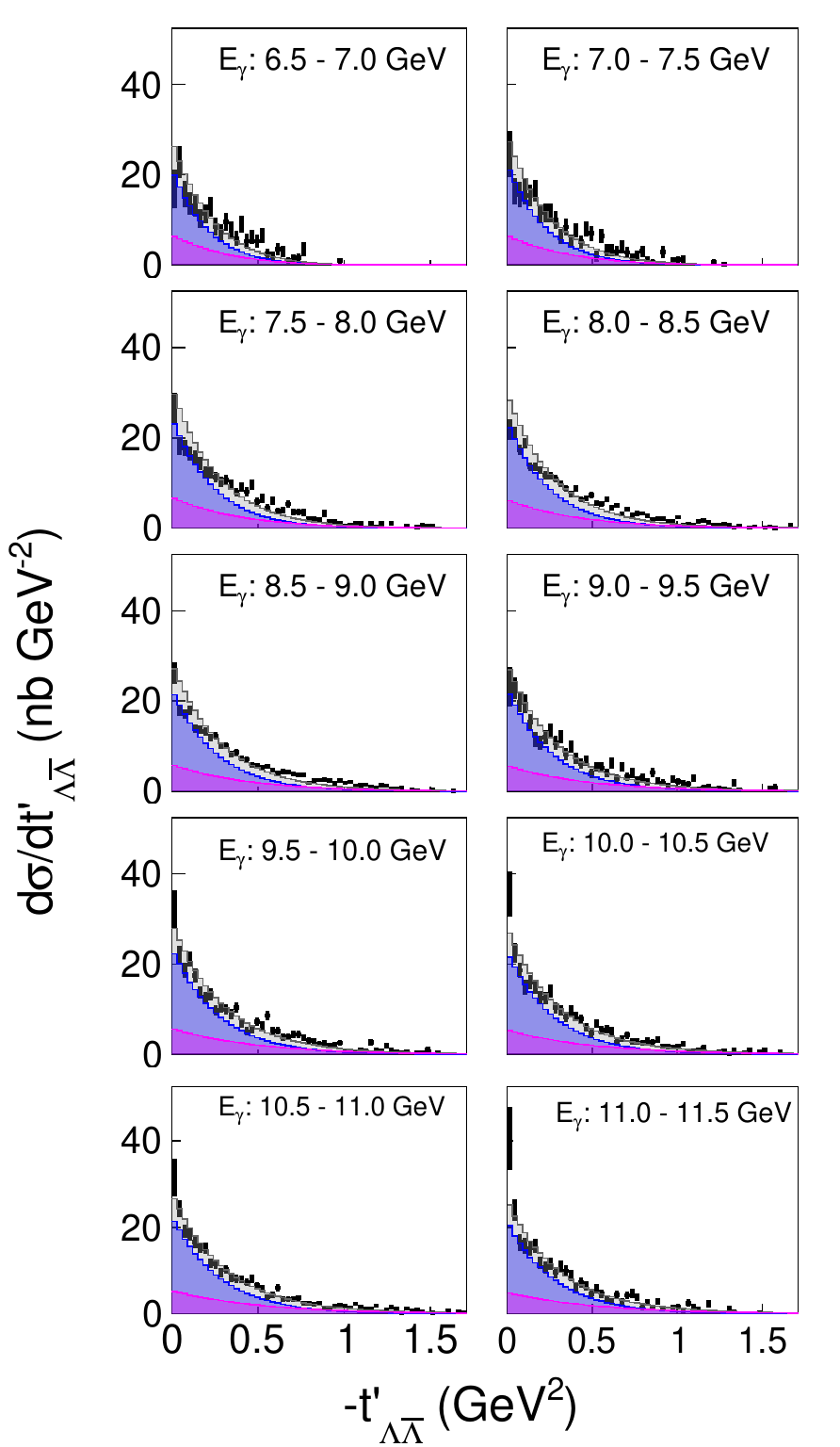}\\
      (b)
    \end{minipage}
    \hspace{0.02\textheight}
    \begin{minipage}[b]{0.30\textheight}
      \includegraphics[width=\linewidth]{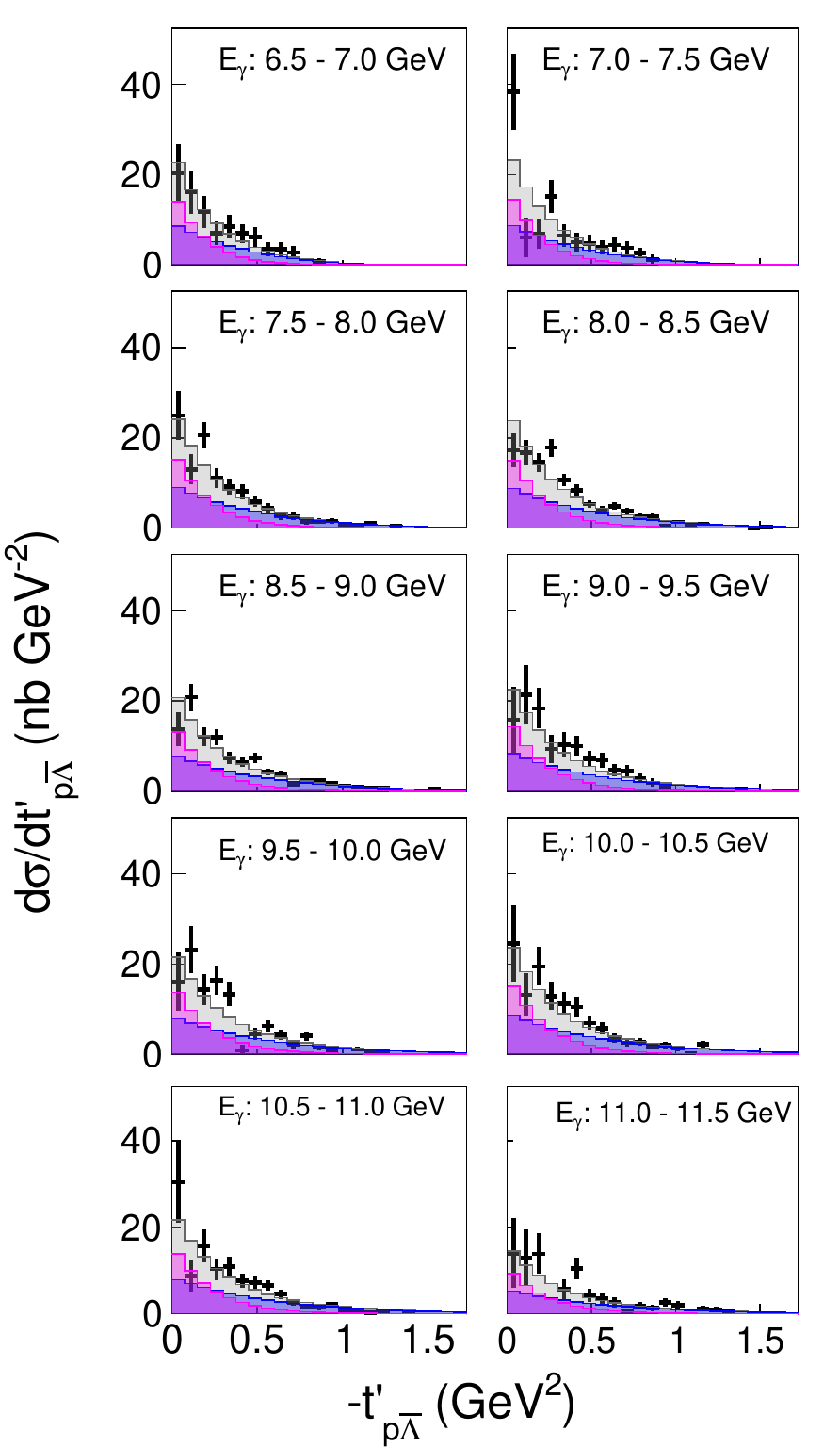}\\
      (c)
    \end{minipage}
  }
  \caption{
    Differential cross section in reduced four-momentum transfer, $t'$, from the beam photon to
    (a) the created $\pbarp$ pair,
    (b) the created $\lamlambar$ pair,
    (c) the created $\plambar$ pair.
    Data points (black markers) are shown with statistical uncertainties only. The panels span the full range of beam energy bin in the experiment, from 3.5 to 11.5 GeV.   Fits from the model defined in Sec.~\ref{sec:fullreactionmodels} are shown for single and double exchanges.   The component color scheme is the same as for the excerpted energy bins in Fig.~\ref{fig:differential_tprime}. 
  }
  \label{fig:difftprime}
\end{sidewaysfigure}

\begin{sidewaysfigure}[htbp]
    \centering
    \begin{minipage}[b]{.375\textwidth}
        \includegraphics[width=\textwidth]{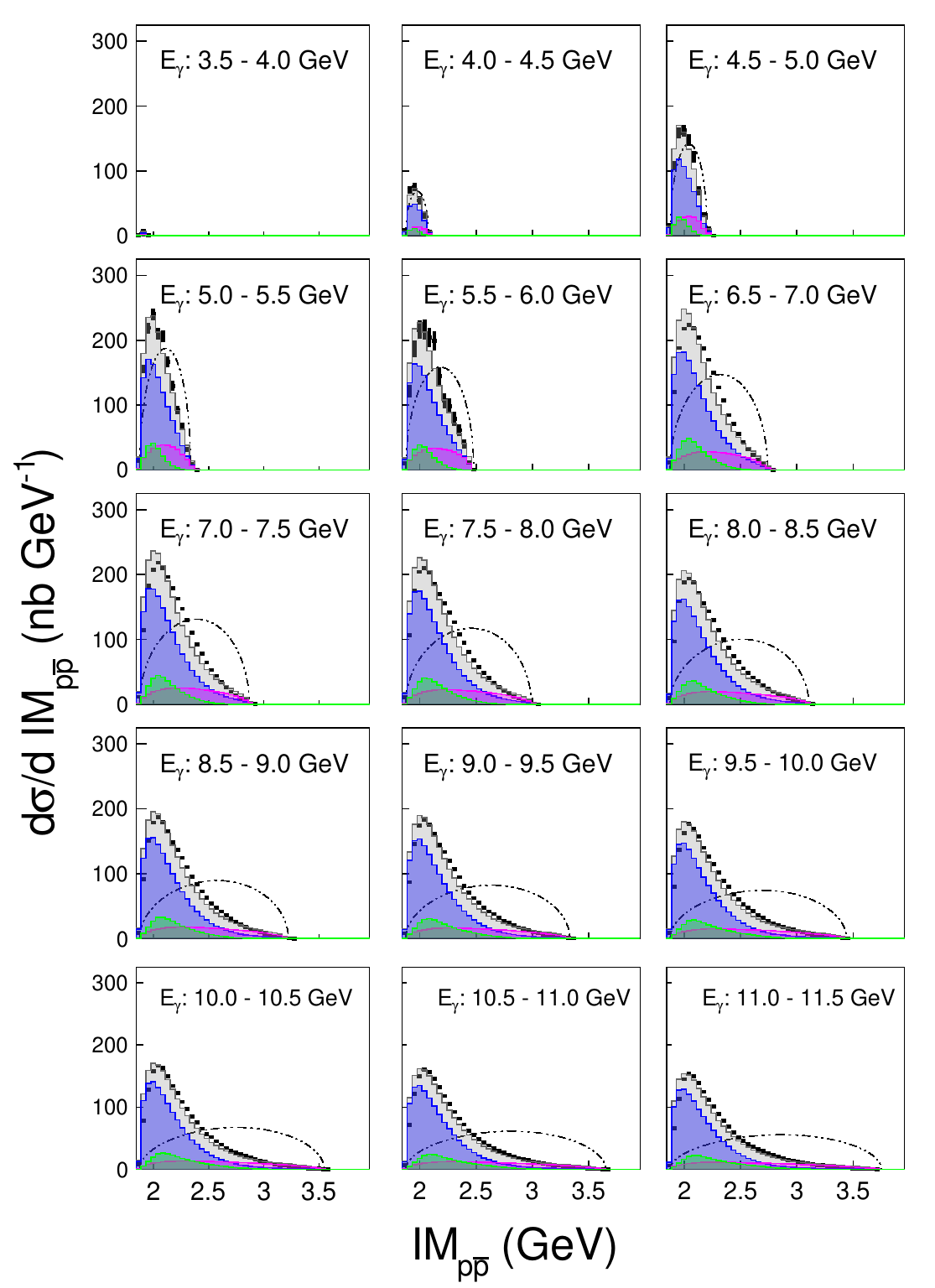}\\ 
        (a)
    \end{minipage}
    \hfill 
    \begin{minipage}[b]{.30\textwidth}
        \includegraphics[width=\textwidth]{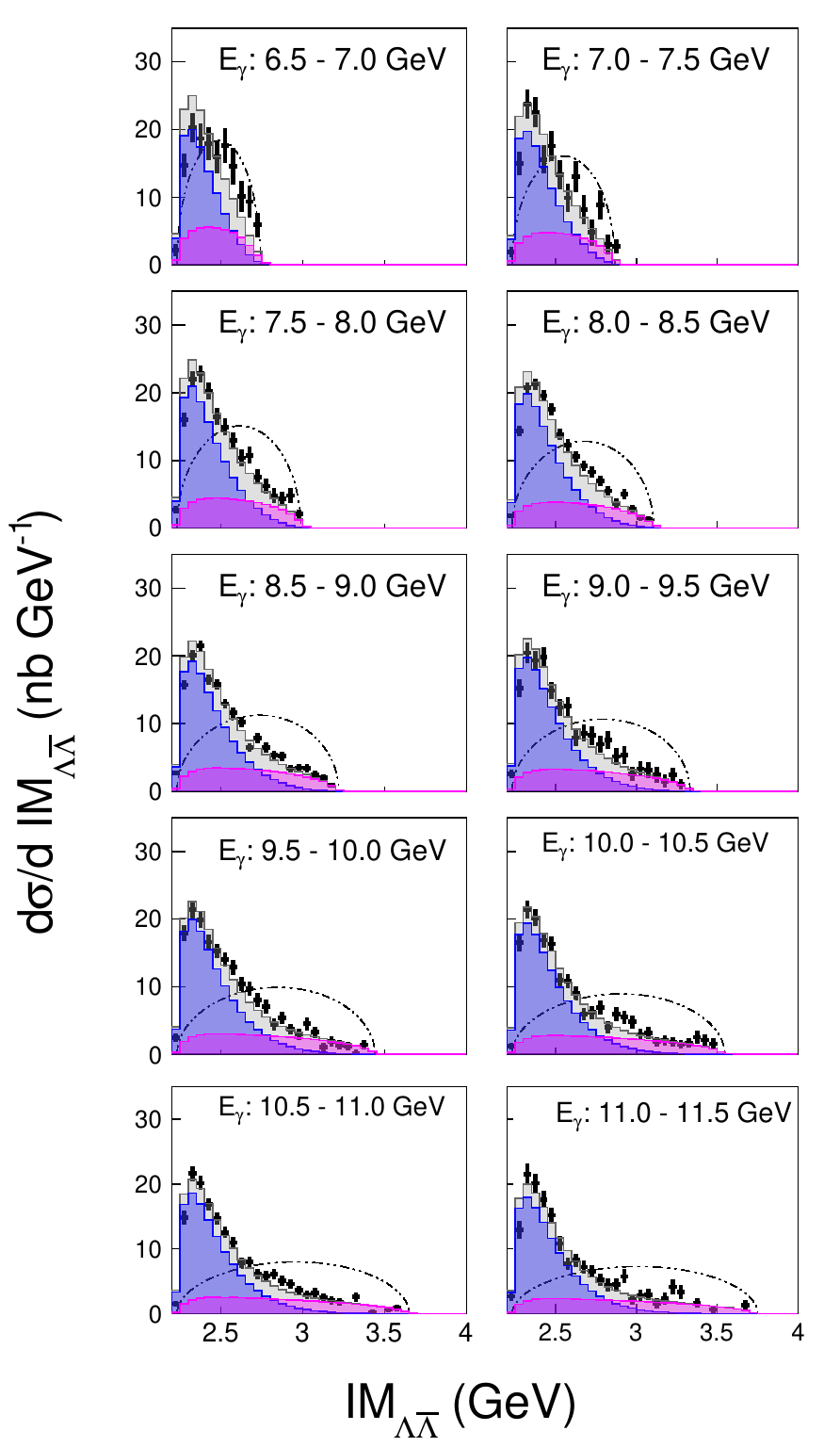}\\ 
        (b)
    \end{minipage}
    \begin{minipage}[b]{.30\textwidth}
        \includegraphics[width=\textwidth]{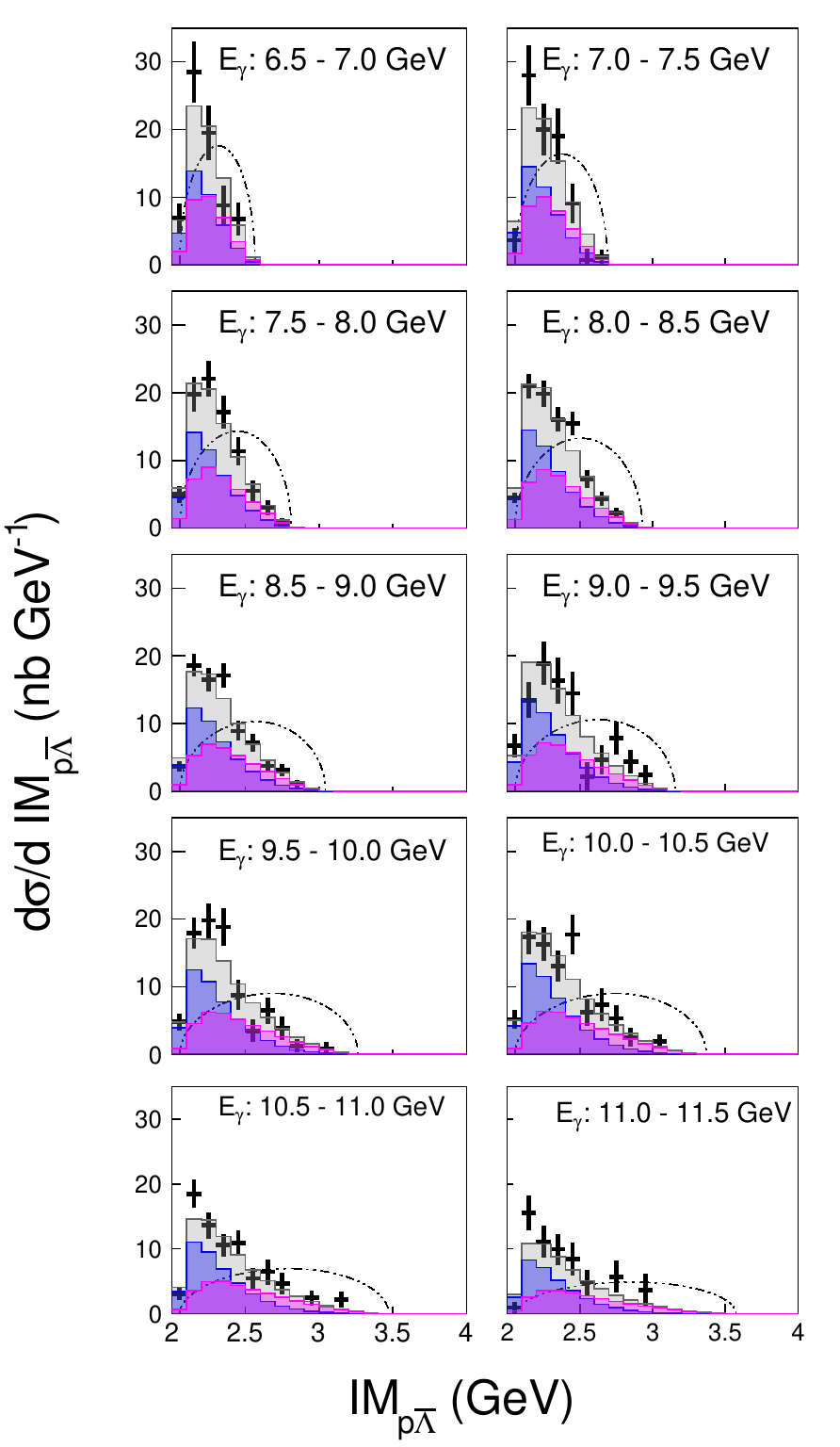}\\ 
        (c)
    \end{minipage}
    
    \caption{Differential cross sections with respect to the invariant mass of the baryon-antibaryon systems for:
    (a) $\pbarp$ pairs,
    (b) $\lamlambar$ pairs, and
    (c) $\plambar$ pairs, binned by photon beam energy. 
    Data points (black markers) are shown with statistical uncertainties only.
    Fits from the model defined in Sec.~\ref{sec:fullreactionmodels} are shown for single and double exchanges. 
    The component color scheme is the same as for the excerpted energy bins in Fig.~\ref{fig:differential_IM}.
    Phase space distribution areas (dot-dashed curves) are set equal to the clustering model curve areas for comparison. 
    \label{fig:diff_IM} }
\end{sidewaysfigure}

Selected results were shown in Figs.~\ref{fig:differential_tprime} and \ref{fig:differential_IM}. Plots of the complete set of results, across all beam energies, are included here in Figs.~\ref{fig:difftprime} and \ref{fig:diff_IM}.

\onecolumngrid

\twocolumngrid

\clearpage
\section{Numerical results}
\label{appendix:numericalresults}

Selected tabulated results for these measurements are included here.  Machine-readable files of these numbers and the full set of differential cross sections are available from the contact authors or from the archive Ref.~\cite{dataarchive}.
The estimated global systematic uncertainties are discussed in Sec.~\ref{sec:systematics}.


\begin{table}[htpb]
\begin{minipage}{\linewidth}
\caption{\label{tab:totalcrosssections_pbarp}%
Total cross section results as a function of photon beam energy for the $\gamma p \rightarrow \{\pbarp\} p$ reaction.  The uncertainties are statistical only.
}
\begin{ruledtabular}
\begin{tabular}{ccc}
\textrm{Beam Energy (GeV)}&
\textrm{Cross Section (nb)}&
\textrm{Uncertainty (nb)}\\
\colrule
   3.8 -    3.9 &        0.2 &        0.2 \\
   3.9 -    4.0 &        1.7 &        0.4 \\
   4.1 -    4.2 &        5.5 &        2.0 \\
   4.2 -    4.3 &        9.8 &        1.5 \\
   4.3 -    4.4 &       13.3 &        1.2 \\
   4.4 -    4.5 &       21.2 &        1.5 \\
   4.5 -    4.6 &       26.5 &        1.6 \\
   4.6 -    4.7 &       33.3 &        2.5 \\
   4.7 -    4.8 &       35.2 &        1.8 \\
   4.8 -    4.9 &       43.2 &        2.0 \\
   4.9 -    5.0 &       49.4 &        2.9 \\
   5.0 -    5.1 &       62.0 &        2.2 \\
   5.1 -    5.2 &       63.2 &        2.3 \\
   5.2 -    5.3 &       67.7 &        2.3 \\
   5.3 -    5.4 &       76.4 &        3.3 \\
   5.4 -    5.5 &       76.4 &        2.4 \\
   5.6 -    5.7 &       83.2 &        2.3 \\
   6.5 -    7.0 &       99.6 &        0.2 \\
   7.0 -    7.5 &      102.3 &        0.2 \\
   7.5 -    8.0 &      101.6 &        0.2 \\
   8.0 -    8.5 &       97.7 &        0.1 \\
   8.5 -    9.0 &       95.8 &        0.1 \\
   9.0 -    9.5 &       95.6 &        0.2 \\
   9.5 -   10.0 &       93.1 &        0.2 \\
  10.0 -   10.5 &       90.6 &        0.2 \\
  10.5 -   11.0 &       87.8 &        0.2 \\
  11.0 -   11.5 &       84.6 &        0.2 \\
\end{tabular}
\end{ruledtabular}
\end{minipage}
\end{table}

\begin{table}[htpb]
\begin{minipage}{\linewidth}
\caption{\label{tab:totalcrosssections_LambarLam}%
Total cross section results as a function of photon beam energy for the $\gamma p \rightarrow \{\LambarLam\} p$ reaction.  The uncertainties are statistical only.
}
\begin{ruledtabular}
\begin{tabular}{ccc}
\textrm{Beam Energy (GeV)}&
\textrm{Cross Section (nb)}&
\textrm{Uncertainty (nb)}\\
\colrule
   5.1 -    5.2 &        0.0 &        1.1 \\
   6.5 -    7.0 &        7.0 &        0.3 \\
   7.0 -    7.5 &        8.0 &        0.3 \\
   7.5 -    8.0 &        8.7 &        0.2 \\
   8.0 -    8.5 &        8.8 &        0.1 \\
   8.5 -    9.0 &        8.7 &        0.1 \\
   9.0 -    9.5 &        9.4 &        0.3 \\
   9.5 -   10.0 &        9.8 &        0.2 \\
  10.0 -   10.5 &        9.5 &        0.2 \\
  10.5 -   11.0 &        9.6 &        0.2 \\
  11.0 -   11.5 &        9.3 &        0.3 \\
\end{tabular}
\end{ruledtabular}
\end{minipage}
\par\vspace{0.2cm} 
\begin{minipage}{\linewidth}
\caption{\label{tab:totalcrosssections_pbarL}%
Total cross section results as a function of photon beam energy for the $\gamma p \rightarrow \{ \pbarL \} \Lambda$ reaction.  The uncertainties are statistical only.
}
\begin{ruledtabular}
\begin{tabular}{ccc}
\textrm{Beam Energy (GeV)}&
\textrm{Cross Section (nb)}&
\textrm{Uncertainty (nb)}\\
\colrule
   5.2 -    5.3 &        3.5 &        3.5 \\
   6.5 -    7.0 &        8.0 &        0.8 \\
   7.0 -    7.5 &        9.2 &        0.9 \\
   7.5 -    8.0 &        9.4 &        0.6 \\
   8.0 -    8.5 &        9.9 &        0.4 \\
   8.5 -    9.0 &        8.9 &        0.4 \\
   9.0 -    9.5 &       10.6 &        0.8 \\
   9.5 -   10.0 &       10.3 &        0.7 \\
  10.0 -   10.5 &       10.7 &        0.7 \\
  10.5 -   11.0 &        9.5 &        0.6 \\
  11.0 -   11.5 &        8.5 &        0.8 \\
\end{tabular}
\end{ruledtabular}
\end{minipage}
\end{table}

\FloatBarrier
\bibliography{b-anti-b_crosssections}
\end{document}

%% file: authors.tex
\affiliation{Polytechnic Sciences and Mathematics, School of Applied Sciences and Arts, Arizona State University, Tempe, Arizona 85287, USA}
\affiliation{Department of Physics, National and Kapodistrian University of Athens, 15771 Athens, Greece}
\affiliation{Ruhr-Universit\"{a}t-Bochum, Institut f\"{u}r Experimentalphysik, D-44801 Bochum, Germany}
\affiliation{Department of Physics, Carnegie Mellon University, Pittsburgh, Pennsylvania 15213, USA}
\affiliation{Department of Physics, The Catholic University of America, Washington, D.C. 20064, USA}
\affiliation{School of Mathematics and Physics, China University of Geosciences, Wuhan 430074, People’s Republic of China}
\affiliation{Department of Physics, University of Connecticut, Storrs, Connecticut 06269, USA}
\affiliation{Department of Physics, Duke University, Durham, North Carolina 27708, USA}
\affiliation{Department of Physics, Florida International University, Miami, Florida 33199, USA}
\affiliation{Department of Physics, Florida State University, Tallahassee, Florida 32306, USA}
\affiliation{Department of Physics, The George Washington University, Washington, D.C. 20052, USA}
\affiliation{Physikalisches Institut, Justus-Liebig-Universit\"{a}t Gie{\ss}en, D-35390 Gie{\ss}en, Germany}
\affiliation{School of Physics and Astronomy, University of Glasgow, Glasgow G12 8QQ, United Kingdom}
\affiliation{GSI Helmholtzzentrum f\"{u}r Schwerionenforschung GmbH, D-64291 Darmstadt, Germany}
\affiliation{Institute of High Energy Physics, Beijing 100049, People's Republic of China}
\affiliation{Department of Physics, Indiana University, Bloomington, Indiana 47405, USA}
\affiliation{National Research Centre Kurchatov Institute, Moscow 123182, Russia}
\affiliation{Department of Physics, Lamar University, Beaumont, Texas 77710, USA}
\affiliation{Department of Physics, University of Massachusetts, Amherst, Massachusetts 01003, USA}
\affiliation{National Research Nuclear University Moscow Engineering Physics Institute, Moscow 115409, Russia}
\affiliation{Department of Physics, Mount Allison University, Sackville, New Brunswick E4L 1E6, Canada}
\affiliation{Department of Physics, North Carolina A\&T State University, Greensboro, North Carolina 27411, USA}
\affiliation{Department of Physics and Physical Oceanography, University of North Carolina at Wilmington, Wilmington, North Carolina 28403, USA}
\affiliation{Department of Physics, Old Dominion University, Norfolk, Virginia 23529, USA}
\affiliation{Department of Physics, University of Regina, Regina, Saskatchewan S4S 0A2, Canada}
\affiliation{Department of Mathematics, Physics, and Computer Science, Springfield College, Springfield, Massachusetts, 01109, USA}
\affiliation{Thomas Jefferson National Accelerator Facility, Newport News, Virginia 23606, USA}
\affiliation{Laboratory of Particle Physics, Tomsk Polytechnic University, 634050 Tomsk, Russia}
\affiliation{Department of Physics, Tomsk State University, 634050 Tomsk, Russia}
\affiliation{Department of Physics and Astronomy, Union College, Schenectady, New York 12308, USA}
\affiliation{Department of Physics, Virginia Tech, Blacksburg, Virginia 24061, USA}
\affiliation{Department of Physics, Washington \& Jefferson College, Washington, Pennsylvania 15301, USA}
\affiliation{Department of Physics, William \& Mary, Williamsburg, Virginia 23185, USA}
\affiliation{School of Physics and Technology, Wuhan University, Wuhan, Hubei 430072, People's Republic of China}
\affiliation{A. I. Alikhanyan National Science Laboratory (Yerevan Physics Institute), 0036 Yerevan, Armenia}


\author{F.~Afzal\orcidlink{0000-0001-8063-6719}} \affiliation{Ruhr-Universit\"{a}t-Bochum, Institut f\"{u}r Experimentalphysik, D-44801 Bochum, Germany}
\author{M.~Albrecht\orcidlink{0000-0001-6180-4297}} \affiliation{Thomas Jefferson National Accelerator Facility, Newport News, Virginia 23606, USA}
\author{M.~Amaryan\orcidlink{0000-0002-5648-0256}} \affiliation{Department of Physics, Old Dominion University, Norfolk, Virginia 23529, USA}
\author{S.~Arrigo} \affiliation{Department of Physics, William \& Mary, Williamsburg, Virginia 23185, USA}
\author{V.~Arroyave} \affiliation{Department of Physics, Florida International University, Miami, Florida 33199, USA}
\author{A.~Asaturyan\orcidlink{0000-0002-8105-913X}} \affiliation{Thomas Jefferson National Accelerator Facility, Newport News, Virginia 23606, USA}
\author{A.~Austregesilo\orcidlink{0000-0002-9291-4429}} \affiliation{Thomas Jefferson National Accelerator Facility, Newport News, Virginia 23606, USA}
\author{Z.~Baldwin\orcidlink{0000-0002-8534-0922}} \affiliation{Department of Physics, Carnegie Mellon University, Pittsburgh, Pennsylvania 15213, USA}
\author{F.~Barbosa} \affiliation{Thomas Jefferson National Accelerator Facility, Newport News, Virginia 23606, USA}
\author{J.~Barlow\orcidlink{0000-0003-0865-0529}} \affiliation{Department of Physics, Florida State University, Tallahassee, Florida 32306, USA}\affiliation{Department of Mathematics, Physics, and Computer Science, Springfield College, Springfield, Massachusetts, 01109, USA}
% ...
\collaboration{The \textsc{GlueX} Collaboration}

%% file: p-anti-p_modeling_diagrams.tex
\renewcommand{\arraystretch}{1.5} 
\begin{table*}[hbtp]
    \centering
    \caption{\label{tab:all_reaction_parameters} Parameters of the components ($k=1,2,3$) which specify the combined model presented in Sec.~\ref{sec:reactionmodel} for each reaction channel: \pbarp, \LamantiLam, \pantiL. The uncertainties are estimated based on statistical modeling studies of the $\ppbar$ channel, discussed in Sec.~\ref{sec:optimizedparameters}. 
    }
    \begin{tabular}{|L|L|c|c|c|c|}
    \hline
    \multicolumn{3}{|c|}{\textbf{Combined Model}} & \multicolumn{3}{c|}{\textbf{Fitted Values}} \\
    \hline
    \textbf{Component } & \textbf{Parameter } &  \textbf{Description} & \textbf{\boldmath$\hat{\theta}(\pantip)$} & \textbf{\boldmath$\hat{\theta}(\LamantiLam)$} & \textbf{\boldmath$\hat{\theta}(\pantiL)$}  \\ 
    
    \hline
    \multirow{3}{*}{\shortstack[c]{~~Single Exchange \\ ~~~($k=1$) \\ ~~~(Eq.~\ref{eq:sR_modeling})}} 
     & $\theta_1$ ($\text{GeV}^{-2}$) & upper vertex slope $(\alpha^{\prime}$ in $d \sigma^2_{single}/dm_{12}dt^\prime)$)  & $.37\pm.02$ & $.40 \pm .02$ & $.20 \pm .02$\\ \cline{2-6} 
     & $\theta_2$ ($\text{GeV}$)      & clustering parameter              ($c_m$ in $d^2\sigma_{single}/dm_{12}dt^{\prime}$)      & $.21\pm.02$ & $.17 \pm .02$ & $.20 \pm .02$\\ \cline{2-6} 
     & $\theta_3$ ($\text{GeV}^2$)    & cutoff of $~t^{\prime}$ ($t_{\text{cutoff}}$ in $d^2\sigma_{single}/dm_{12}dt^{\prime}$) & $.11\pm.03$ & $.00 \pm .03$ & $.00 \pm .03$\\ 
     
     \hline
     \multirow{3}{*}{\shortstack[c]{Double Exchange-I \\ ($k=2$) \\ (Eq.~\ref{eq:dR_modeling})}} 
     & $\theta_4$ ($\text{GeV}^{-2}$)   & upper vertex slope ($\alpha_1^{\prime}$ in $d^2\sigma_{double}/t_1^{\prime}u_3^{\prime}$) & $.40\pm.20$  & $.40 \pm .20$  & $.40 \pm .20$\\ \cline{2-6} 
     & $\theta_5$ ($\text{GeV}^{-2}$)   & lower vertex slope     ($\alpha_3^{\prime}$ in $d^2\sigma_{double}/t_1^{\prime}u_3^{\prime}$) & $.14\pm.03$ & $.10 \pm .03$ & $.40 \pm .03$\\ \cline{2-6} 
     & $\theta_6$ ($\text{GeV}^2$)      & cutoff of $t^{\prime}_1$ ($t_{\text{1 cutoff}}$ in $d^2\sigma_{double}/t_1^{\prime}u_3^{\prime}$) & $.49\pm.20$  & $.00 \pm .20$  & $.50 \pm .20$\\ 

    \hline
    \multirow{3}{*}{\shortstack[c]{Double Exchange-II  \\ ($k=3$) \\ (Eq.~\ref{eq:dR_modeling})}} 
     & $\theta_7$ ($\text{GeV}^{-2}$)   & upper vertex slope ($\alpha_1^{\prime}$ in $d^2\sigma_{double}/t_1^{\prime}u_3^{\prime}$)  & $.80\pm.40$  & - & -\\ \cline{2-6} 
     & $\theta_8$ ($\text{GeV}^{-2}$)   & lower vertex slope     ($\alpha_3^{\prime}$ in $d^2\sigma_{double}/t_1^{\prime}u_3^{\prime}$)  & $1.60\pm.40$ & - & -\\ \cline{2-6} 
     & $\theta_9$ ($\text{GeV}^2$)      & cutoff of $t^{\prime}_1$ ($t_{\text{1 cutoff}}$ in $d^2\sigma_{double}/t_1^{\prime}u_3^{\prime}$)  & $.49\pm.20$  & - & -\\ \hline
    \end{tabular} 
\end{table*}
\renewcommand{\arraystretch}{1.0} 

%% file: baryonium_table.tex
\begin{table*}[htpb]
\caption{\label{tab:earlymeasurements}%
Selected early measurements of $p\antip$ production tabulated by year, including searches for narrow baryonium states. 
Counts of accumulated $p\antip$ pairs, resonance peak measurements, as well as width estimates are listed. A more comprehensive enumeration and discussion can be found in Ref~\cite{MONTANET1980201}.
Several unpublished studies by the CLAS Collaboration~\cite{  
bstokesPhD2006, wphelpsPhD2017} and \cite{Phelps:2016huv} with negative results are not included here.}
{\small
  \begin{tabular}{lllll}
    \toprule
    Experiment (year) & \# of $\ppbar$ & Mass (MeV)~~ & Width (MeV)~~ & Comment \\
    \midrule
    \midrule
    CERN (1977) \cite{BENKHEIRI1977483} & 6000 & $2020\pm3$ & $24\pm12$ & $\pi^-p\rightarrow p_{\text{fwd}} p\bar{p}\pi^-$ \\
     &   & $2204\pm5$  & $16^{+20}_{-16}$  &  \\
    \midrule
    Cornell (1979) \cite{PhysRevLett.42.1593} & 65 &  &  &Virtual photoproduction ($ep\rightarrow e p p \bar{p}$) \\
       &                                & 2020 & $<40$ & Signal cross section $6.6\pm2.2$ nb   \\
       &                                & 2200 & 60 & Signal cross section $5\pm2.5$ nb   \\
    \midrule
    LAMP2 (1979) \cite{Barber:1979ah} & 137 & - & - & Photoproduction; No evidence \\
    \midrule
                                       &     &            &           &  Photoproduction \\
    DESY(1984) \cite{Bodenkamp:1984dg} & 200 & $1939\pm6$ & $52\pm16$ &  Signal cross section $21\pm4$ nb\\
                                       &     & $2024\pm5$ & $29\pm13$ &  Signal cross section $14\pm5$ nb\\
    \midrule
    GlueX (this work)  & 10 million & - & - & Photoproduction; No evidence  \\
    \bottomrule
  \end{tabular}
}
\end{table*}